\documentclass[aps, pre, groupedaddress, twocolumn]{revtex4-1}
\usepackage{graphicx}
\usepackage{tabularx}
\newcolumntype{Y}{>{\centering\arraybackslash}X}
\pdfoutput=1 
\usepackage{amsmath}
\usepackage{natbib}

\usepackage{color}
\usepackage{subfigure}
\usepackage{url}
\usepackage{verbatim}
\usepackage{lipsum}
\usepackage{braket}
\usepackage{float}
\usepackage{soul,xcolor}

\begin{document}
\setstcolor{red}

\title{2D Implosion Simulations with a Kinetic Particle Code}
\author{I. Sagert$^{1,2,3}$, W. P. Even$^{2,3,4}$, T. T. Strother$^{1,3}$}
\affiliation{$^1$XTD-IDA, Los Alamos National Laboratory, Los Alamos, New Mexico, 87545, USA\\
$^2$CCS-2, Los Alamos National Laboratory, Los Alamos, New Mexico, 87545, USA\\
$^3$Center for Theoretical Astrophysics, Los Alamos National Laboratory, Los Alamos, New Mexico, 87545, USA\\
$^4$Department of Physical Science, Southern Utah University, Cedar City, Utah, 84720, USA}
\date{\today}
\pacs{}
\begin{abstract}
We perform two-dimensional (2D) implosion simulations using a Monte Carlo kinetic particle code. The application of a kinetic transport code is motivated, in part, by the occurrence of non-equilibrium effects in inertial confinement fusion (ICF) capsule implosions, which cannot be fully captured by hydrodynamic simulations. Kinetic methods, on the other hand, are able to describe both, continuum and rarefied flows. We perform simple 2D disk implosion simulations using one particle species and compare the results to simulations with the hydrodynamics code \textsc{RAGE}. The impact of the particle mean-free-path on the implosion is also explored. In a second study, we focus on the formation of fluid instabilities from induced perturbations. We find good agreement with hydrodynamic studies regarding the location of the shock and the implosion dynamics. Differences are found in the evolution of fluid instabilities, originating from the higher resolution of \textsc{RAGE} and statistical noise in the kinetic studies.      
\end{abstract}
\maketitle
\section{Introduction}
\label{intro}
Astrophysics and laboratory plasma physics problems often contain flows at different Knudsen numbers $K$. The latter can be defined as the ratio of the interaction mean-free-path $\lambda$ of particles in the system and a problem-specific hydrodynamic length scale. For $K \leq 1.0$, the dynamical evolution of a system can be described through fluid dynamics equations \cite{Burnett36, Chapman70, Foch73, Grad49}. Transport equations, like the Boltzmann equation, are applied for larger $K$ or for flows containing components with small and large Knudsen numbers. They describe the change in the phase space density function of a system due to forces and particle interactions. One approach to solve them numerically is by kinetic particle methods, such as Direct Simulation Monte Carlo (DSMC) \cite{Bird2013} and Particle-in-Cell (PIC) \cite{Hockney88, Birdsall04,Bowers09}. Here, $N$ simulation particles approximate the density function in position and momentum space \cite{Wong82}. The particles move according to their $6N$ coupled equations of motion:
\begin{align}
\frac{d \vec{x}_i}{d t} = \frac{\vec{p}_i}{m_i}, \:\: \frac{d \vec{p}_i }{d t} = -\vec{\nabla} U , \:\: i = 1 \ldots N ,
\end{align}
where $U$ is a mean-field potential. The particle interaction depends on the collisional integral and can be e.g. elastic scattering (DSMC) or Coulomb collisions (PIC). Despite their computational cost (for small $K$, the number of particle interactions is large while time scales become small), particle methods are attractive tools as they can model non-equilibrium and continuum flows \cite{Gallis17, Gallis04}.\\
Our code is such a kinetic particle code. Its development is motivated by flows that are transient and move between small and large $K$ or contain multiple interacting components with different Knudsen numbers. One possible future application lies in inertial confinement fusion (ICF) capsule implosion studies \cite{Craxton15, Betti16, ICF}. Although the implosion dynamics is governed by hydrodynamic phenomena, the fusion fuel ions (deuterium and tritium (D/T)) can have large mean-free-paths, which leads to kinetic effects and might impact ignition  \cite{Amendt11, Molvig12, Bellei13, Rosenberg14, Kagan15, Hsu16, Goldstein_2012}. At present, our code does not have ICF capabilities since the required physics input (e.g. electric field, electrons, and physical cross-sections) is yet to be implemented. The purpose of the current work is to test our code in simple implosion scenarios (beyond the standard shock wave and fluid instability simulations \cite{Sagert14, Sagert15}) and to compare with results from hydrodynamics codes. The majority of studies are therefore performed close to the continuum regime but we also test large values of $\lambda$. All simulations are done in 2D. Although our code can run in 3D, small time steps and long implosion timescales require an efficient distributed memory parallelization scheme which has not been implemented yet \cite{Howell2015}. Since there are no analytic solutions to the implosion problems in this work we compare our results to the Radiation Adaptive Grid Eulerian hydrodynamics code \textsc{RAGE} \cite{Gittings08}.\\  
In the following, we give a brief description of our code in Sect.\:\ref{kinetic_code} together with a rough overview of ICF implosions to motivate our simulation setups. The first studies explore imploding disks with homogeneous initial density (2-zone simulations) in Sect.\:\ref{simple_implosion}. These include comparisons between \textsc{RAGE} and the kinetic code in the continuum limit and tests on the impact of different particle numbers, mean-free-paths and resolution. An analysis of non-equilibrium phenomena is also performed. In Sect.\:\ref{implosion_perturbation}, we study fluid instabilities in disks with different density layers (3-zone simulations). These calculations are done with RAGE and the kinetic code in the continuum limit. A summary is given in Sec.\:\ref{summary}.
\section{Code and Simulation Overview}
\label{kinetic_code}
Our current code is similar to DSMC methods, the main difference lying in the way we pick interaction partners. In traditional DSMC, particles are sorted into spatial cells and those within the same cell are randomly selected as scattering partners. This usually leads to a dependence between the particle mean-free-path $\lambda$ and the cell size $\Delta x$. Typically, both have to be of the same size. In our code, we search for interaction partners by calculating the point of closes approach (PoCA) \cite{Bertsch88, Bonasera94}, which decouples $\lambda$ and $\Delta x$. This scattering partner search adds to the computational time, however it can improve spatial resolution and reduce acausality effects in relativistic simulations.\\
In the current study, particles undergo elastic collisions. Since these are short-range interactions, we can sort the particles into a spatial grid and select collision partners from the same or neighboring cells only. We use a cartesian grid for collision partner search that is equidistant and fixed in size. Initially, we set the cell size $\Delta x$ so that a cell contains several particles and the calculation is computationally efficient. Furthermore, $\Delta x$ together with the maximum particle velocity $v_\mathrm{max}$ defines the time step via $\Delta t = \Delta x / v_\mathrm{max}$. Once interactions partners are identified, the collision is performed by choosing the directions of the outgoing particle velocities randomly in the center-of-mass frame. For elastic collisions, $\lambda$ is an input parameter of our simulations. As for hard spheres, it is connected to the 2D effective particle radii via $r_\mathrm{eff, 2D} = \left(2 \: \lambda \: n \right)^{-1} $ where $n$ is the number density. To reach a regime that is close to the continuum, $\lambda$ should be minimized. However, due the finite number of particles per grid cell $N_\mathrm{cell}$, $\lambda$ has a smallest possible value of $\lambda_\mathrm{min} \sim 2 \: \sqrt{ \pi / (4 N_\mathrm{cell})} \: \Delta x$, which results in a finite viscosity and diffusivity \cite{Sagert15}. Both can be reduced by increasing the value of $N_\mathrm{cell}$ or decreasing $\Delta x$. This generally requires a large total particle number and we typically use $N \sim 10^7 - 10^8$. To ensure that a system is modeled as close to the continuum limit as possible (even for varying values of $N_\mathrm{cell}$) we usually set $\lambda$ to a very small value, e.g. $10^{-5}\:\Delta x$. Although this can result in $r_\mathrm{eff} \gg \Delta x$, we still only consider particles in the neighboring cells for interactions. In the continuum limit, we usually analyze thermodynamic variables, such as density and pressure, which we determine as average properties per output grid cell \cite{Sagert14}.\\
\newline
In ICF capsule implosions, a capsule can generally be divided into three regions \cite{Craxton15}. The inner most region typically contains a mixture of D/T gas that is surrounded by a dense shell of D/T ice. The outer region is the ablator. As the capsule is irradiated by lasers or X-rays, material from the ablator expands outwards forcing the remaining matter to move inwards to conserve momentum (rocket effect). This initializes the capsule implosion and launches shock waves that propagate toward the capsule center, rebound on themselves and interact with the converging cold D/T ice shell. The interaction decelerates and halts the shell while the enclosed compressed D/T gas reaches high temperatures. A hotspot is created in the center, surrounded by cold D/T fuel; it becomes the starting point for fusion reactions. Fluid instabilities play an important role in ICF capsule implosion studies. They generally have a negative impact on the fusion yield as they mix cold and hot fuel and induce deformations of the capsule \cite{Regan12, Ma13}. The main instabilities that arise in ICF are Rayleigh-Taylor (RTIs) and Richtmyer-Meshkov instabilities (RMIs). The first are caused by opposite density and pressure gradients when two fluids of different densities are accelerated into each other \cite{Rayleigh82, Taylor50}. RMIs arise due to the passage of a shock wave through the interface of two fluids \cite{Richtmyer60,Meshkov69}. So-called Kelvin-Helmholtz instabilities (KHIs) are created by a velocity difference across a fluid interface \cite{Chandra61, Chou98}. They can form at the edges of RTIs and RMIs.\\
In the past, we simulated simple converging and blast problems \cite{Sagert14} and studied 2D single-mode RTIs, all with good agreement with analytic predictions. However, numerical studies of implosions are especially challenging: As the initial configuration converges, small perturbations can grow into large instabilities and shocks can become affected by non-physical grid effects \cite{Gunther02, Hopkins15}. At the center of the simulation space, grid-based methods have to increase the resolution in order to capture small structures while kinetic approaches have to deal with an increasing number of particles. In the current work, we therefore extend our previous studies of standard shocks and fluid instabilities to implosion scenarios. We also perform a check whether our code can evolve 2D KHIs in the continuum limit (see appendix \ref{kelvinhelmholtz}).
\section{2-zone Implosion Simulations}
\label{simple_implosion}
\subsection{Simulation Setup and Shock Launch}
\begin{figure}
\begin{center}
\includegraphics[width = 0.47\textwidth]{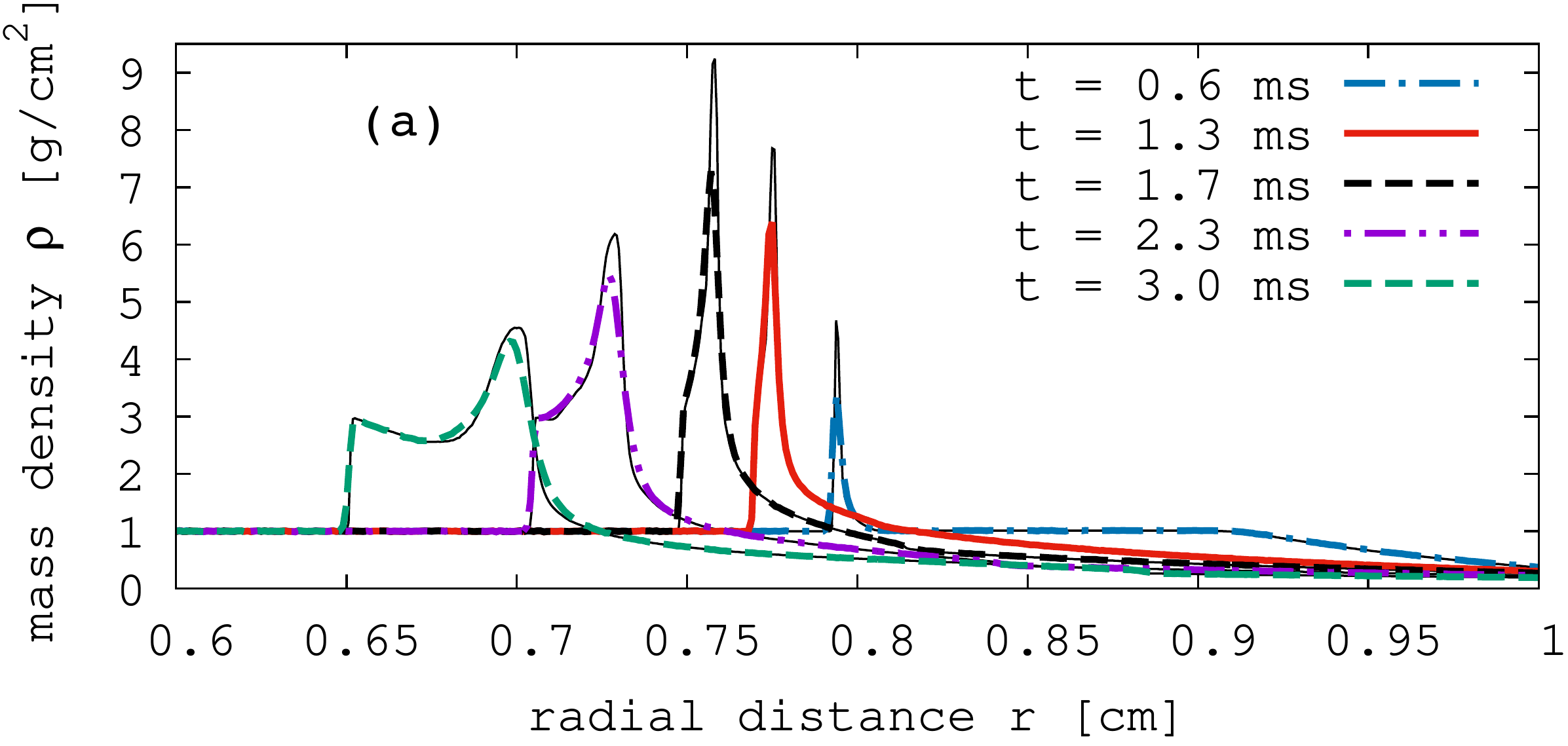}
\includegraphics[width = 0.47\textwidth]{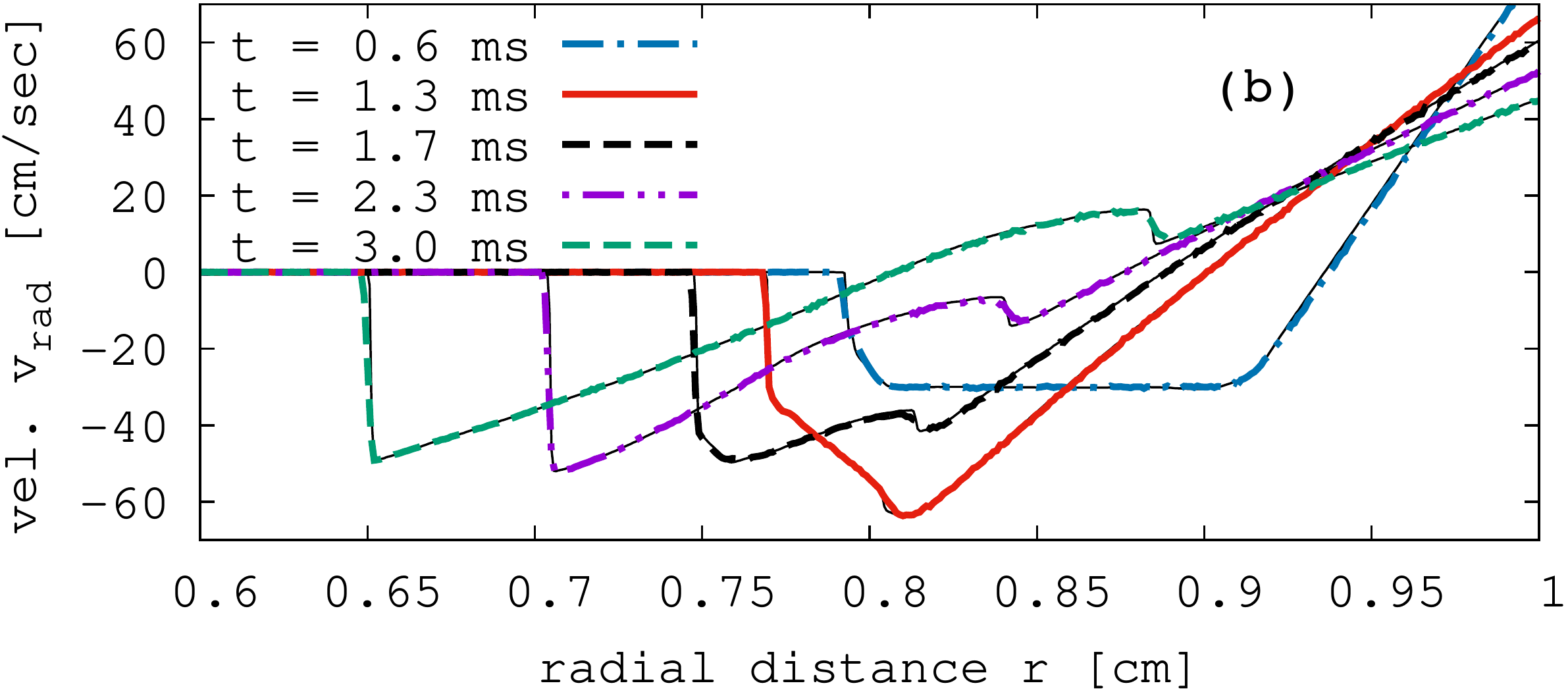}
\includegraphics[width = 0.47\textwidth]{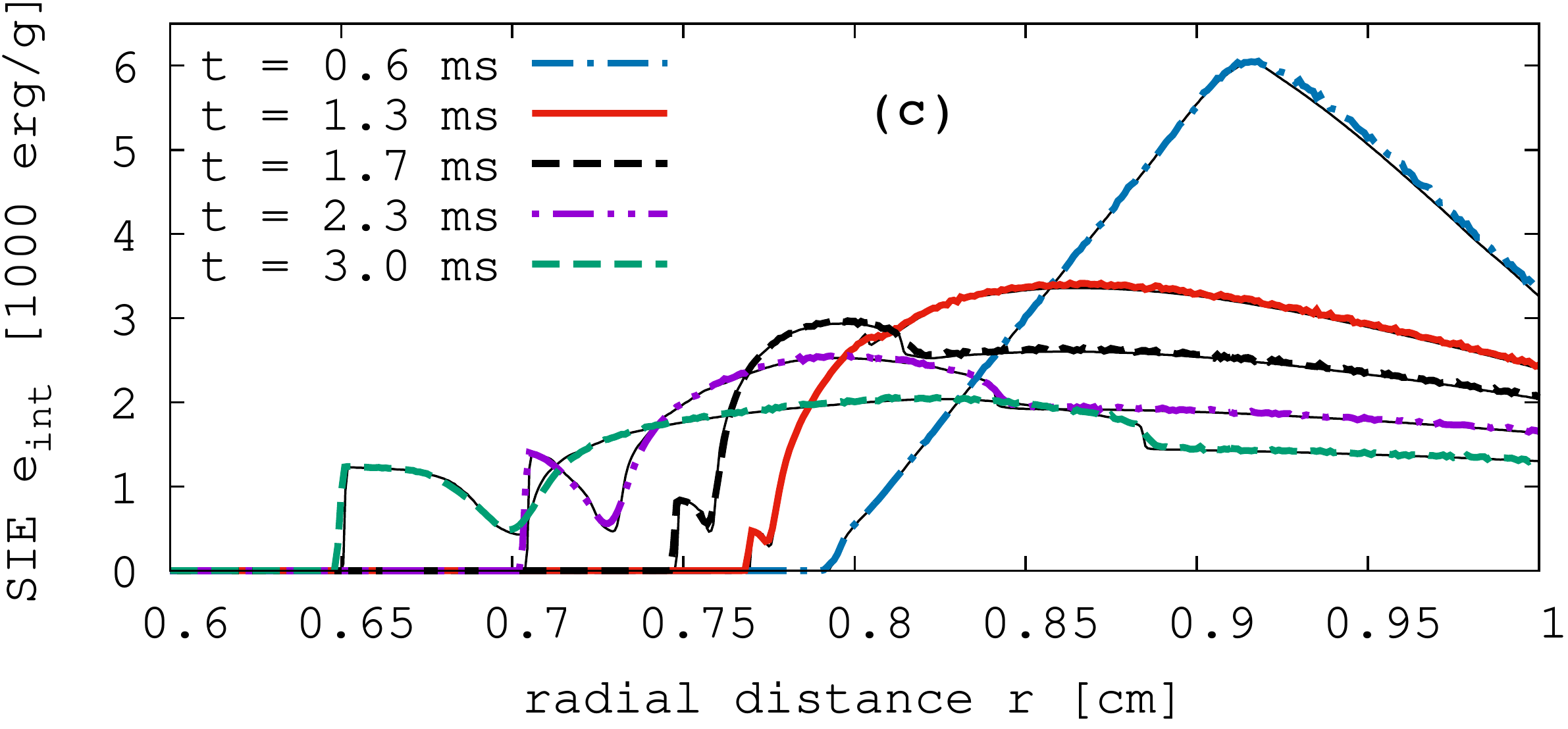}
\caption{Mass density $\rho$ (a), radial velocity $v_\mathrm{rad}$ (b), and SIE $e_\mathrm{int}$ (c) radial profiles for Kinetic-Q-60 (colored thick lines) and hr-RAGE (thin solid black lines) at early times.}
\label{density_shock_launch}
\end{center}
\end{figure}

Here, we discuss 2D implosion simulations following the setup in Garc{\'{\i}}a-Senz et al.\:\cite{Garcia09}. A disk with radius $r_2 = 1.0\:$cm and homogeneous density $\rho = 1\:\mathrm{g/cm^2}$ is divided into two zones. Zone-1 extends up to a radial distance of $r_1 = 0.8\:$cm with specific internal energy (SIE) of $e_\mathrm{int} = 1\:$erg/g everywhere. Zone-2 lies between $r_1$ and $r_2$. Its SIE increases linearly from $e_\mathrm{int} (r_1) = 1\:$erg/g to $e_\mathrm{int} (r_2) = 10^4 \:$erg/g. The energy deposition is instantaneous at $t = 0$ and the resulting surface ablation and rocket effect compress the disk and launch a shock wave. The shock converges towards the center, increasing the density and temperature of matter, rebounds on itself and propagates again outwards. In all simulations of this paper, an imploding 2D disk can be understood as a slice of an infinite cylinder with the simulated spatial domain lying perpendicular to the cylinder axis. This is of course different from imploding spherical ICF capsules and the main motivation to use this geometry is to do a direct comparison with the hydrodynamic studies in \cite{Garcia09}. Anticipated differences to spherical setups are implosion time scales and details in the compression, e.g. the evolution of the density (for a comparison between cylindrical and spherical implosion simulations see \cite{Joggerst2014}).\\ 
We perform a high-resolution \textsc{RAGE} simulation with $1000 \times 1000$ grid points and two levels of refinement (hr-RAGE). A low-resolution version uses $400 \times 400$ grid points and no refinement (lr-RAGE). The simulation space has a size of $0\:\mathrm{cm} \leq x,y \leq  5 \: \mathrm{cm}$ and contains one quarter of the disk. For the kinetic studies, we test different particle numbers and resolutions, summarized in Table \ref{simulation_parameters}.
\begin{table*}
\centering
\begin{tabular}{ | l | c | c | c | r | }                    
\hline
Simulation & Particle number &  Calculation bins &  Output bins &  Domain size \\
\hline 
\hline 
Kinetic-Q-60      & $6 \times 10^7$    & $6000 \times 6000$  & $2000 \times 2000$ & $0 \:\mathrm{cm} \leq x,y \leq 2\:\mathrm{cm}$  \\
Kinetic-F-100    & $1 \times 10^8$    & $8000 \times 8000$  & $2000 \times 2000$  & $-2\:\mathrm{cm} \leq x,y \leq 2\:\mathrm{cm}$  \\
Kinetic-F-20      & $2 \times 10^7$    & $4000 \times 4000$  & $500 \times 500$      & $-2\:\mathrm{cm} \leq x,y \leq 2\:\mathrm{cm}$  \\
\hline  
\hline
\end{tabular}
\caption{Parameters of the kinetic simulations in the 2-zone implosion setup.}
\label{simulation_parameters}
\end{table*}
The highest resolution is achieved with $6 \times 10^7$ particles, simulating one quarter of the disk with $0\:\mathrm{cm} \leq x,y \leq 2\:\mathrm{cm}$ (Kinetic-Q-60). Boundary conditions are reflective for $x,y < 0\:$cm while particles with coordinates $x,y > 2\:$cm are ignored. We will refer to the latter as \textit{free} boundary conditions. 
\begin{figure}
\begin{center}
\includegraphics[width = 0.45\textwidth]{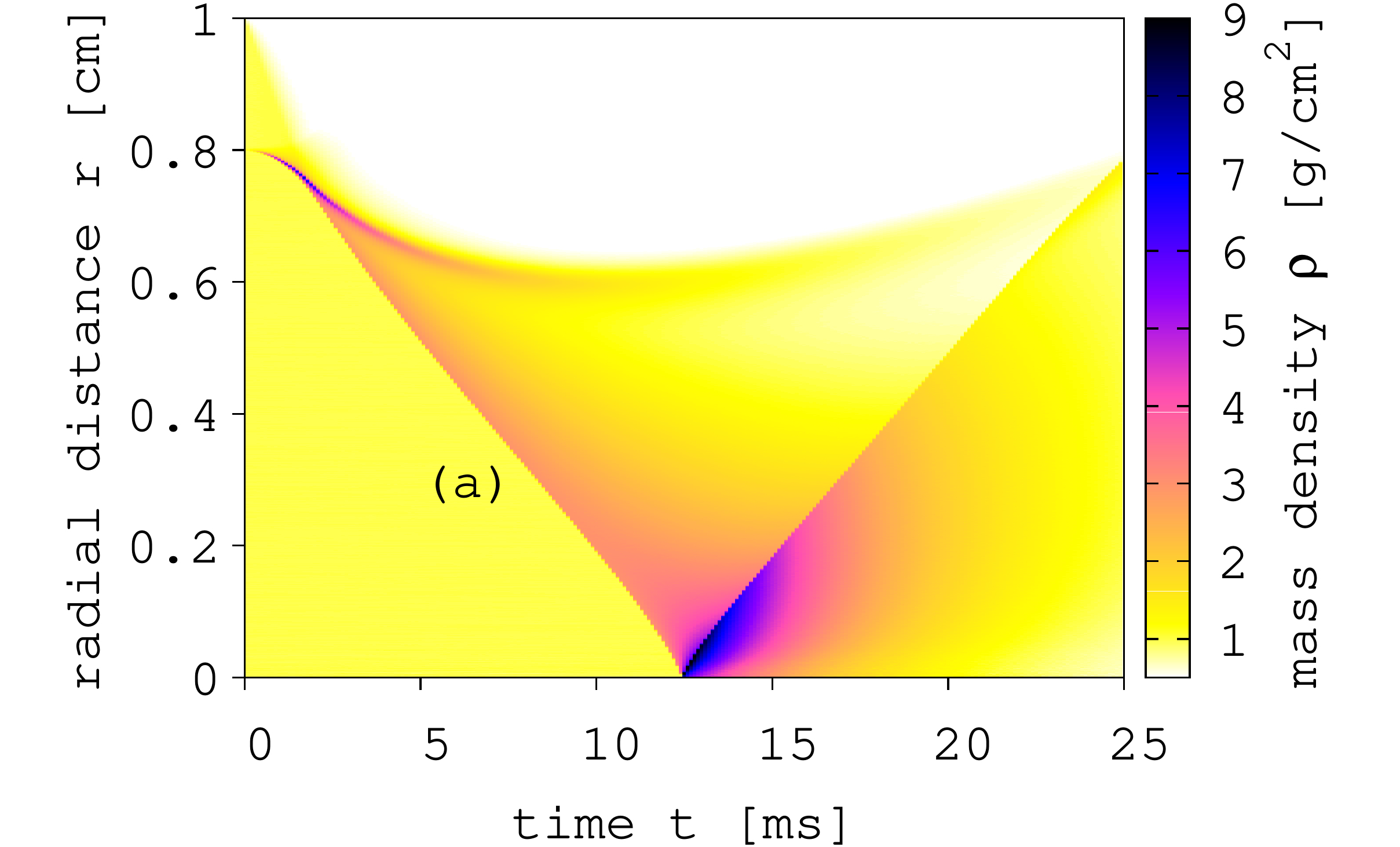}\hfill
\includegraphics[width = 0.45\textwidth]{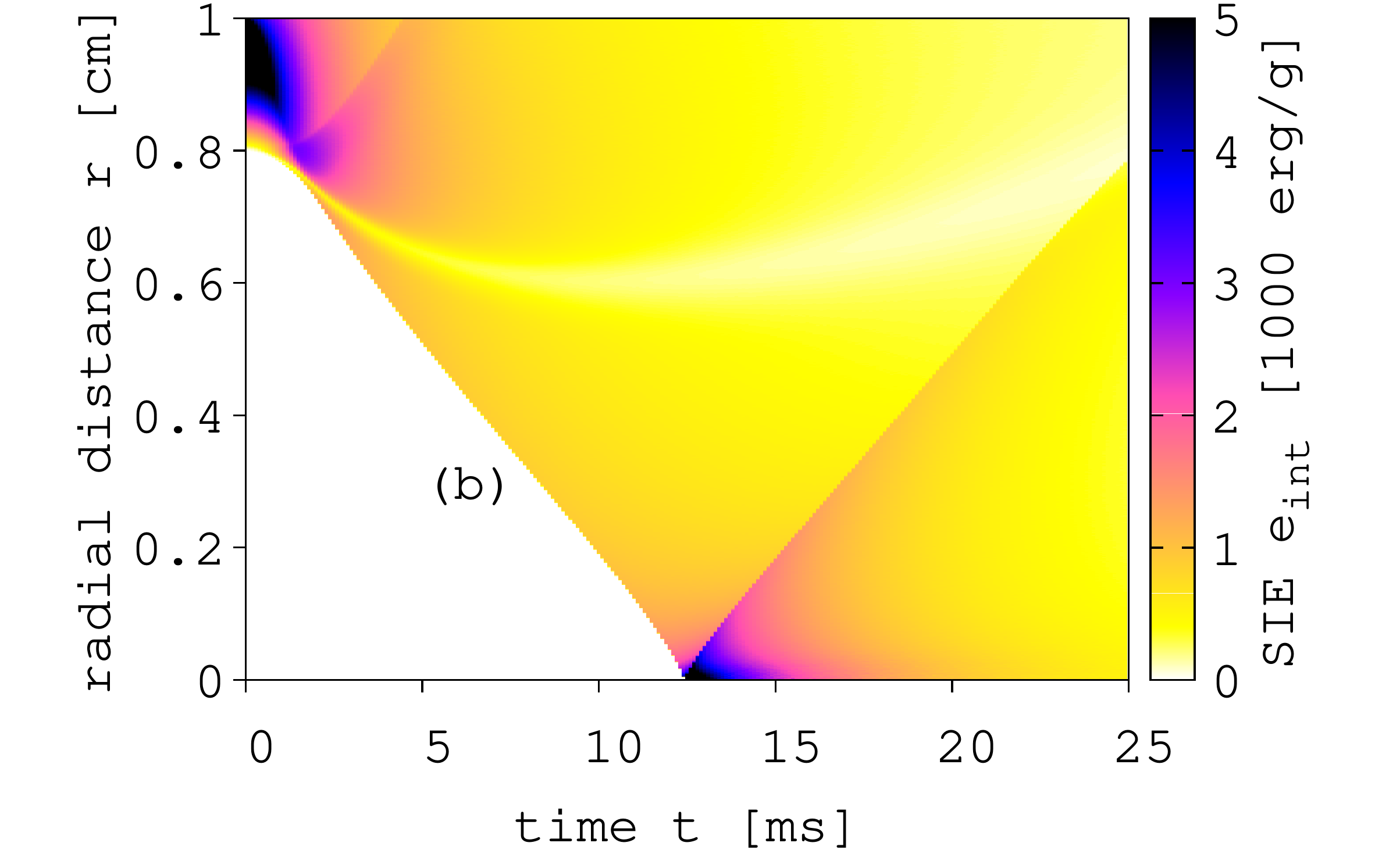}\hfill
\caption{Time evolution of the mass density $\rho$ (a) and SIE $e_\mathrm{int}$ (b) radial profiles for Kinetic-Q-60.}
\label{time_s}
\end{center}
\end{figure}
The other kinetic simulations evolve the full disk with $-2\:\mathrm{cm} \leq x,y \leq 2\:\mathrm{cm}$ and free boundary conditions everywhere. Kinetic-F-100 uses $10^8$ particles in $8000 \times 8000$ grid cells and Kinetic-F-20 has $2 \times 10^7$ particles in $4000 \times 4000$ cells. Kinetic-F-100 can be regarded as a simulation with intermediate resolution while Kinetic-F-20 has low resolution. To achieve a regime that is close to the continuum limit, we set $\lambda = 10^{-5} \: \Delta x$. The implosion is evolved over $25\:$ms, which marks the point when the rebound shock wave reaches $\sim r_0$. In both, \textsc{RAGE} and the kinetic simulations, matter is an ideal gas with an adiabatic index of $\gamma = 2$, i.e. two velocity degrees of freedom ($f=2$). With the internal energy $E_\mathrm{int} = f N \: k_b T/2$, the SIE is given by the root-mean-square velocity $v_\mathrm{rms}$:
\begin{align}
e_\mathrm{int} = E_\mathrm{int} / (N \: m) = N\: k_b T/(N\: m) =  v^2_\mathrm{rms} / 2, 
\label{eint1}
\end{align}
where the temperature is $k_b  T = m \: v^2_\mathrm{rms} / 2$. Particles are initialized with equal masses $m$ while their velocities are determined from a 2D Maxwell-Boltzmann (MB) distribution using $v_\mathrm{rms}$. The pressure can be calculated either from the ideal gas equation of state:
\begin{align}
P = e_\mathrm{int} \: \rho \:  ( \gamma - 1)
\label{eint1}
\end{align}
or the stress tensor \cite{Irving50, Mulero08}:
\begin{align}
P  = - \frac{1}{f A}  \sum_{i=1}^N m_i \: \left( \vec{v}_i - \vec{v}_{b} \right)^2 .
\label{stress0}
\end{align}
The above sum runs over all particles $i$ with velocities $\vec{v}_i$, while $\vec{v}_b$ is the bulk velocity. Fig.\:\ref{time_s}\: shows the time evolution of the mass density and SIE radial averages. The shock launches at the interface of zone-1 and 2, converging to the center at $t \sim 12.5\:$ms. As it propagates inward (and outward after the reflection) the shock leads to a strong compression and heating of matter.\\ 
The details of the shock launch are given in Fig.\:\ref{density_shock_launch} via mass density, radial velocity and SIE profiles for $t \leq 3.0\:$ms. Results are taken from Kinetic-Q-60 and hr-RAGE. At $t \sim 0.6\:$ms, high-energy particles are leaving the disk surface causing a rocket effect. Matter in zone-2 start to move inwards with a homogeneous but steadily increasing radial velocity (see Fig.\:\ref{density_shock_launch}(b)). As particles converge onto the stationary matter at the interface with zone-1, a density peak forms around $r_1$ and moves inwards. By $t \sim 1.3\:$ms, the velocity plateau has steepened into a maximum that sits right behind the density peak. At this point, as Garc{\'{\i}}a-Senz et al.\:\cite{Garcia09}, we observe the formation of two shock waves - one that is moving outwards and one that is moving inwards. The first can be seen in Fig.\:\ref{density_shock_launch}(b) at $r \sim 0.81\:$cm for $t = 1.7\:$ms. It eventually leaves the simulation space. At the same time, the inward shock wave ($r \sim 0.75\:$cm) passes through the compressed matter. 
Its breakout leads to a double peak structure in the density which can be seen for $t \geq 2.3\:$ms in Fig.\:\ref{density_shock_launch}(a). The first peak (at smaller $r$) is the shock wave while the second one is the remaining compressed matter.  The higher resolution in \textsc{RAGE} is clearly visible via the larger peak densities and sharper shock profiles. However, in general, the agreement between Kinetic-Q-60 and hr-RAGE is very good.
\subsection{Fluid Instabilities and Implosion Symmetry}
\begin{figure*}
\begin{center}
\includegraphics[width = 0.47\textwidth]{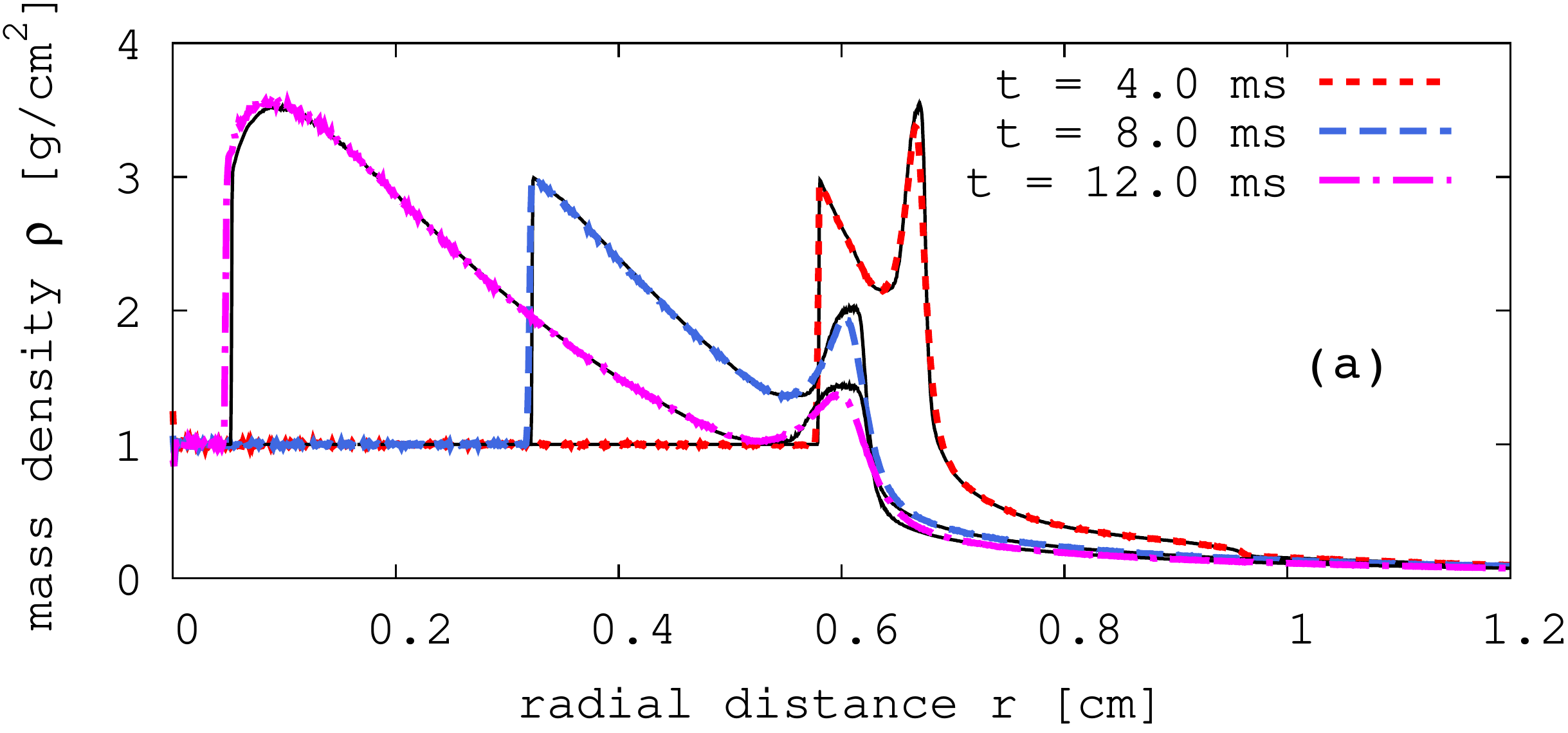}\hfill
\includegraphics[width = 0.47\textwidth]{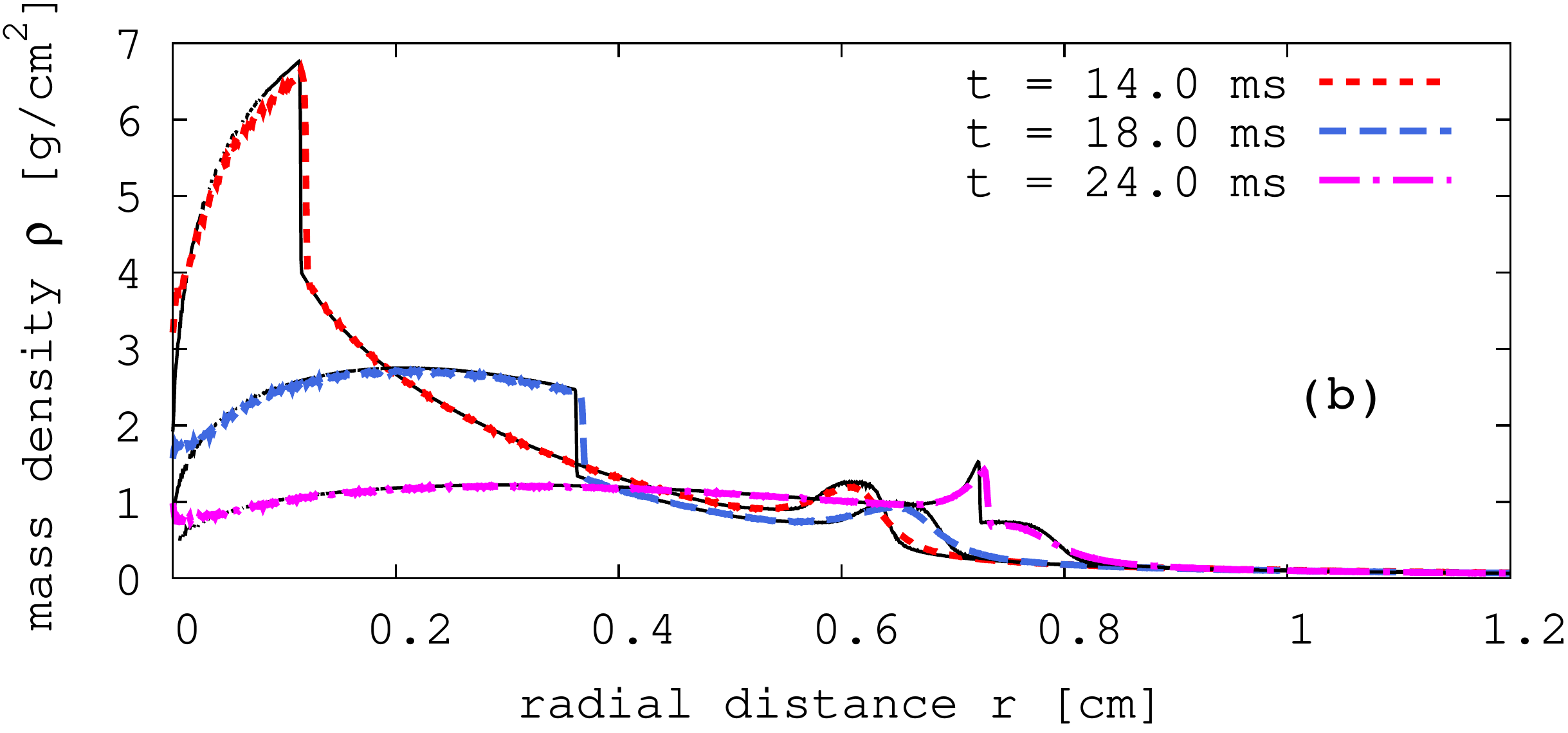}\hfill
\includegraphics[width = 0.47\textwidth]{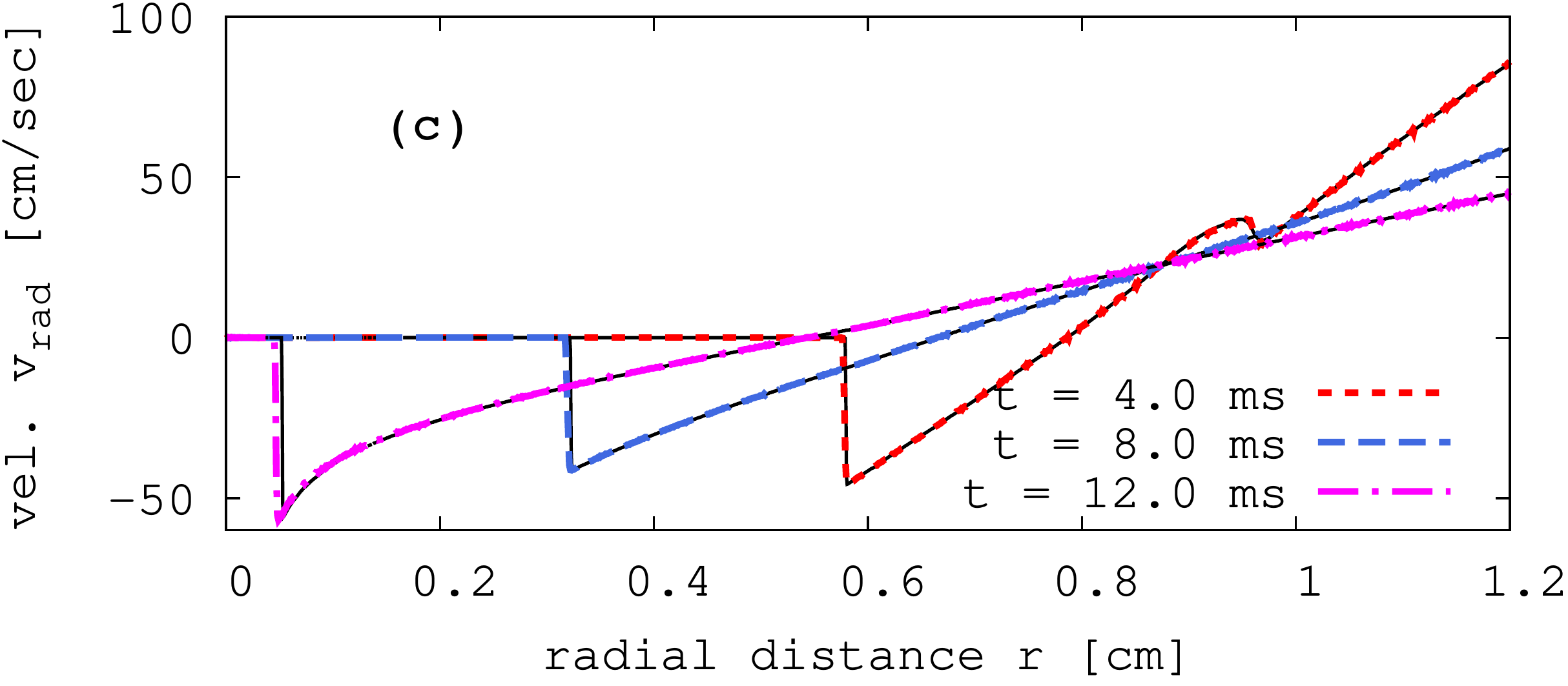}\hfill
\includegraphics[width = 0.47\textwidth]{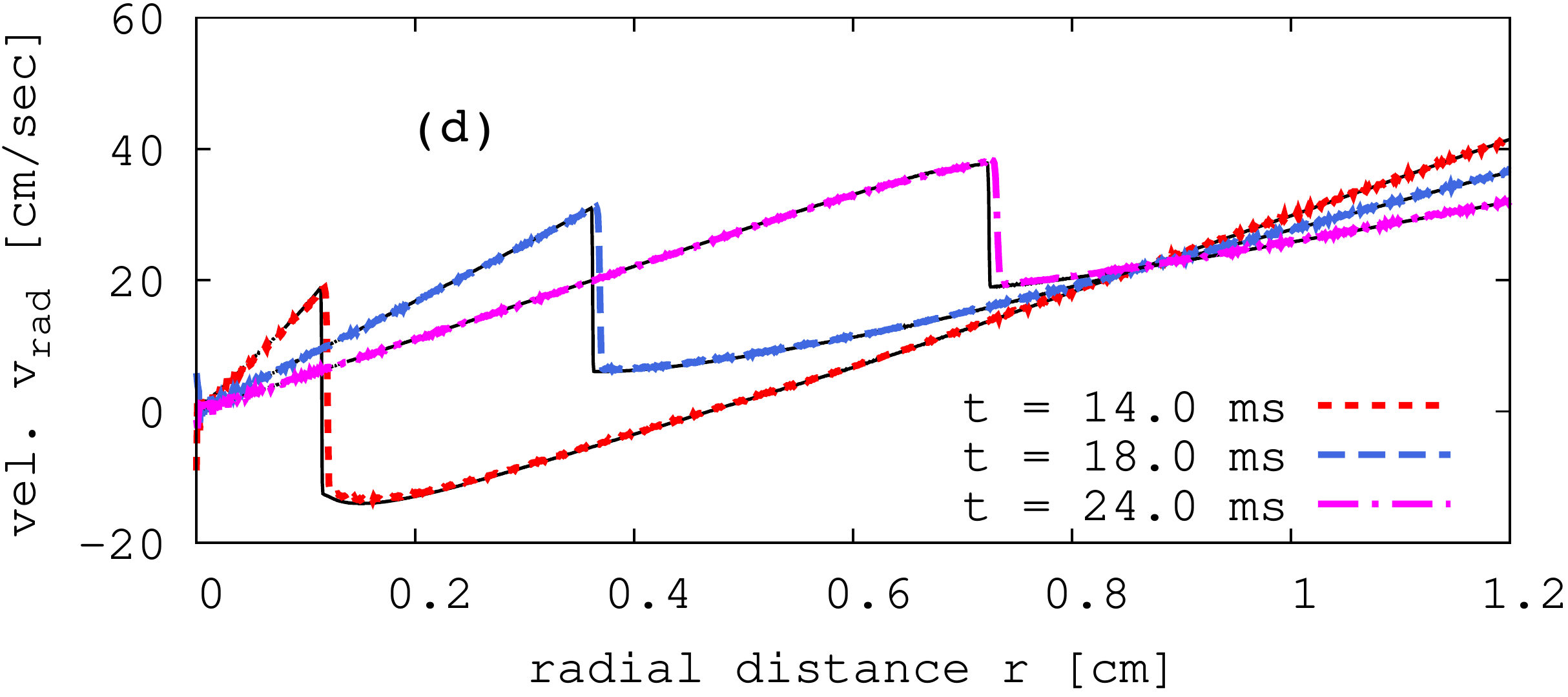}\hfill
\includegraphics[width = 0.47\textwidth]{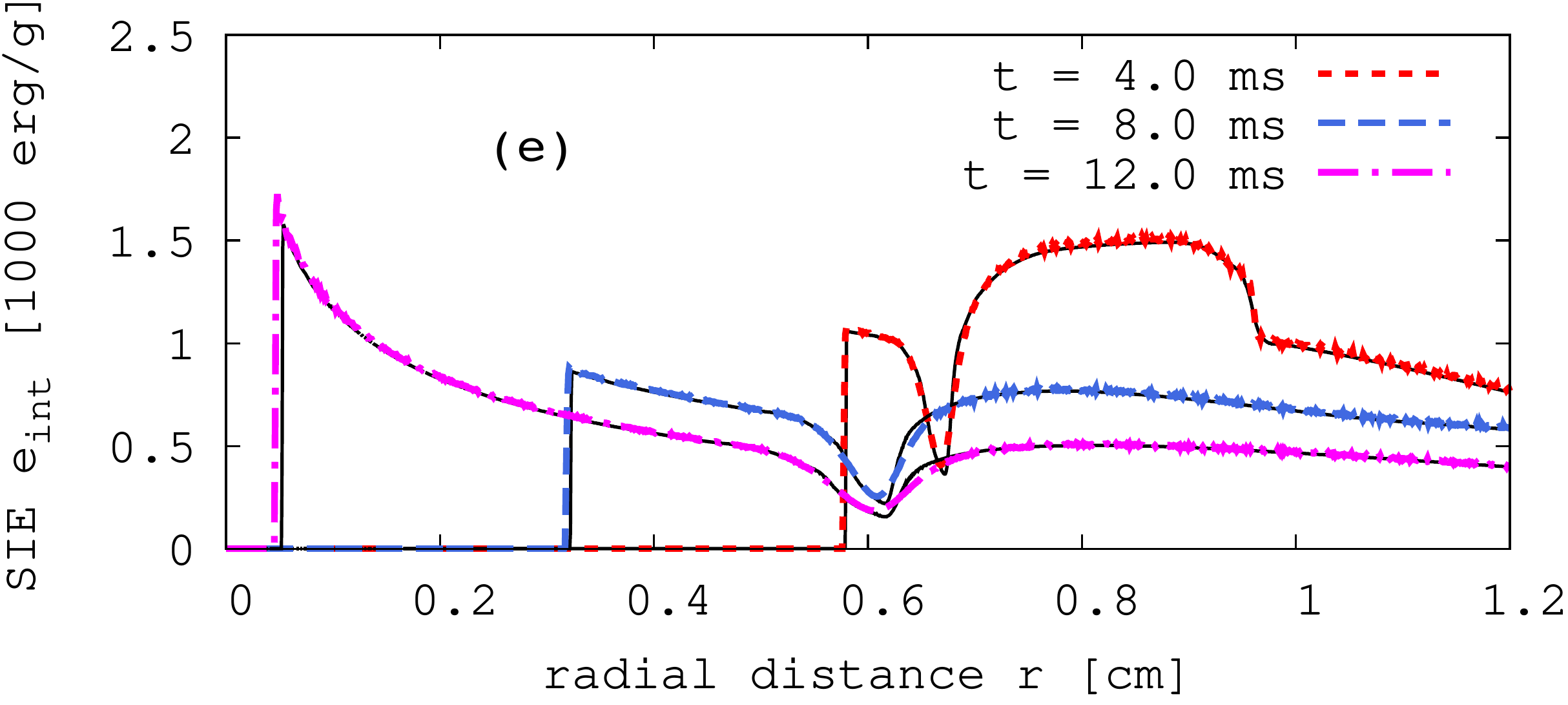}\hfill
\includegraphics[width = 0.47\textwidth]{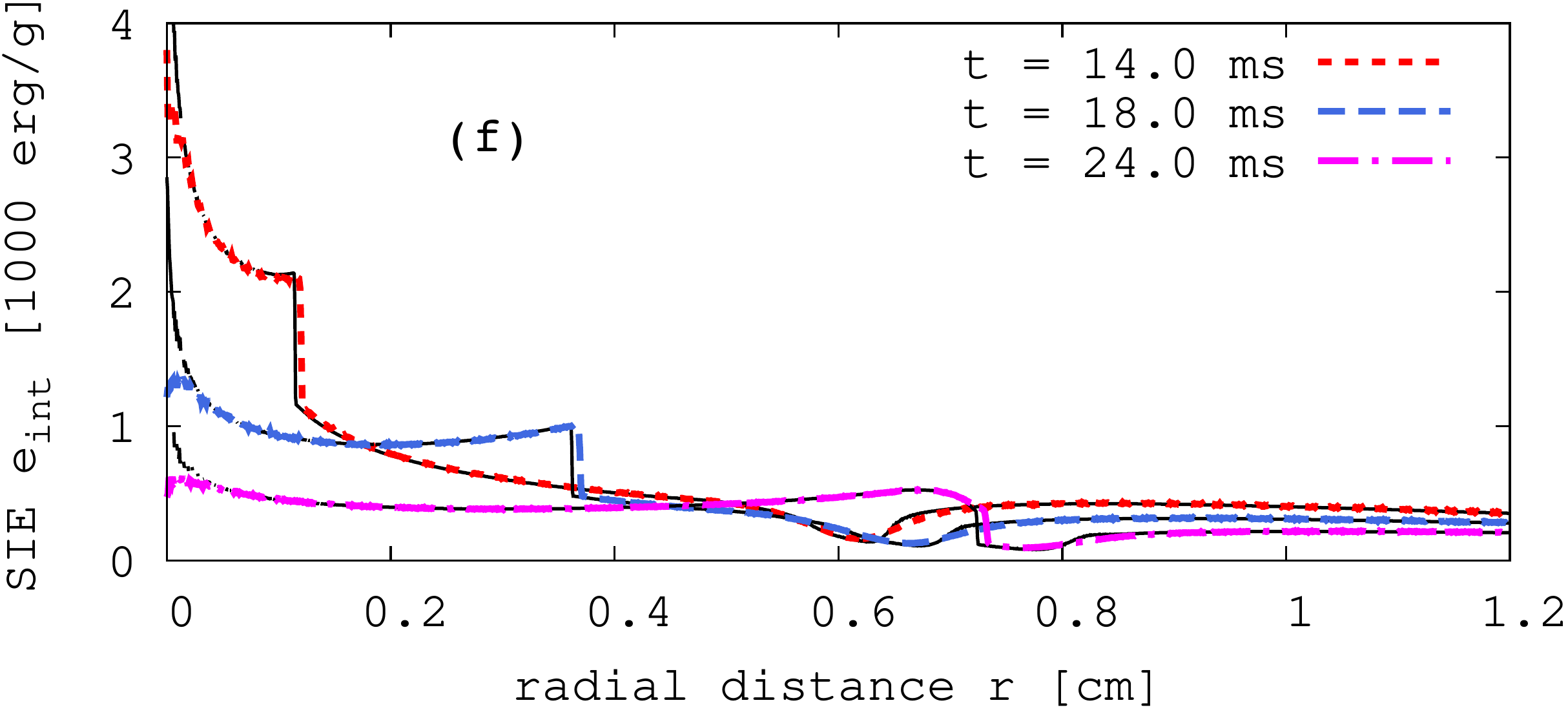}\hfill
\caption{Time snapshots of mass density $\rho$ (a,b), radial velocity $v_\mathrm{rad}$ (c,d) and SIE $e_\mathrm{int}$ (e,f) profiles of Kinetic-Q-60 (thick colored lines) and hr-RAGE (think solid lines).}
\label{sequence}
\end{center}
\end{figure*}

\begin{figure}
\begin{center}
\includegraphics[width = 0.47\textwidth]{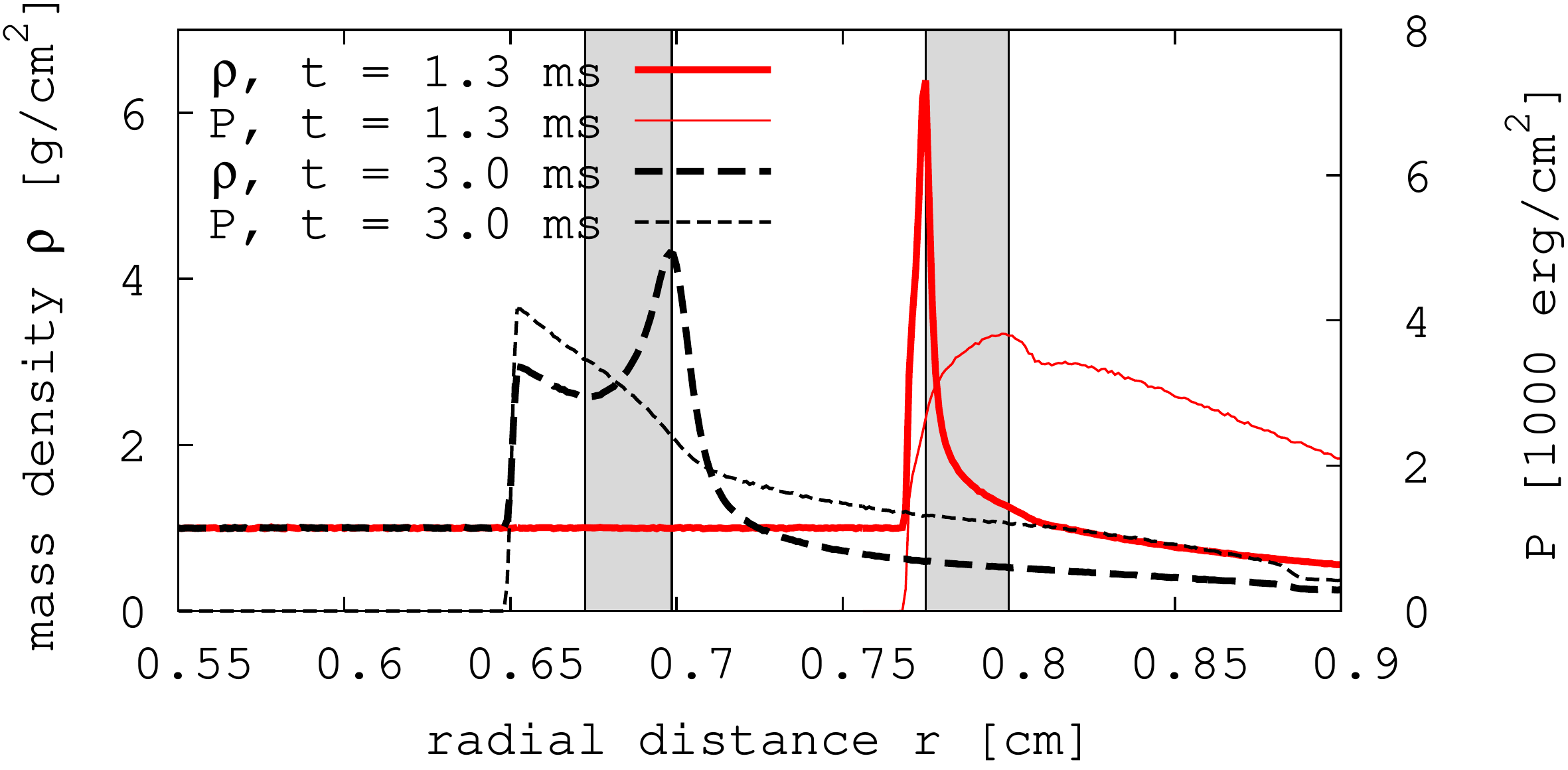}
\caption{Mass density $\rho$ (thick lines) and pressure $P$ (thin lines) radial profiles for Kinetic-Q-60 at $t = 1.3\:$ms (red solid lines) and $t = 3.0\:$ms (dashed black lines). Regions with opposite $\rho$ and $P$ gradients are marked in grey.}
\label{density_pressure_comp}
\end{center}
\end{figure}

\begin{figure}
\begin{center}
\includegraphics[width = 0.4\textwidth]{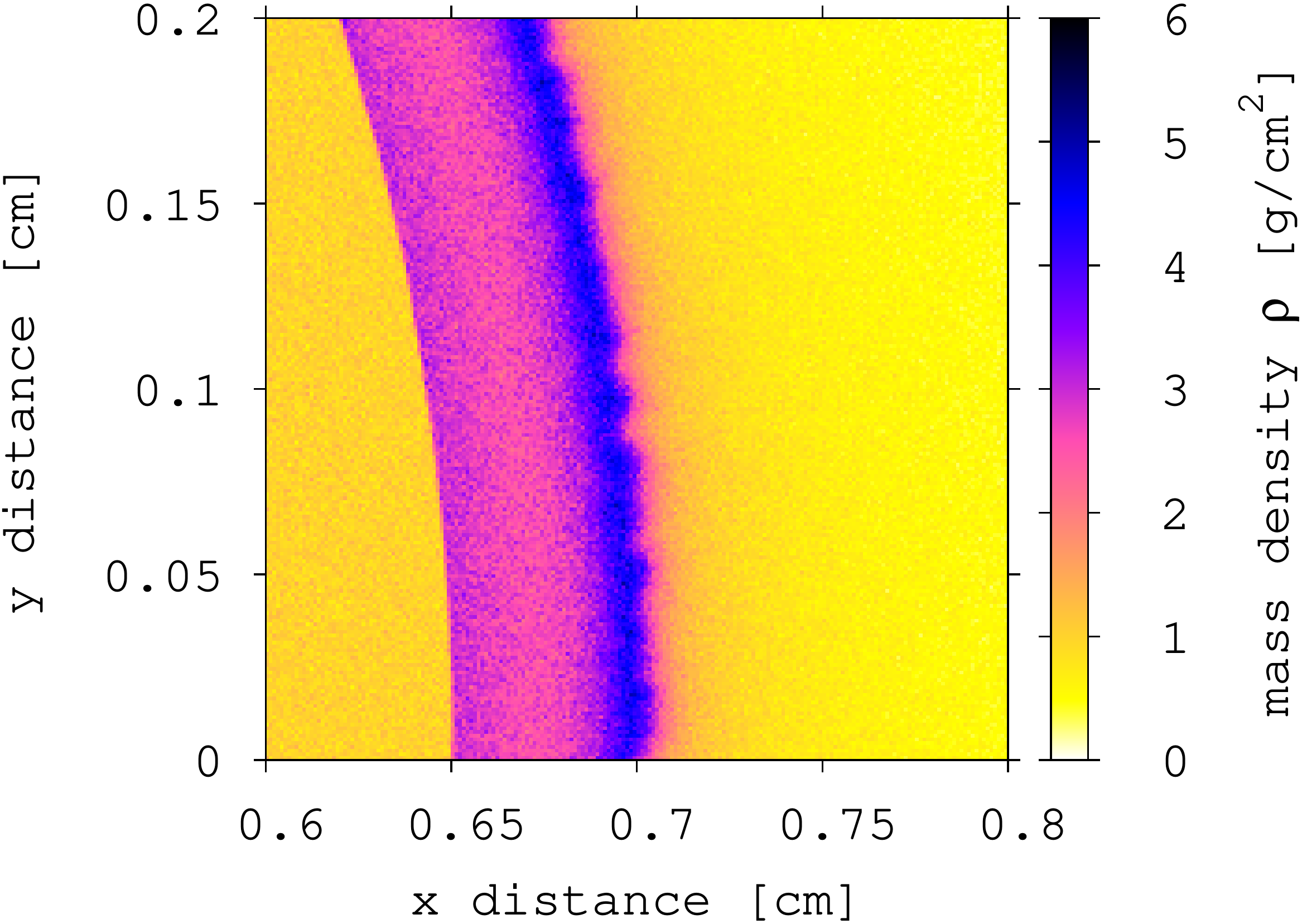}\hfill
\caption{Zoom into the two-peak density configuration for Kinetic-Q-60 at $t = 3.0\:$ms. We can see the formation of RTIs at the edge of the outer density peak.}
\label{perturbation}
\end{center}
\end{figure}

\begin{figure*}
\begin{center}
\includegraphics[width = 0.33\textwidth]{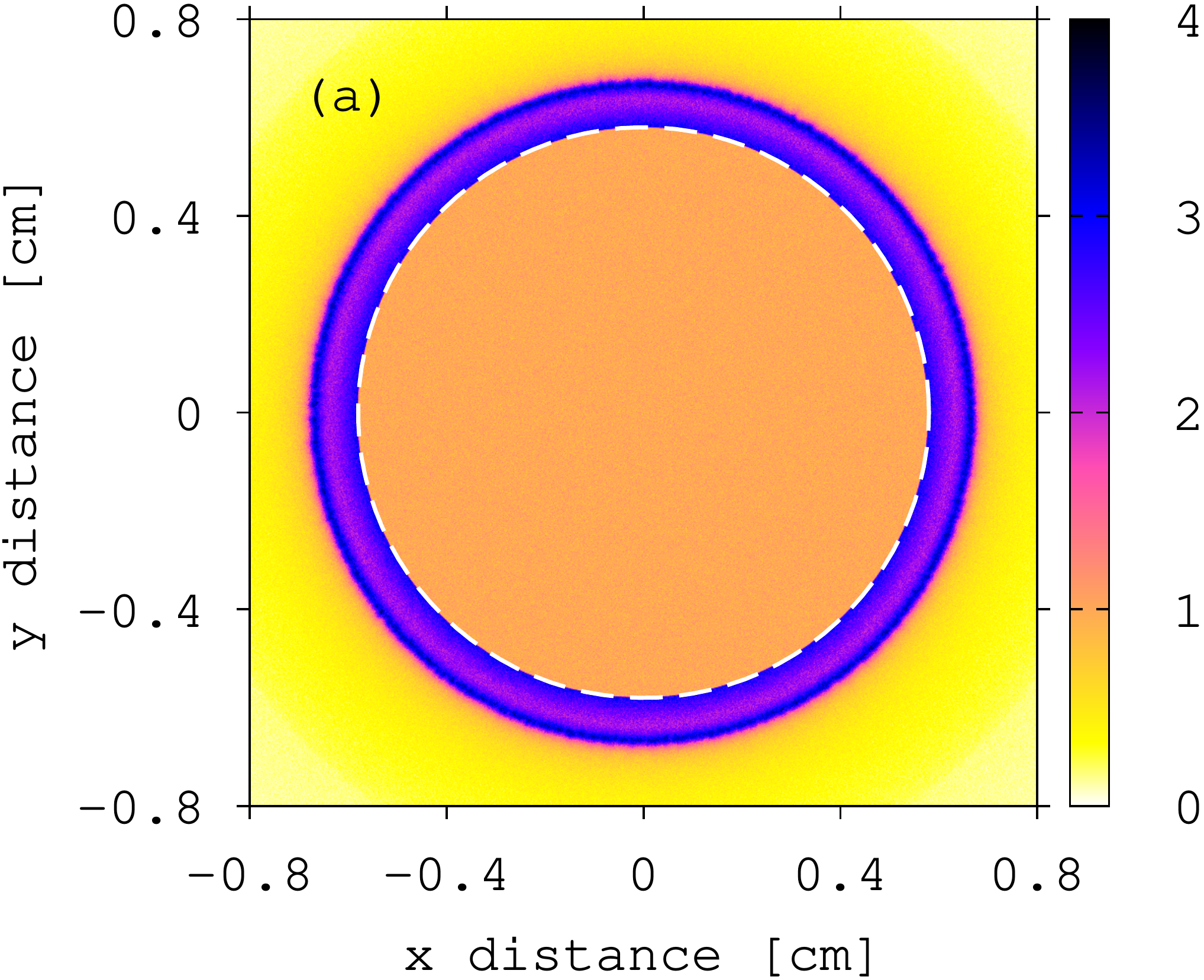}\hfill
\includegraphics[width = 0.312\textwidth]{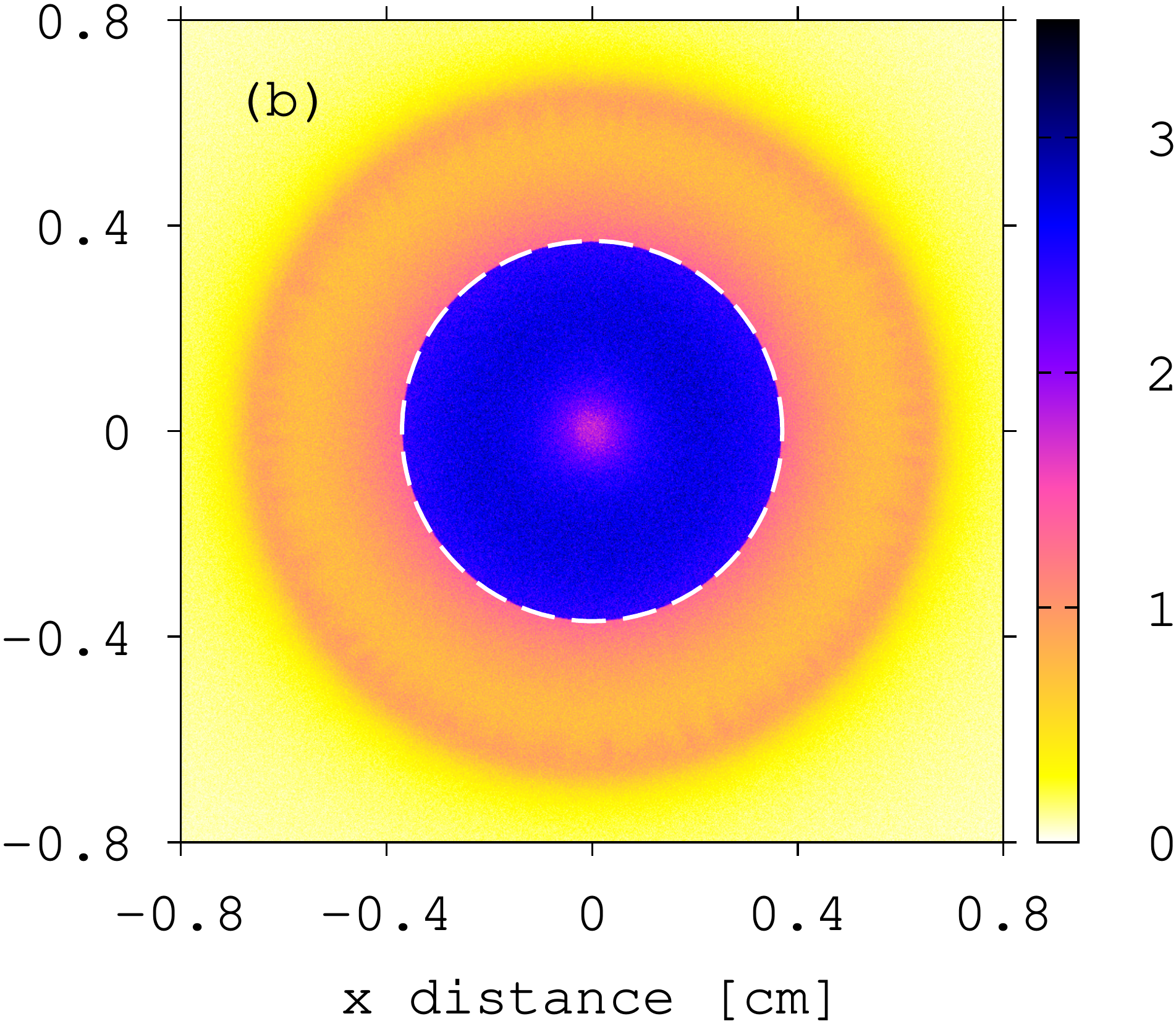}\hfill
\includegraphics[width = 0.3425\textwidth]{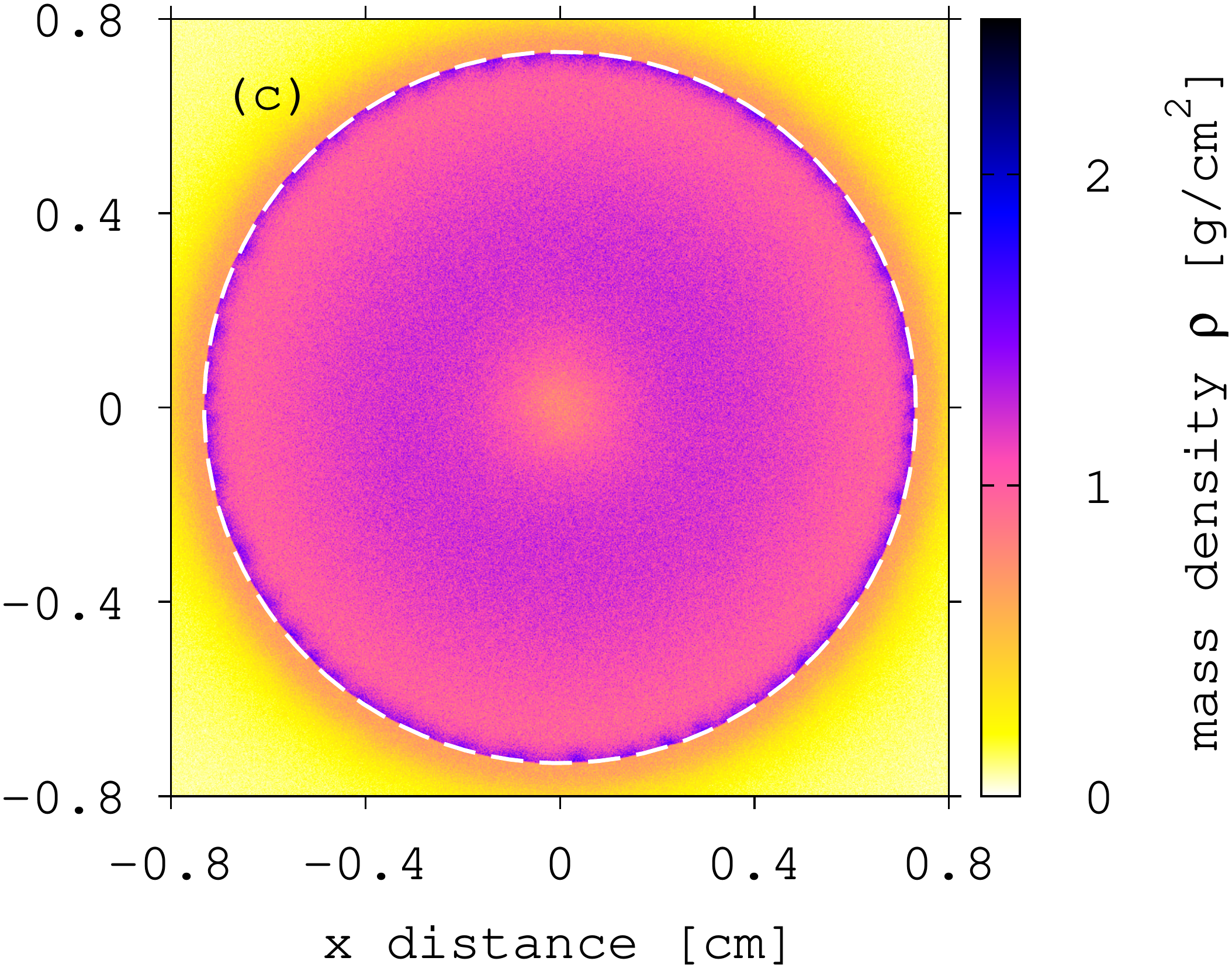}\hfill
\caption{Mass density $\rho$ for Kinetic-F-100 at $t = 4.0\:$ms (a), $18.0\:$ms (b) and $24,0\:$ms (c). The white dashed lines mark the shock positions as obtained from radial profiles in Fig.\:\ref{sequence}.}
\label{2D_density_100M}
\end{center}
\end{figure*}

Fig.\:\ref{sequence} shows mass density and radial velocity profiles of Kinetic-Q-60 and hr-RAGE for $t \geq 4.0\:$ms. Both calculations agree well, especially in the beginning of the implosion. As in all particle-based methods, statistical noise is present in the thermodynamic properties of the kinetic simulations. The smooth profiles at large $r$ are due to radial averaging over many cells. At the center, fluctuations become more pronounced since only a few output cells are present. We notice that for $t \geq 12.0\:$ms, the shock in the kinetic simulation is slightly ahead of the one in \textsc{RAGE}. Fluctuations in the thermodynamic properties might lead to some local acceleration of the shock. However, unless the effect is systematic, we would expect it to average out with time. The difference in the shock positions might also originate in the initial stage of the implosion. hr-RAGE has a significantly higher resolution than Kinetic-Q-60 and can therefore resolve the shock formation much better, which could lead to deviations between the two approaches later on. 
However, the most likely reason for the disagreement between the shock positions seems to be the presence of RTIs in the kinetic simulation. Fig.\:\ref{density_shock_launch}(c) shows the evolution of the SIE during shock launch. In combination with Fig.\:\ref{density_shock_launch}(a) and (b), we see that in the initial stages of the implosion, hot matter from zone-2 is accelerated into the cold denser matter of zone-1. This usually favors the formation of RTIs. Indeed, an examination of the density and pressure profiles at $t=1.3\:$ms and $t=3.0\:$ms in Fig.\:\ref{density_pressure_comp} reveals opposite $\rho$ and $P$ gradients. The corresponding unstable regions (marked by gray areas) lie behind the compressed matter at $t = 1.3\:$ms and between the shock front and second density peak at $t=3.0\:$ms. The resulting instabilities can be seen in a density map at $t=3.0\:$ms in Fig.\:\ref{perturbation}.
They are seeded by small perturbations due to the finite particle number in our code \cite{Sagert15}. The opposite pressure and density gradients persist until late times and result in filament-like structures, seen in e.g. Fig.\:\ref{2D_density_100M}(b) at $r \sim 0.6\:$cm. For \textsc{RAGE}, we do not see such phenomena, which is most likely due to the lack of perturbations. The difference in the shock locations might be either initialized early, during the passage of the shock through the RTIs, or occur at later times and be caused by e.g. a slightly different compression of matter due to the presence of RTIs.\\ 
\newline
Unlike in realistic ICF simulations, instabilities in the 2-zone setup are unlikely to lead to large-scale deformations. However, as the shock converges and rebounds, grid effects can impact its shape. We test how well symmetry is preserved by plotting the density distribution for Kinetic-F-100 in Fig.\:\ref{2D_density_100M} for $t = 4.0\:$ms, $18.0\:$ms and $24.0\:$ms, together with the shock positions from the radial profiles in Fig.\:\ref{sequence}.
With $10^8$ test-particles, we achieve a good resolution and see excellent agreement between the shock position from radial averaging and the 2D density map, with well preserved spherical symmetry. At $t = 18.0\:$ms, the outgoing shock encounters and compresses the filament-like structures. However, the interaction does not impact the shock propagation.
\subsection{Particle Number Dependence}
\begin{figure}
\begin{center}
\includegraphics[width = 0.47\textwidth]{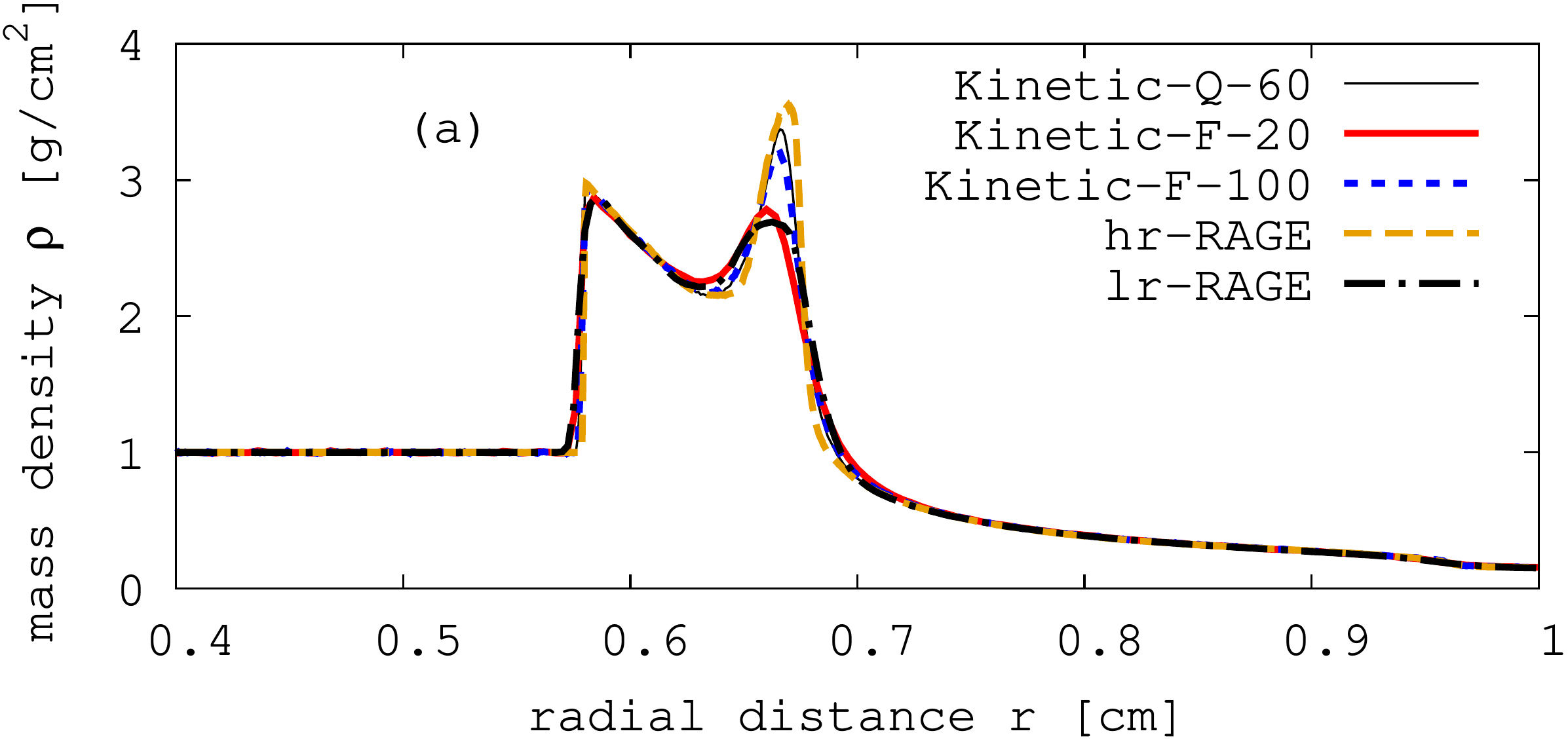}\hfill
\includegraphics[width = 0.47\textwidth]{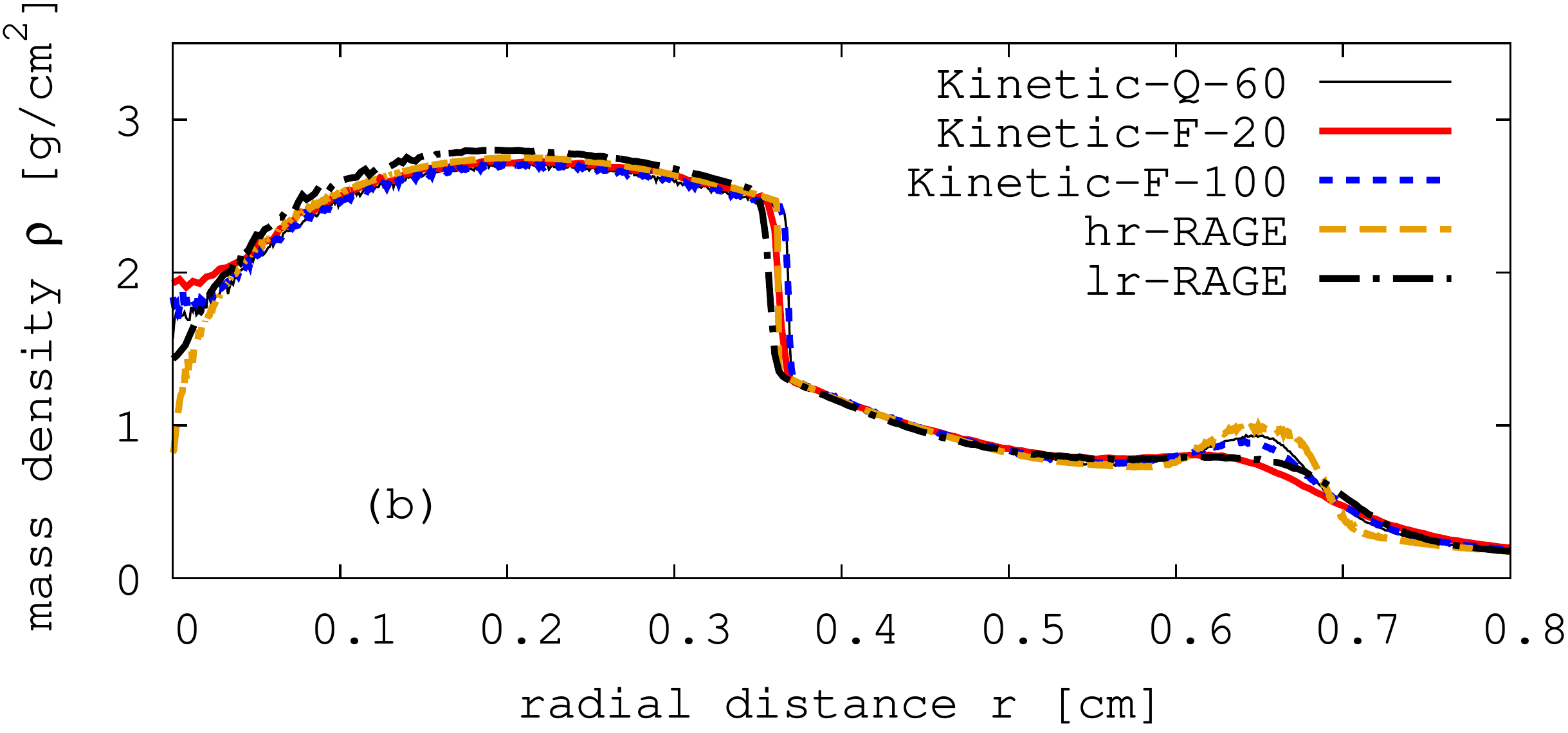}\hfill
\includegraphics[width = 0.47\textwidth]{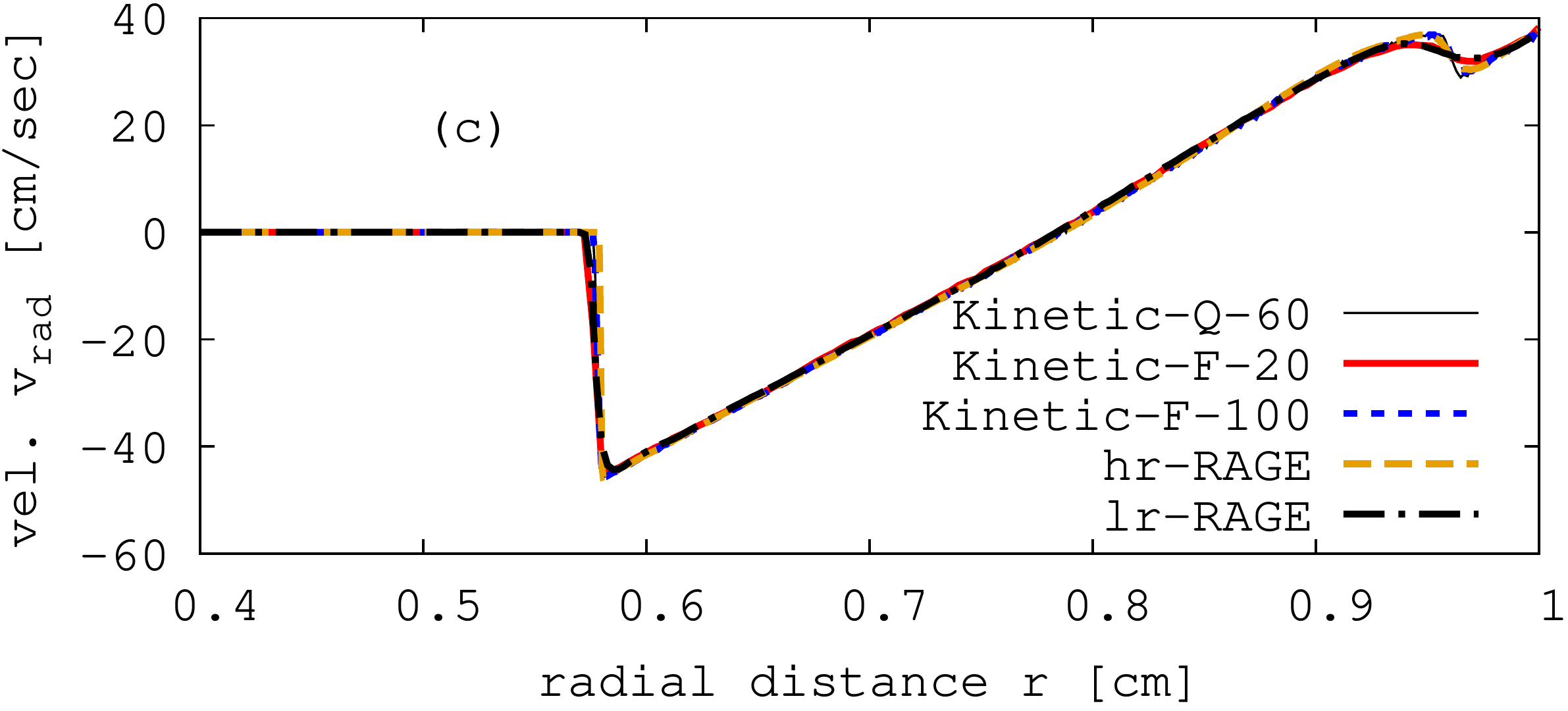}\hfill
\includegraphics[width = 0.47\textwidth]{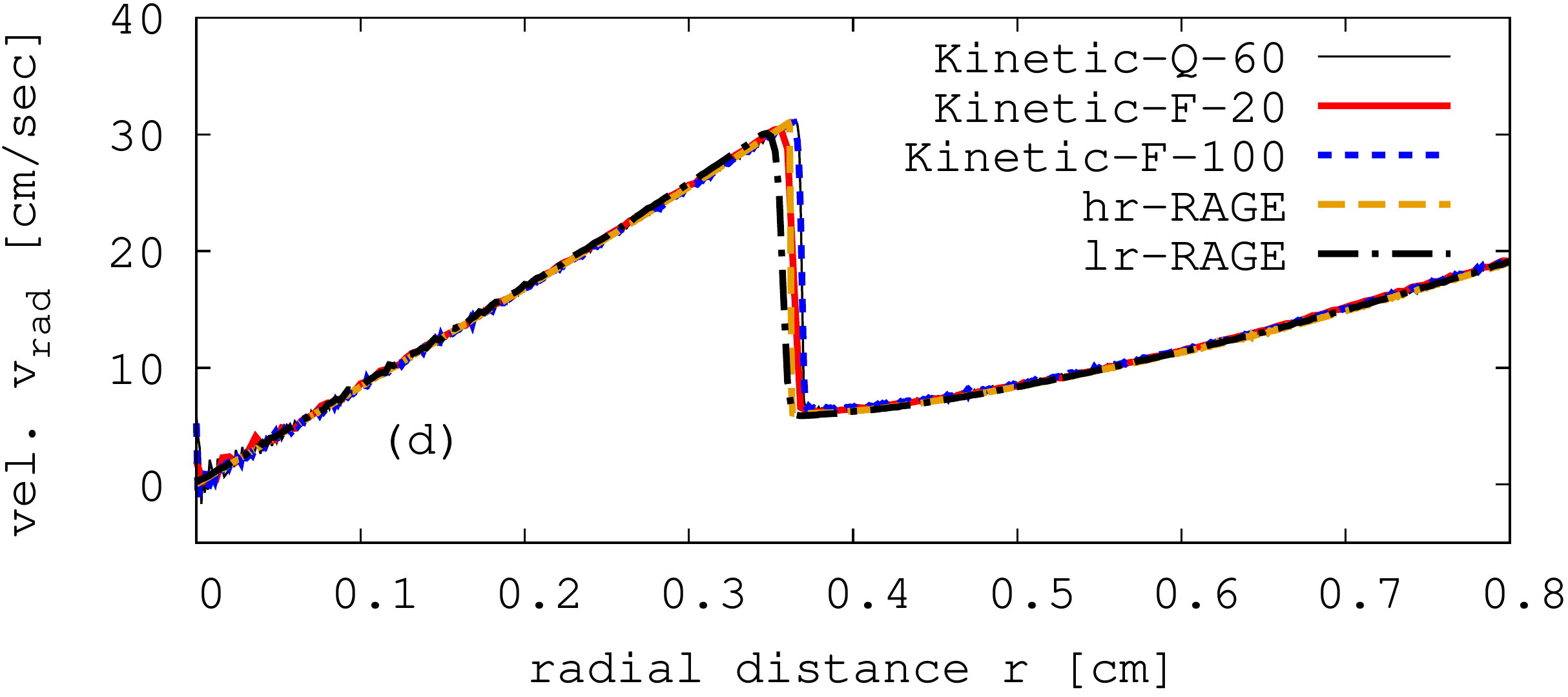}\hfill
\caption{Density $\rho$ (a,b) and radial velocity $v_\mathrm{rad}$ (c,d) profiles for all simulations at $t = 4.0\:$ms (a,c) and $18.0\:$ms (b,d).}
\label{part_num_comp}
\end{center}
\end{figure}

We now test the dependence of the implosion dynamics on the particle number and resolution. Fig.\:\ref{part_num_comp} shows density and radial velocity profiles of all kinetic and hydrodynamic models for $t = 4.0\:$ms and $t = 18.0\:$ms. At $t = 4.0\:$ms, the shock location is consistent between all models, with only Kinetic-F-20 and lr-RAGE having slightly broader fronts. Interestingly, the height of the second density peak is much more sensitive to resolution and particle number. The highest density is achieved with hr-RAGE followed by Kinetic-Q-60. For lower particle number and resolution, the height of the peak decreases while its width becomes larger. The radial velocity, on the other hand, does not show any significant dependence on either, $N$ or the resolution. \\
At $t = 18\:$ms, the shock in lr-RAGE lags behind the one in hr-RAGE. lr-RAGE also has slightly higher densities for $r < 0.35\:\mathrm{cm}$ while the kinetic simulations agree well with hr-RAGE for $0.025 \: \mathrm{cm} < r < 0.6\:\mathrm{cm}$. However, they show differences for smaller $r$: at the disk center, their densities are around $\rho_c \sim 1.8 \:\mathrm{g/cm^2}$, while hr-RAGE tends towards $\rho_c \sim 0.8 \: \mathrm{g/cm^2}$ and lr-RAGE has $\rho_c \sim 1.5 \:\mathrm{g/cm^2}$. 
This could be caused by so-called wall-heating, enhanced temperature and decreased density at the origin that many fluid dynamics codes are subject to and that is not seen in the kinetic approach \cite{Noh87, Marti99,Sagert14}. On the other hand, the disagreement between the kinetic and the hydrodynamic studies could also originate from finite particle numbers in the kinetic code or grid effects in \textsc{RAGE}.
\subsection{Mean-Free-Path and Non-Equilibrium Study}
\label{simple_implosion_mfp}
\begin{figure*}
\begin{center}
\includegraphics[width = 0.47\textwidth]{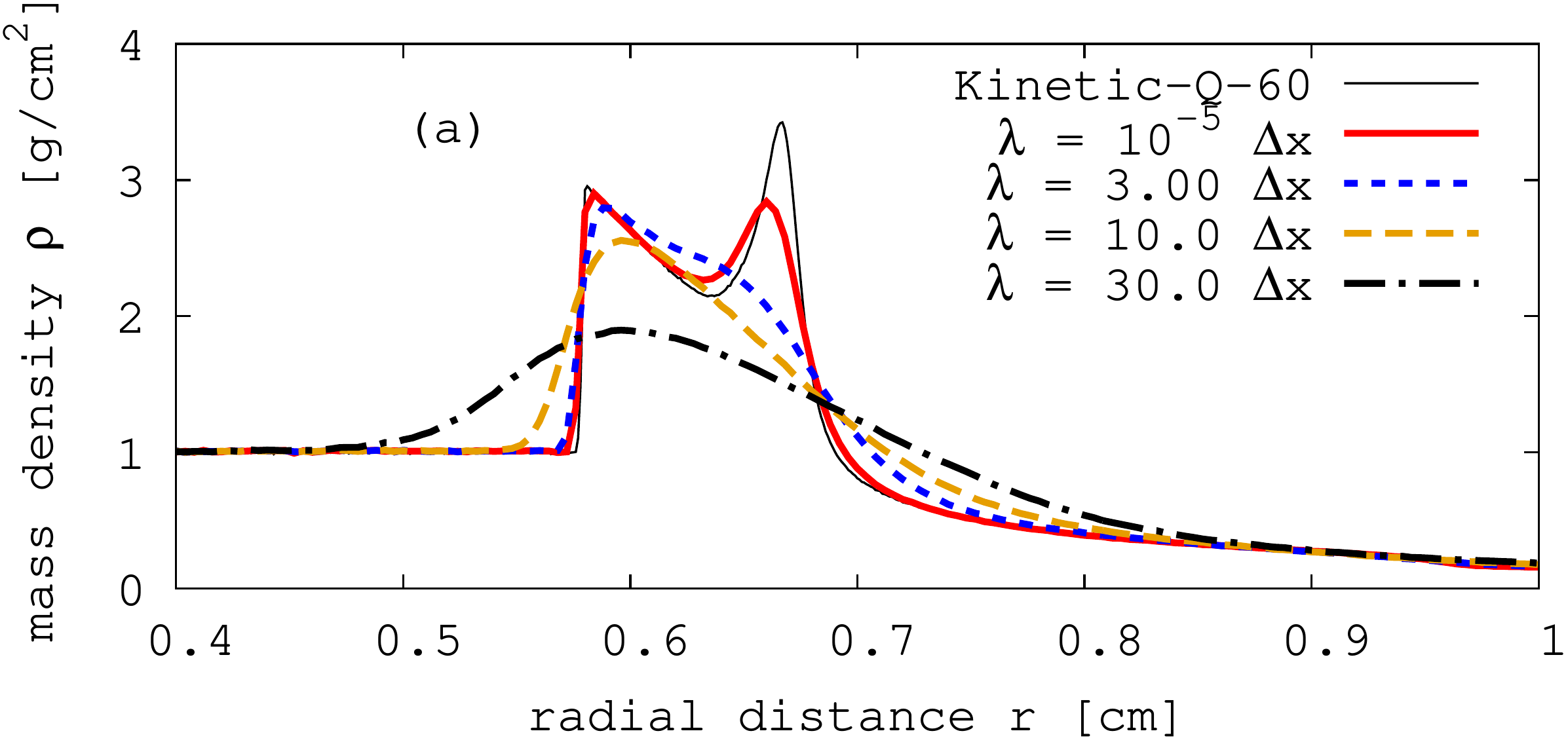}\hfill
\includegraphics[width = 0.47\textwidth]{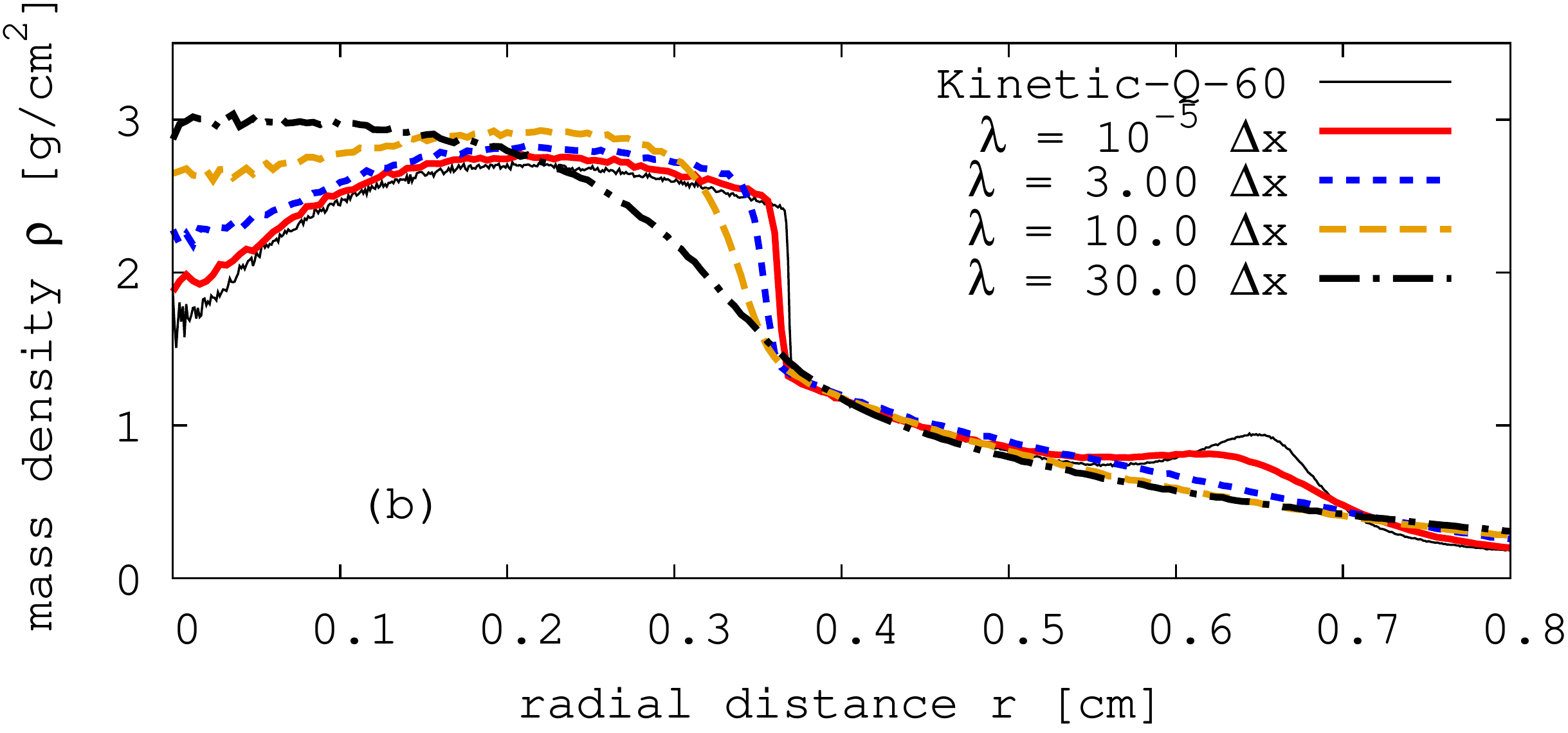}\hfill
\includegraphics[width = 0.47\textwidth]{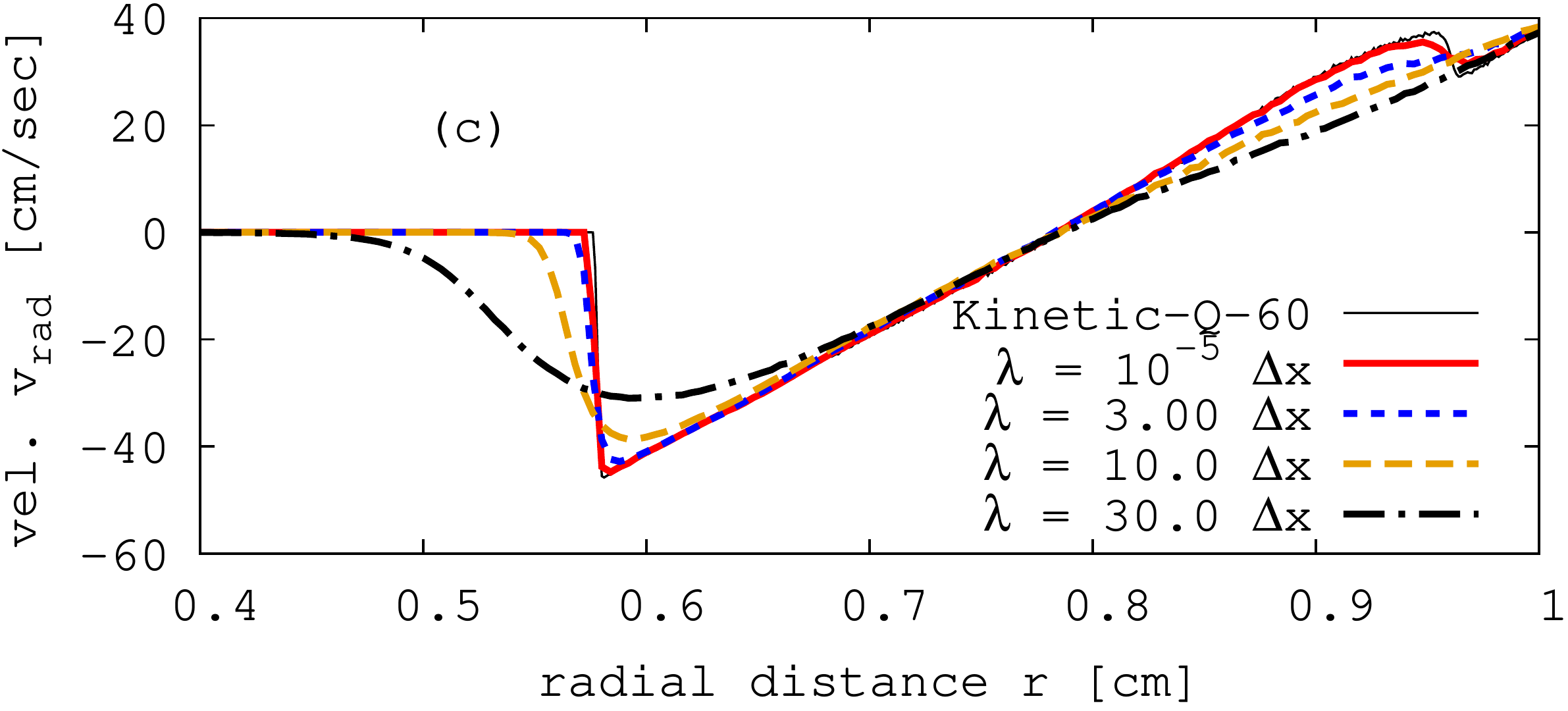}\hfill
\includegraphics[width = 0.47\textwidth]{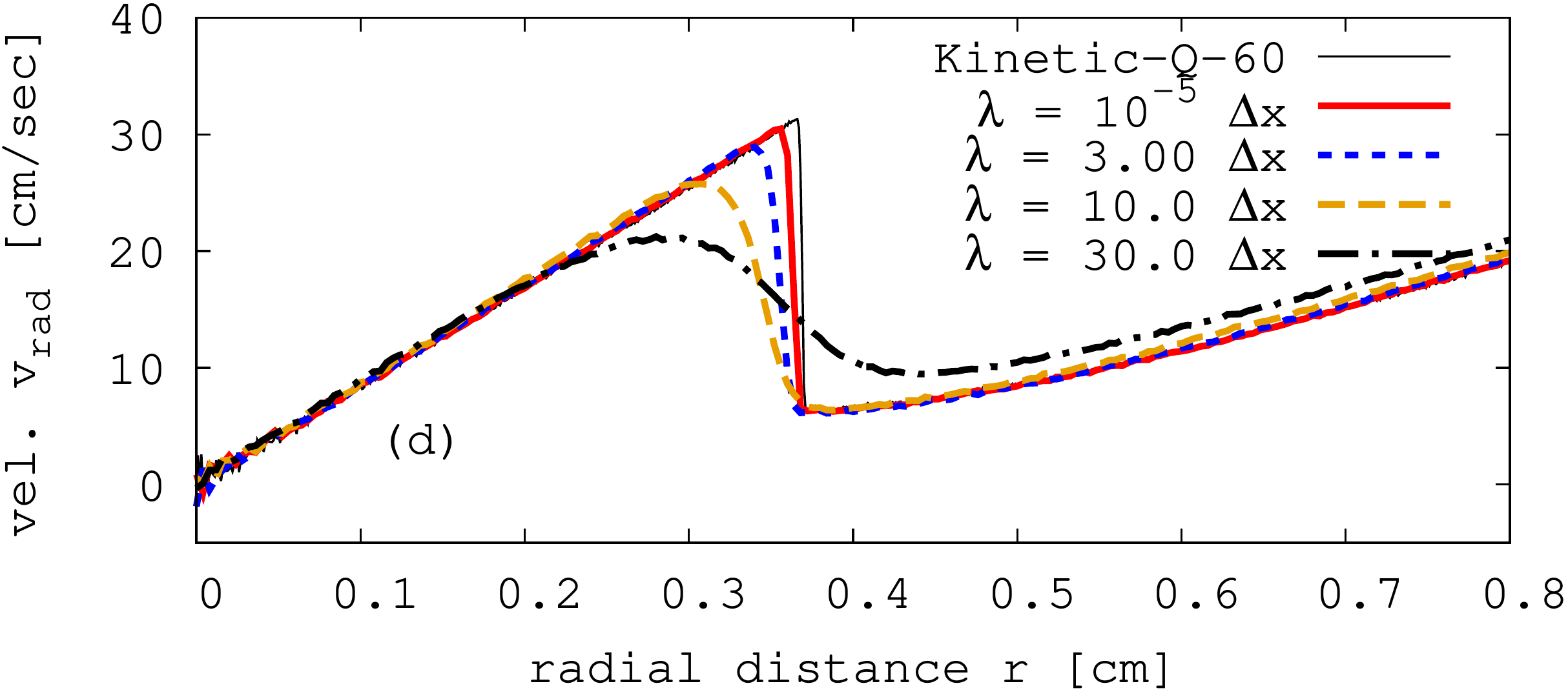}\hfill
\includegraphics[width = 0.47\textwidth]{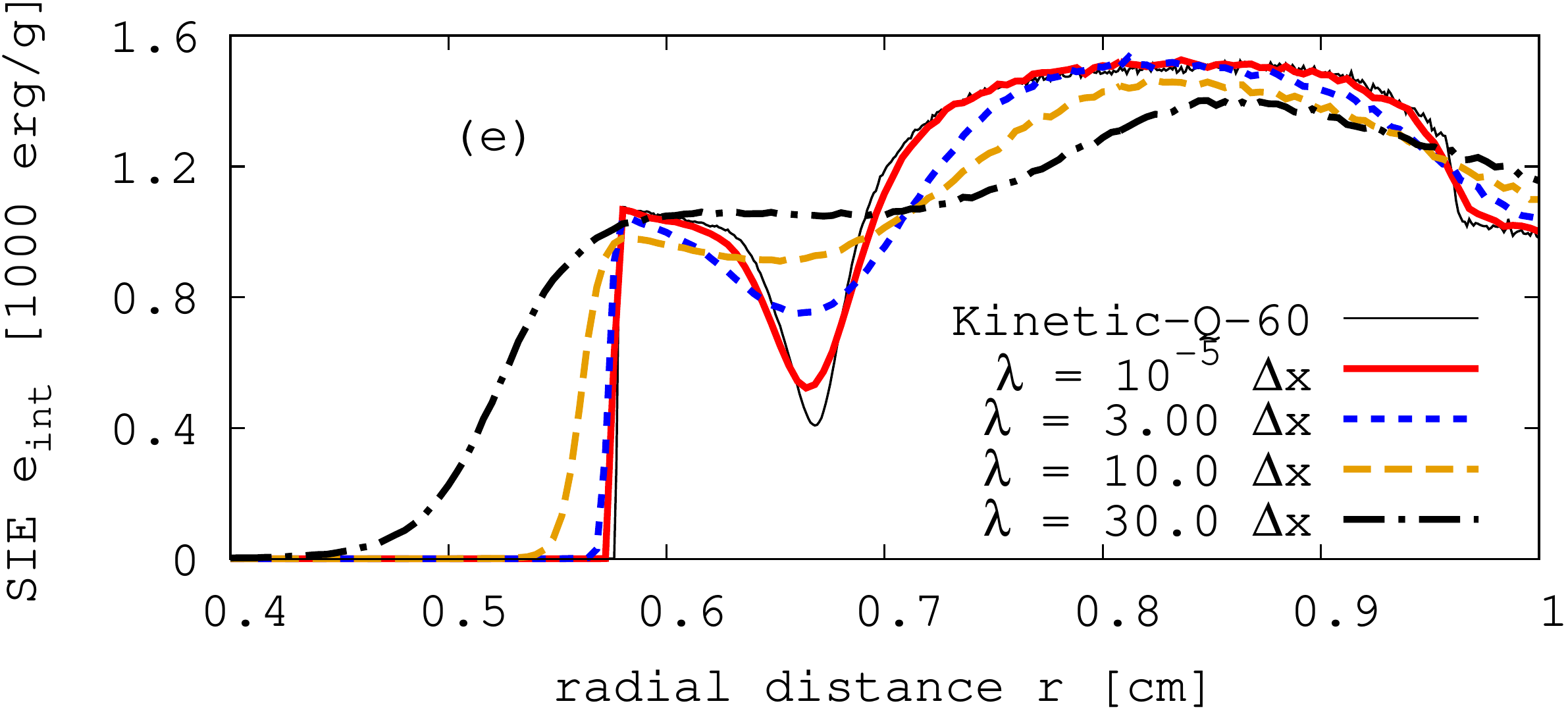}\hfill
\includegraphics[width = 0.47\textwidth]{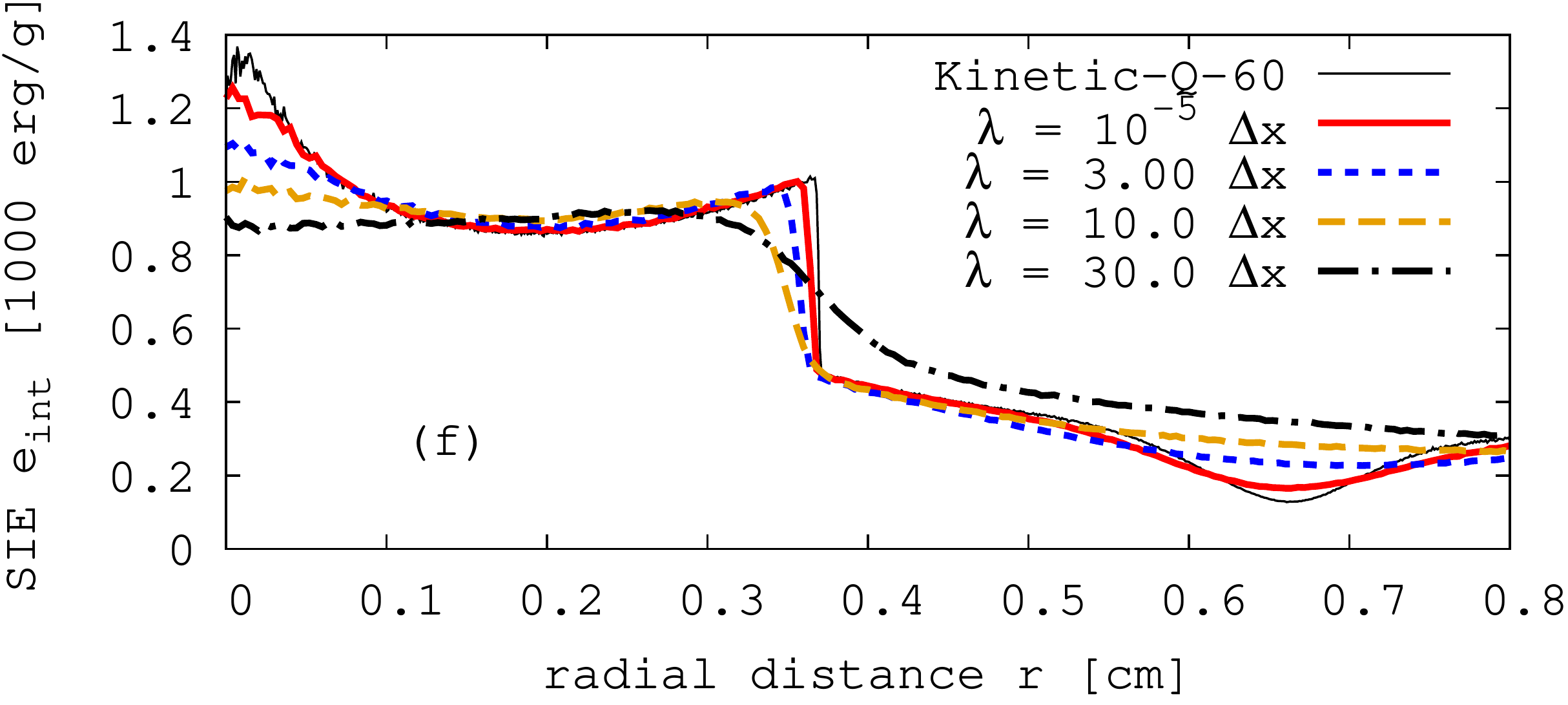}\hfill
\caption{Mass density $\rho$ (a,b), radial velocity $v_\mathrm{rad}$ (c,d), and SIE $e_\mathrm{int}$ (e,f) profiles for Kinetic-F-20 and different $\lambda$ at $t = 4.0\:\mathrm{ms}$ (a,c,e) and $t = 18.0\:\mathrm{ms}$ (b,d,f). Profiles for Kinetic-Q-60 with $\lambda = 10^{-5} \: \Delta x$ (thin black line) are added for comparison.}
\label{part_mfp_comp_den}
\end{center}
\end{figure*}
\begin{figure}
\begin{center}
\includegraphics[width = 0.47\textwidth]{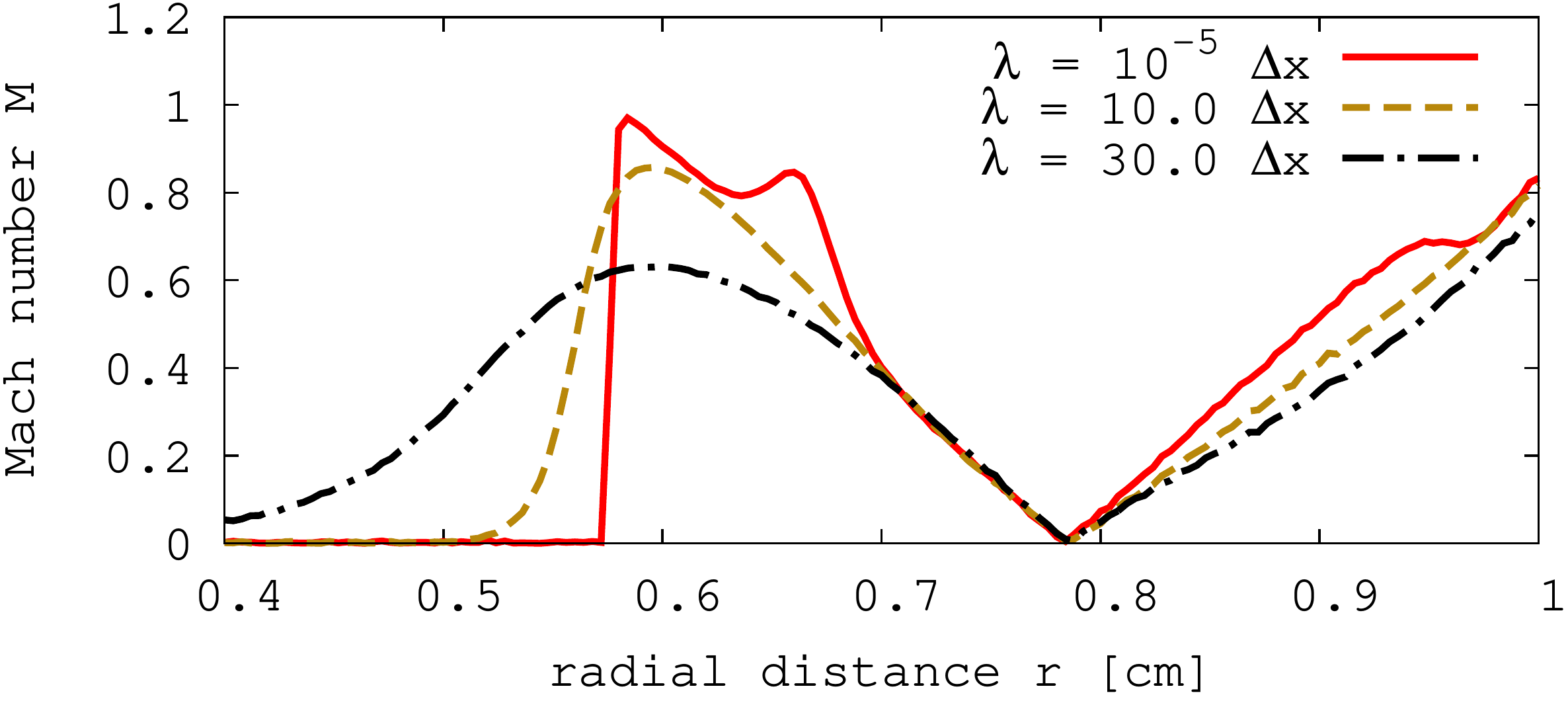}
\caption{Mach number $M$ radial profiles at $t=4.0\:$ms for different mean-free-paths $\lambda$.}
\label{mach_temp_40}
\end{center}
\end{figure}

In this section, we explore non-equilibrium phenomena in our simulations using large particle mean-free-paths. Since our previous studies found that particles with $\lambda < 3\:\Delta x$ behave similar to matter in the continuum limit, while $\lambda > 30\:\Delta x$ results in particle behavior close to free-streaming \cite{Sagert15, Sagert14}, we run Kinetic-F-20 with $\lambda = 10^{-5}\:\Delta x$, $3\:\Delta x$, $10\:\Delta x$ and $30\:\Delta x$. The low particle number is chosen for fast computation. As seen in the previous section, the density profile is slightly different from Kinetic-Q-60. However, the general features and response to $\lambda$ should be similar.\\
During the initial stages of the implosion, we see a strong dependence of matter compression on $\lambda$ with the width and height of the forming density peak being very sensitive to the mean-free-path. Fig.\:\ref{part_mfp_comp_den} shows the mass density, radial velocity, and SIE profiles at $t = 4.0\:$ms and $18.0\:$ms. At $t = 4.0\:$ms, the shock broadening is clearly visible for increasing $\lambda$. As for different values of $N$, the second density peak and corresponding dip in the SIE are very sensitive to the mean-free-path. The radial velocity, on the other hand, is smooth at the density peak but shows a dependence on $\lambda$ in the outgoing shock at $r \sim 0.95\:$cm. At $t = 18.0\:$ms, we can see that after shock rebound, the central density remains high for longer times if $\lambda$ is large. This is a consequence of lower particle velocities as the slow particles linger at the disk center. Since the SIE is connected to the particle speed, its values at the center are also lower for large $\lambda$.\\
\newline
Shocks, especially at high Mach numbers, can experience non-equilibrium behavior such as deviations from MB velocity distributions and temperature anisotropies \cite{Maricante17, Zhak99, Anisimov97, Candler94}. One example are two-component velocity distributions found in kinetic ICF implosion simulations that are caused by energetic run-away ions and the mixture of matter upstream and downstream of the shock \cite{Larroche12, Le16, Larroche16}. Another example are rarefied hypersonic flows that are prone to anisotropies in $T$ (or $P$) components longitudinal and transverse to the  direction of shock propagation. The longitudinal components experience an overshoot at the shock front, which is due to a slow transformation of the ordered longitudinal motion into thermal transverse random motion \cite{Anisimov97}. \\
We will explore whether the above or similar effects also appear in our simulations. First, we determine the Mach numbers $M$ that are reached during the implosions. These are defined as $M = v_\mathrm{rad} / v_s$, where $v_\mathrm{rad}$ is the radial velocity of matter and $v_s$ is the speed of sound. The latter can be calculated via $v_s = v_\mathrm{rms} \sqrt{\gamma / 2.0}$ for a 2D ideal gas. We find that in our simulation setup the Mach numbers rarely reach $M >1$, the highest values being present at the beginning of the implosion. Furthermore, $M$ decreases for larger $\lambda$, which can be seen in Fig.\:\ref{mach_temp_40} where we plot radial profiles of the Mach numbers for different $\lambda$ at $t=4.0\:$ms. However, note that $M$ is calculated using the bulk radial velocity while 
particles in the high-velocity tail of the MB distribution can have much larger speeds.\\ 
\begin{figure}
\begin{center}
\includegraphics[width = 0.47\textwidth]{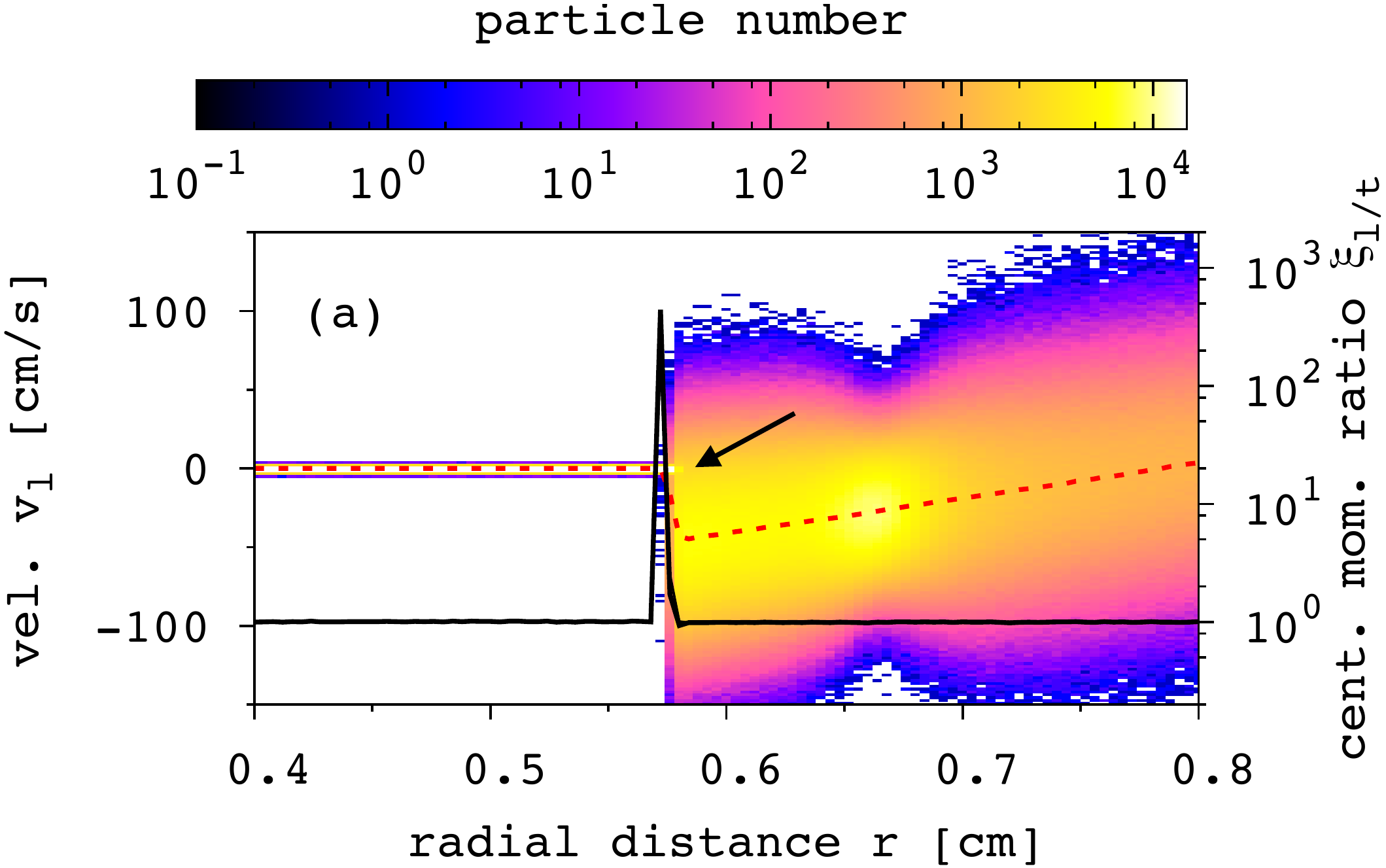}\hfill
\includegraphics[width = 0.47\textwidth]{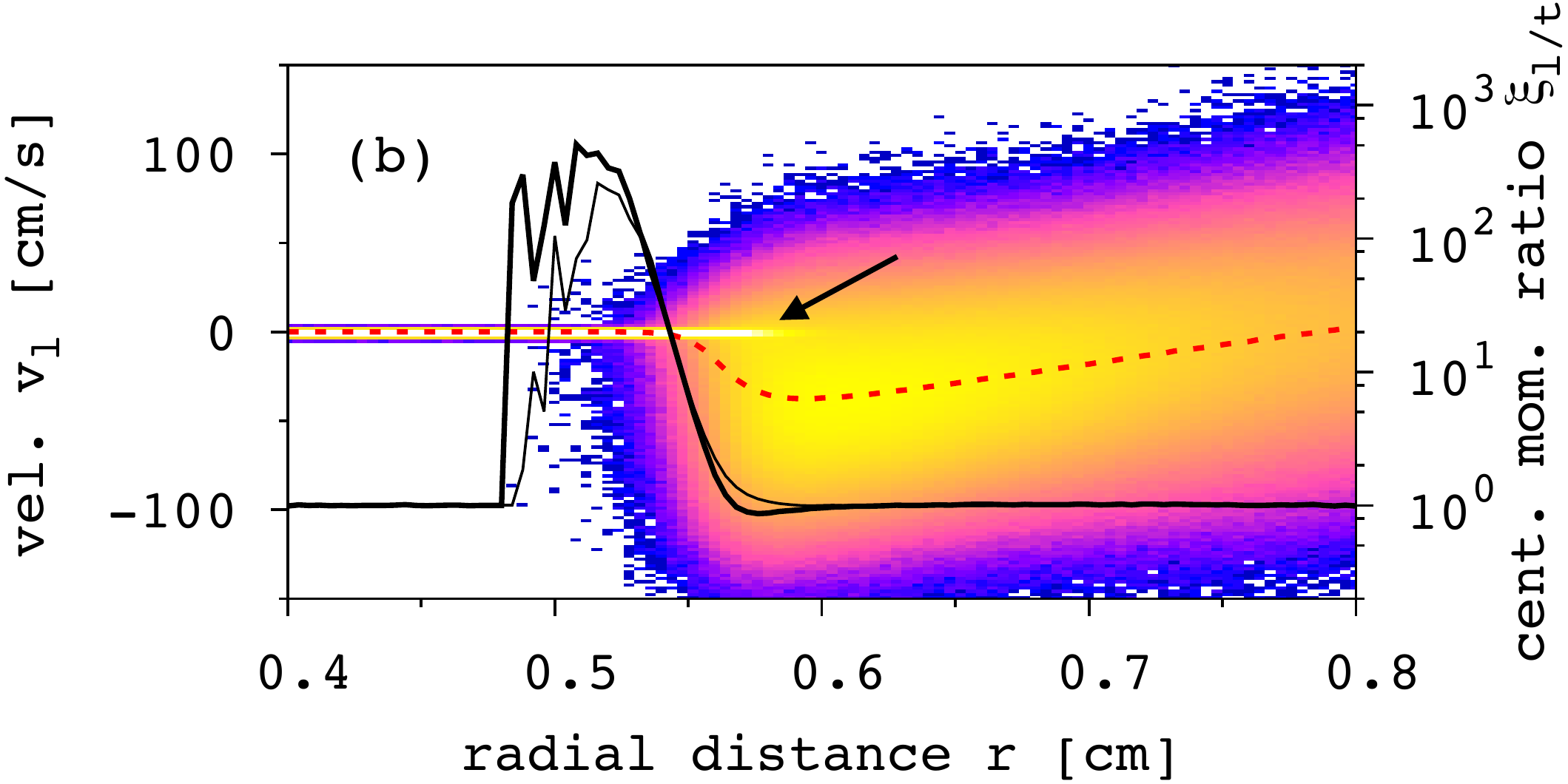}\hfill
\caption{Distribution of longitudinal particle velocities $v_{l}$ for $\lambda = 10^{-5}\: \Delta x$ (a) and $\lambda = 10\: \Delta x$ (b) at $t = 4.0\:$ms. The radial velocity is plotted as a (red) dashed line. The ratios of the 4th and 2nd central velocity moments are given by a thick solid line for $\xi_l$ and thin solid like for $\xi_t$.}
\label{fv_lambda_40}
\end{center}
\end{figure}
\begin{figure}
\begin{center}
\includegraphics[width = 0.47\textwidth]{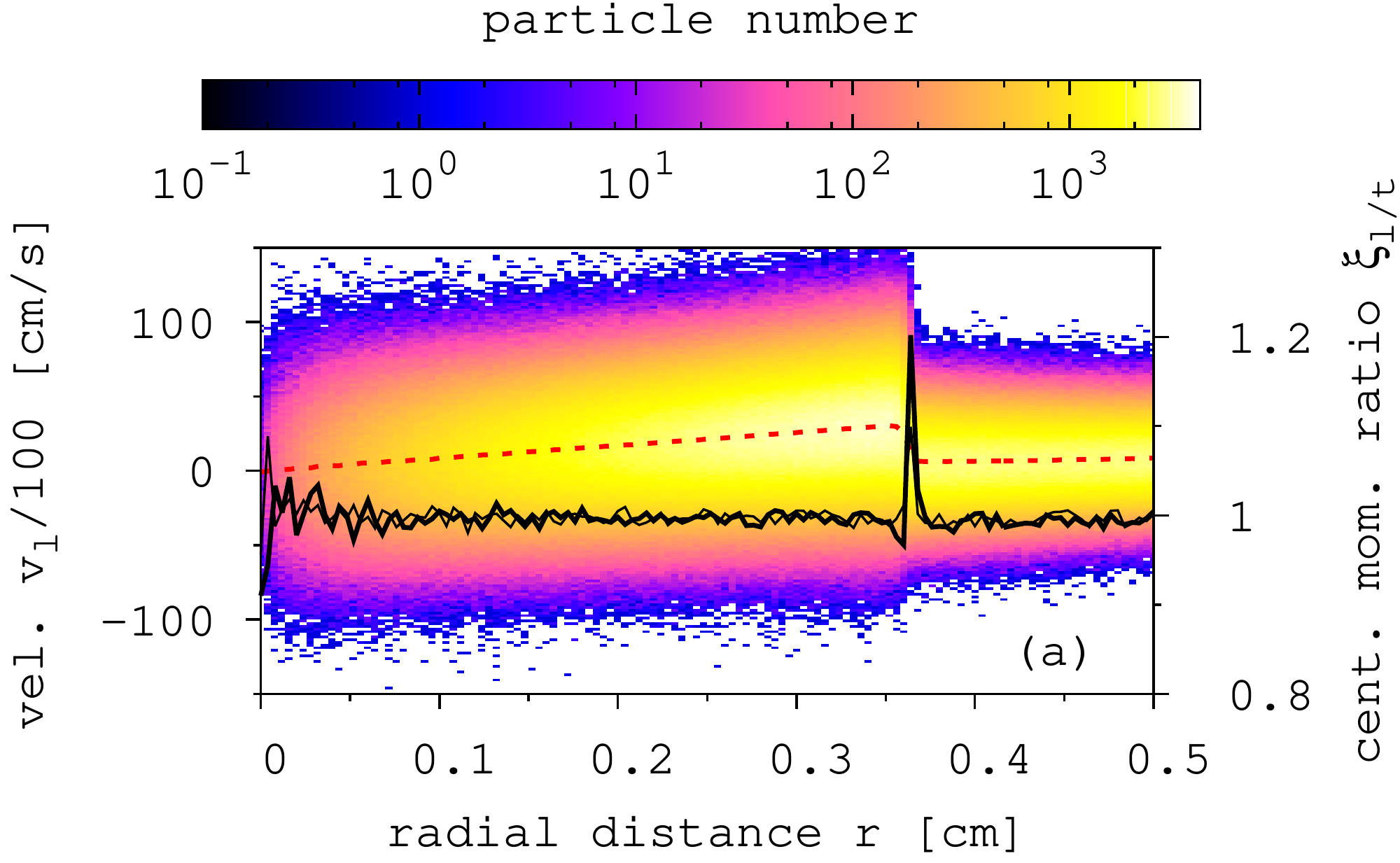}\hfill
\includegraphics[width = 0.47\textwidth]{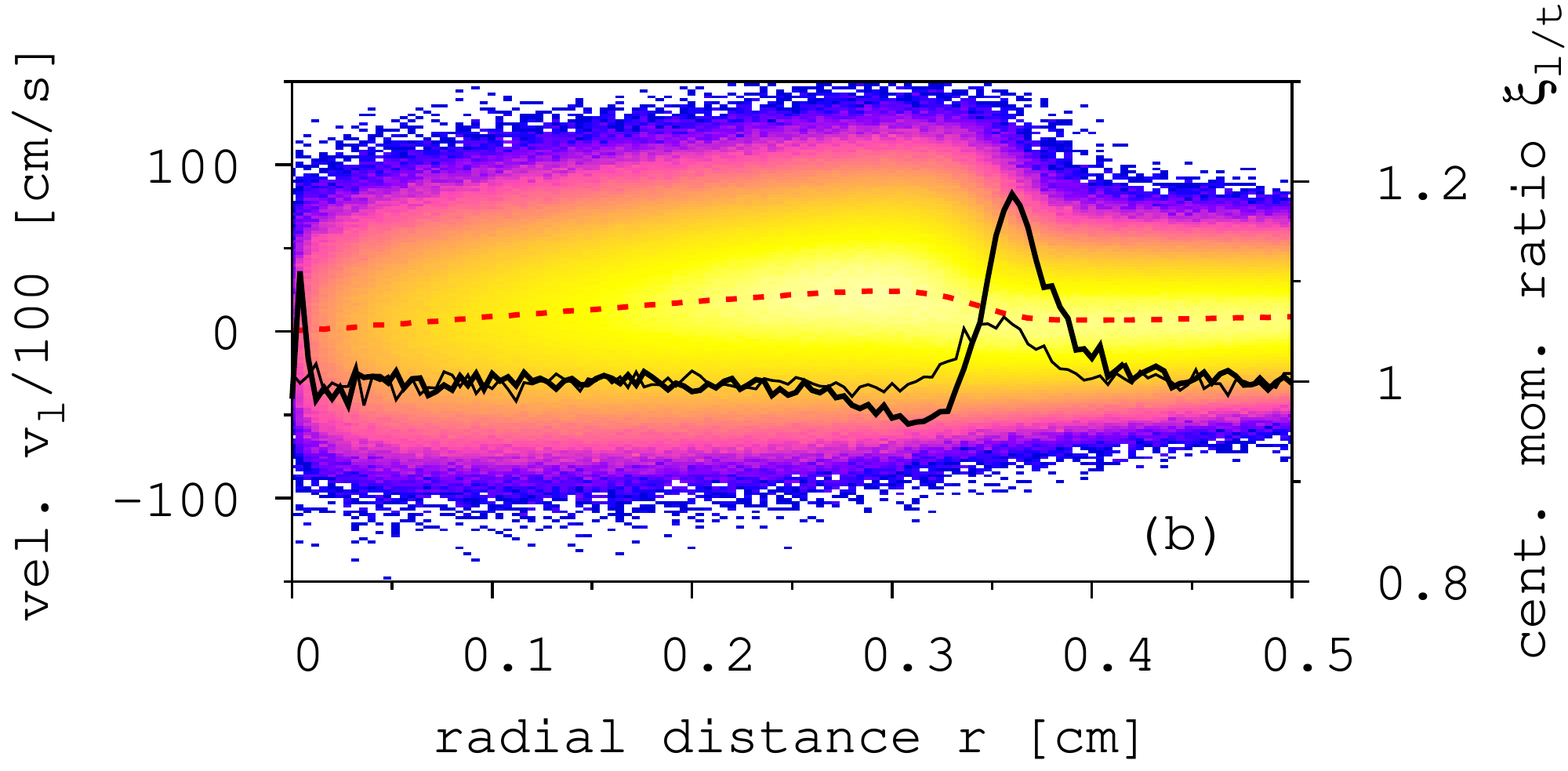}\hfill
\caption{Same as in Fig.\:\ref{fv_lambda_40} at $t = 18.0\:$ms.}
\label{fv_lambda_180}
\end{center}
\end{figure}
Figs.\:\ref{fv_lambda_40} and \ref{fv_lambda_180} show the distributions of longitudinal and transverse particle velocities, $v_l$ and $v_t$, for $\lambda = 10^{-5} \: \Delta x$ and $10 \: \Delta x$ at $t = 4.0\:$ms and $18.0\:$ms. The corresponding velocity components for a particle $i$ at a radial distance $r$, $v_{l,i}$ and $v_{t,i}$, are determined by:
\begin{align}
v_{l,i} &=  v_{il} - \langle v_l \rangle,\:\: v_{il} = \vec{v}_i \cdot \vec{\hat{r}}_i,\:\: \langle v_l \rangle = \sum\limits_j^{N_r}  \frac{\vec{v}_j \cdot \vec{\hat{r}}_j}{N_r},\\
v_{t,i} &= \pm \sqrt{\sum\limits_\alpha v_{t,i\alpha}^2}, \:\: v_{t,i\alpha} = v_{i\alpha} - v_{il} \: \hat{r}_{i\alpha} - \langle v_{t\alpha} \rangle, 
\end{align}
\begin{align}
\langle v_{t\alpha} \rangle = \sum\limits_j^{N_r} \frac{v_{j\alpha} - v_{jl} \: \hat{r}_{j\alpha}}{N_r},\:\: \alpha = x,y, 
\end{align}
where $N_r$ is the number of particles at $r$ and $\vec{\hat{r}}_i = \vec{r}_i / | \vec{r}_i |$ is the normalized distance vector. For $v_{t,i}$ we have to choose a sign. We select $v_{t,i} < 0$ if $v_{t,i\alpha} < 0$. To locate the position of the shock, we also plot the radial velocity. We find two distinct features connected to non-equilibrium behavior. First, instead of immediately equilibrating with the hot matter upstream of the shock, the cold shocked matter seems to retain its MB distribution for a short time. This is especially visible for $\lambda = 10\:\Delta x$ and we mark the distributions by arrows in Fig.\:\ref{fv_lambda_40}\:(a) and (b). Furthermore, we find that for both, $\lambda = 10^{-5}\:\Delta x$ and $\lambda = 10\:\Delta x$, a few fast particles are moving ahead of the shock. To quantify the resulting deviations of the velocity distributions from equilibrium, we follow Marciante et al.\:\cite{Maricante17} and calculate the relation between the 4th and 2nd central velocity moments, $\langle \Delta v^4 \rangle$ and $\langle \Delta v^2 \rangle$, for $v_l$ and $v_t$:
\begin{align}
\langle \Delta v^k_{l/t} \rangle = \sqrt{\frac{m}{2 \pi kT}} \:  \frac{1}{N_r} \sum\limits_i^{N_r} \left(v_{l/t, i} \right)^k.
\end{align}
For MB, the moments are expressed as
\begin{align}
\langle \Delta v^k \rangle = \sqrt{\frac{m}{2 \pi \: kT}} \int \left( v - \langle v \rangle \right)^k \exp \left(- \frac{m v^2 }{2 \: kT} \right),
\end{align}
and should fulfill $\xi_{l/t} = \langle \Delta v^4_{l/t} \rangle / 3 \langle \Delta v^2_{l/t} \rangle^2 = 1$. \\
In Fig.\:\ref{fv_lambda_40}, we see that at $t = 4.0\:$s, $\xi_l$ and $\xi_t$ can be much larger than one and the deviations are present for both, $\lambda = 10^{-5}\:\Delta x$ and $10\:\Delta x$. For $\lambda = 10^{-5}\:\Delta x$, $\xi_l \gg 1$ only in a small region right at the shock front. The non-equilibrium behavior is a little surprising due to the low Mach numbers and the small mean-free-path. A likely cause is the finite minimal value of $\lambda$ (see discussion in Sect.\:\ref{kinetic_code}), which could lead to small non-continuum effects. For $\lambda = 10\:\Delta x$, the fast shock-heated particles can move farther into the cold matter. As a consequence, the regions with $\xi_{l/t} \gg 1$ are more extended and located ahead of the shock. Although for both, the longitudinal and transverse velocities, the deviations from MB are large, $\xi_t$ is smaller than $\xi_l$ and its deviation sets in a little later. This behavior is due to a higher velocity of the shock-heated particles in the direction of shock propagation. Furthermore, the larger density at the shock in the transverse direction can lead to more particle interactions that equilibrate matter and keep $\xi_t$ lower. \\
The velocity distributions at $t = 18.0\:$ms are plotted in Fig.\:\ref{fv_lambda_180}. As for $t = 4.0\:$s, $\xi_l$ and $\xi_t$ exceed one at the shock front. However, this time, their values are significantly smaller due to the relatively similar SIEs in matter upstream and downstream of the shock. 
\begin{figure}
\begin{center}
\includegraphics[width = 0.47\textwidth]{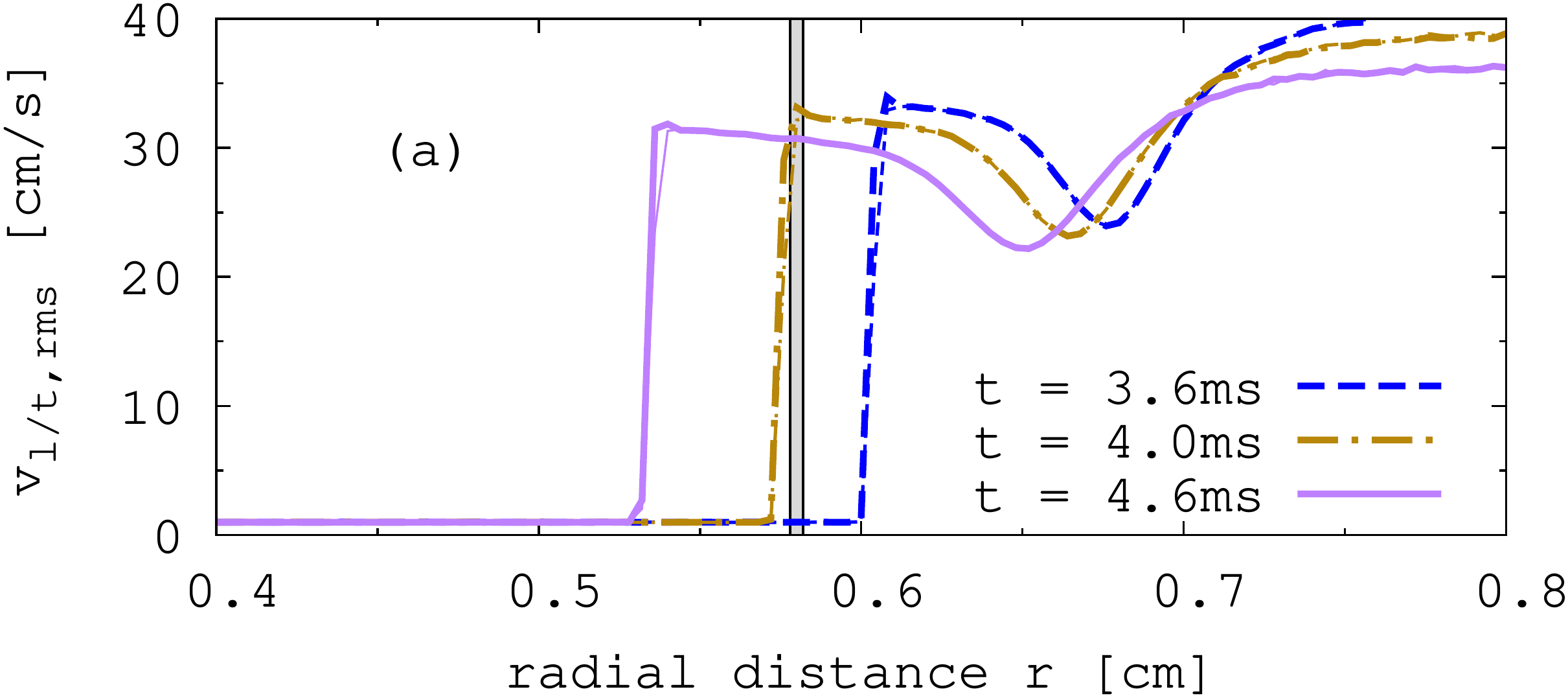}
\includegraphics[width = 0.47\textwidth]{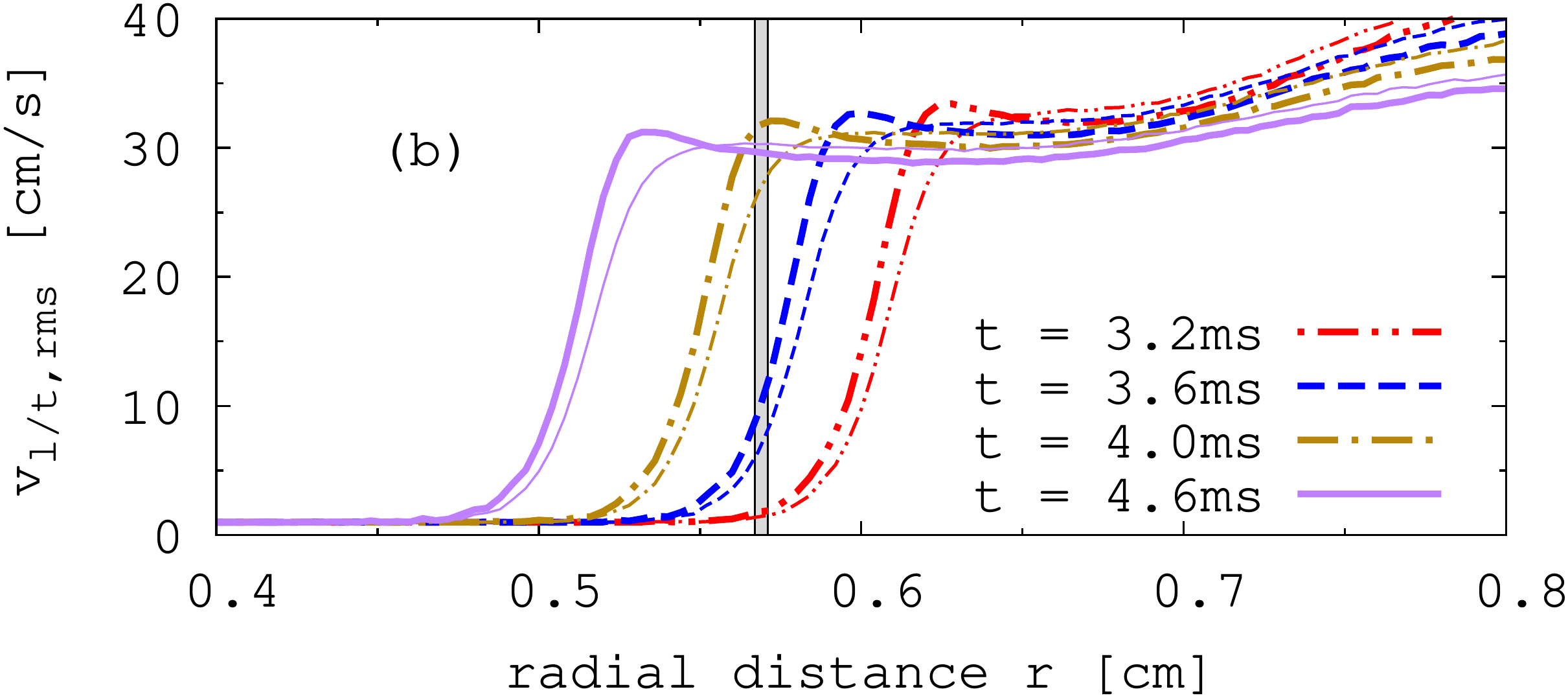}
\caption{Longitudinal (thick lines) and transverse (thin lines) velocities $v_{l,\mathrm{rms}}$ and $v_{t,\mathrm{rms}}$ for $\lambda = 10^{-5} \: \Delta x$ (a) and $\lambda = 10 \: \Delta x$ (b) at different times. Particles with a radial distance $0.578\: \mathrm{cm} \leq r \leq 0.582\:$cm (a) and $0.567\: \mathrm{cm} \leq r \leq 0.571\:$cm (b) (gray bands) are used to determine velocity distributions.}
\label{temperature_neq}
\end{center}
\end{figure}
\begin{figure}
\begin{center}
\includegraphics[width = 0.47\textwidth]{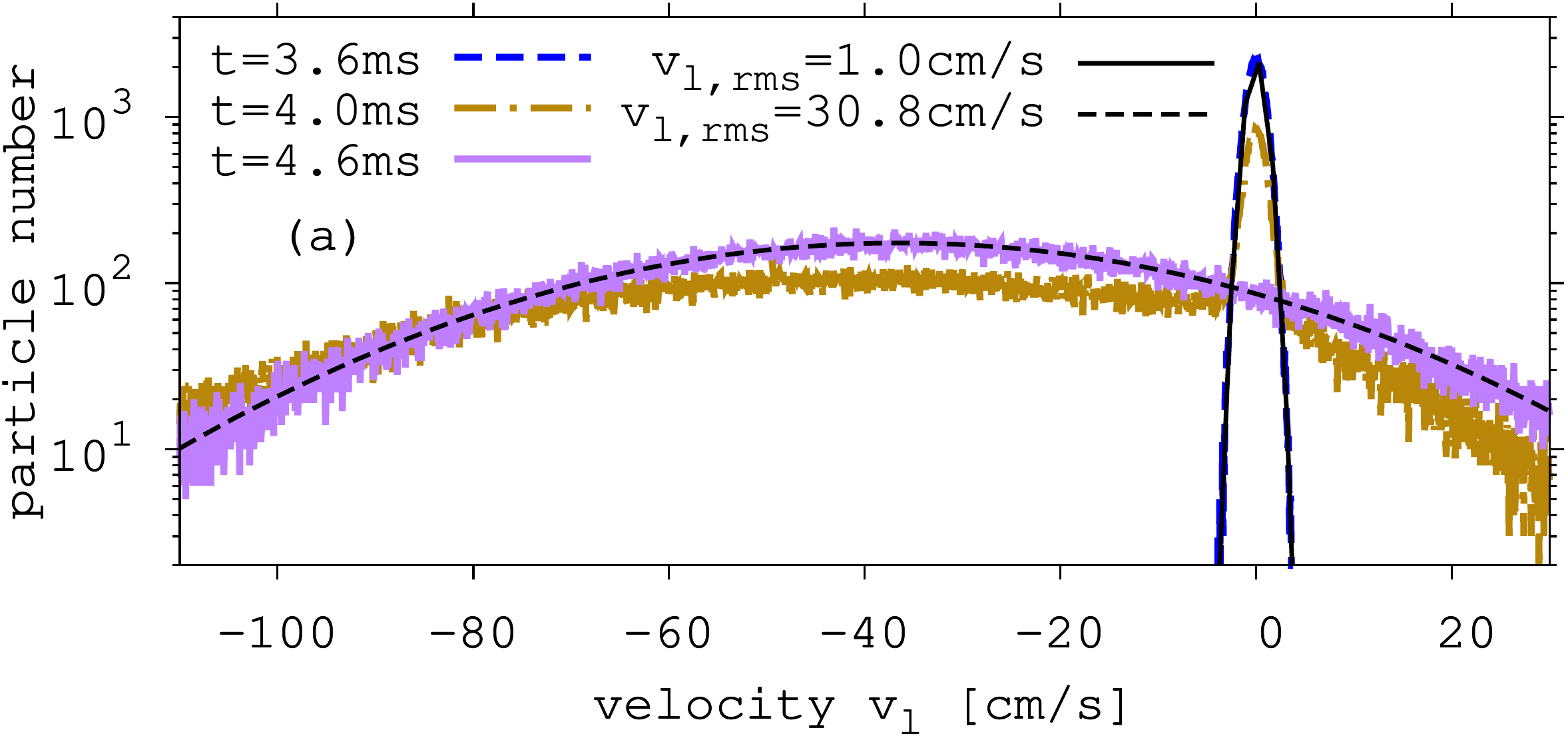}
\includegraphics[width = 0.47\textwidth]{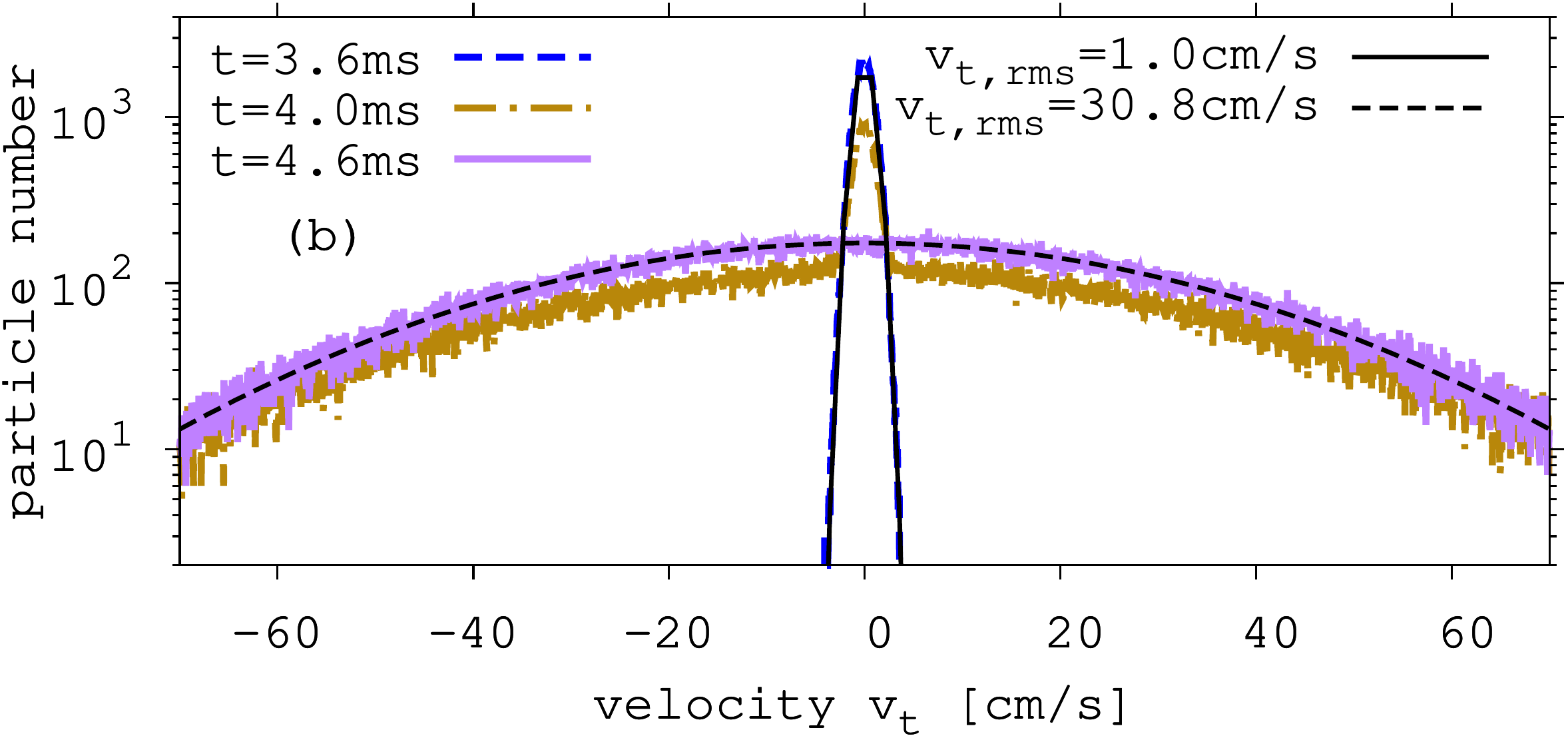}
\caption{Distributions of longitudinal (a) and transverse (b) velocities for particles with $0.578\: \mathrm{cm} \leq r \leq 0.582\:$cm and $\lambda = 10^{-5}\: \Delta x$. Before and after shock wave passage the distributions follow MB with given $v_{l/t,\mathrm{rms}}$ (thin lines). Non-equilibrium behavior is present at $t=4.0\:$ms.} 
\label{distributions_neq_h}
\end{center}
\end{figure}
\begin{figure}
\begin{center}
\includegraphics[width = 0.47\textwidth]{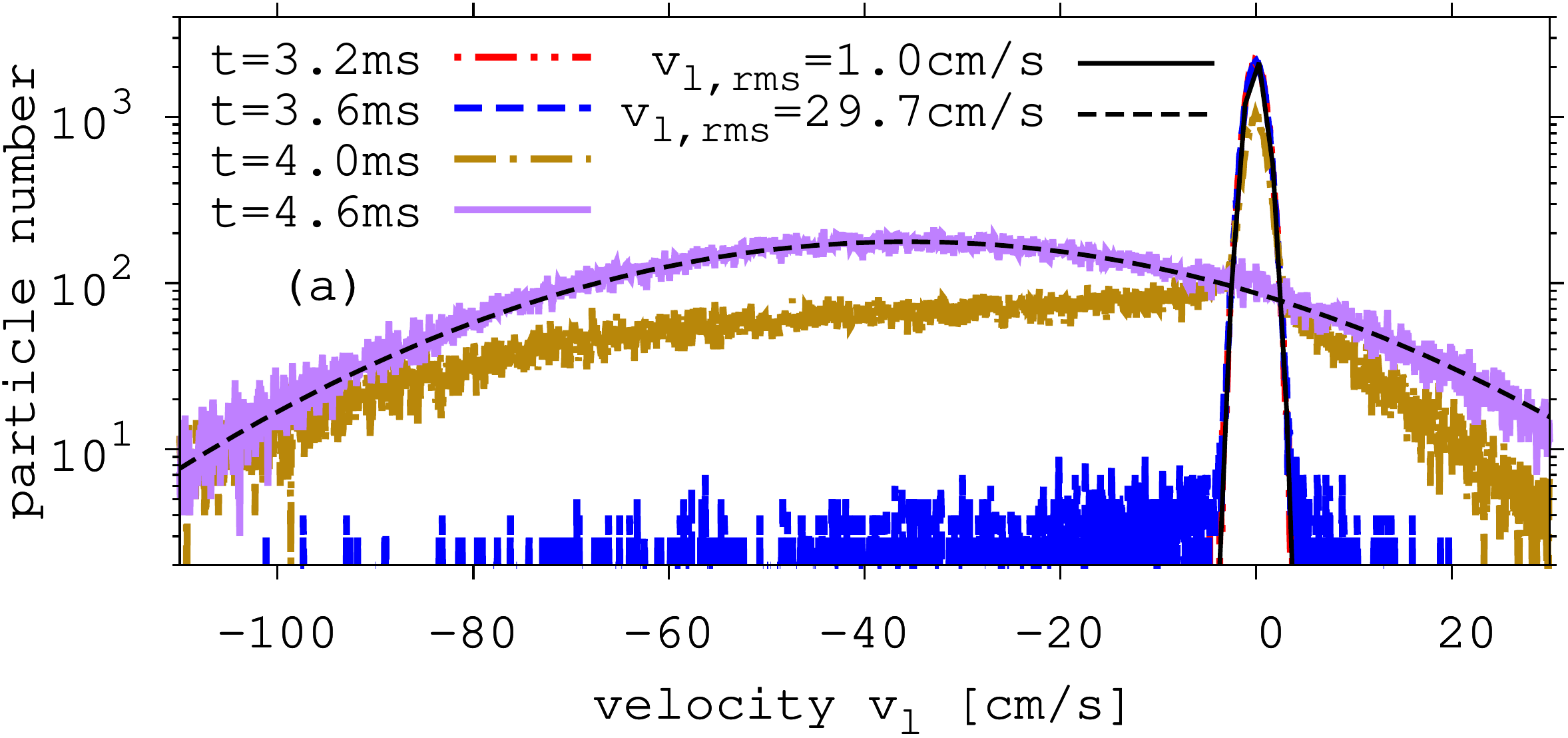}
\includegraphics[width = 0.47\textwidth]{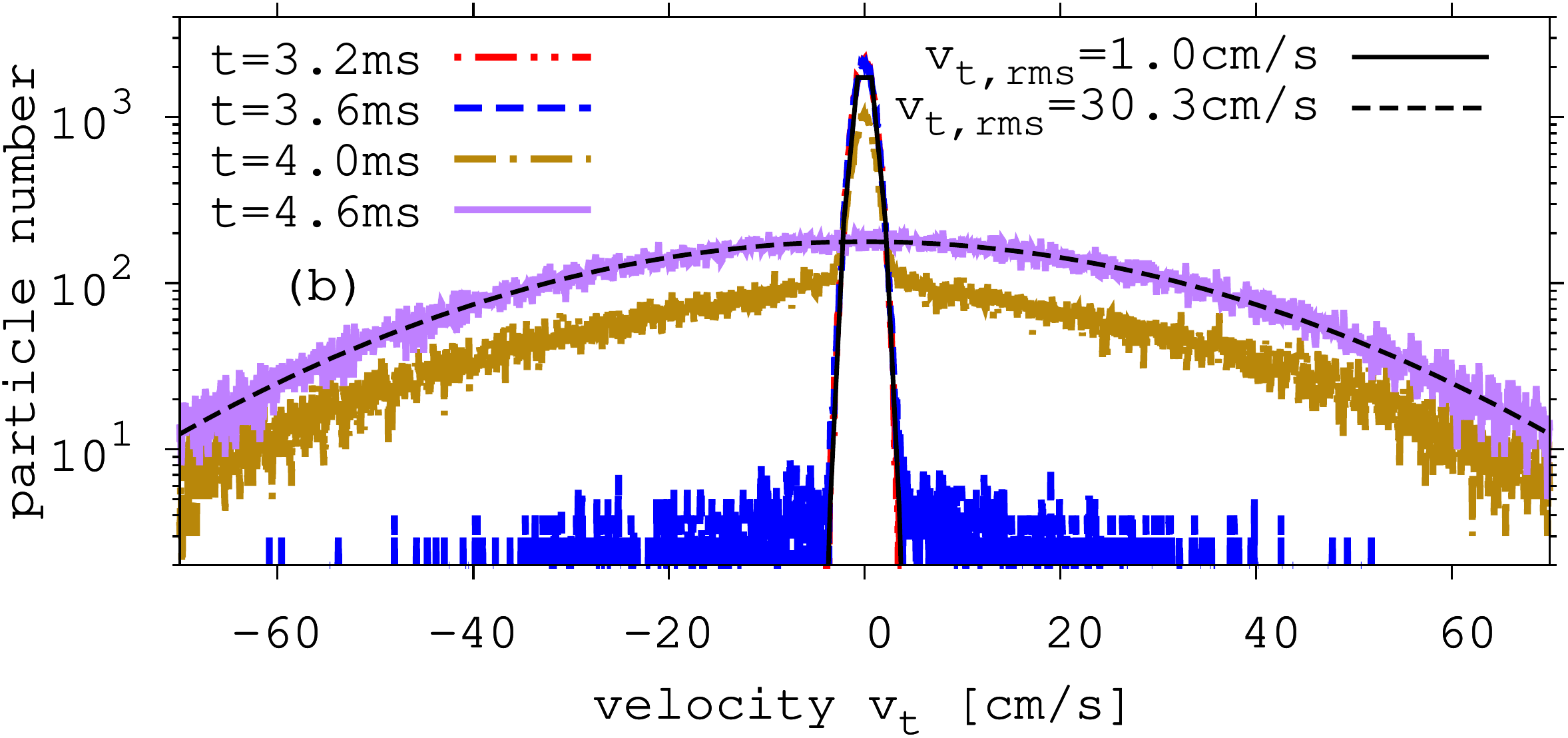}
\caption{Same as in Fig.\:\ref{distributions_neq_h} but for particles with $0.567\: \mathrm{cm} \leq r \leq 0.571\:$cm and $\lambda = 10\: \Delta x$. Non-equilibrium behavior is visible via the high-velocity tail of the MB distributions at $t=3.6\:$ms and the two-component structure at $t=4.0\:$ms.} 
\label{distributions_neq_10}
\end{center}
\end{figure}
This leads to faster particle equilibration and therefore smaller deviations from MB. As before, we find that while  for $\lambda = 10^{-5}\:\Delta x$ the longitudinal and transverse velocities are very similar, $\xi_t$ is smaller than and lags behind $\xi_l$ for $\lambda = 10\:\Delta x$, showing again that the non-equilibrium behavior is more pronounced in the direction of shock propagation.\\
\newline
We will now look in more detail at the velocity distributions as particles are accelerated by the shock. Fig.\:\ref{temperature_neq} shows the radial profiles of the longitudinal and transverse root-mean-square velocities $v_{l,\mathrm{rms}}$ and $v_{t,\mathrm{rms}}$ at different times. The velocities are determined from the 2nd central moments via $v_{l/t,\mathrm{rms}} = \langle v_{l/t}^2 \rangle^{1/2}$. Interestingly, despite the small values of $M$, we find an overshoot of $v_{l,\mathrm{rms}}$ at and around the shock. The overshoot is barely visible for $\lambda = 10^{-5}\:\Delta x$ but strongly pronounced for $10\:\Delta x$. Furthermore, for the latter, while $v_{l,\mathrm{rms}} > v_{t,\mathrm{rms}}$ at the shock, the relation inverses for larger $r$. Although the difference between  $v_{l,\mathrm{rms}}$ and $v_{t,\mathrm{rms}}$ is small, it points to separate equilibrations of the longitudinal and transverse particle velocities.\\
To analyze the velocity distributions, we use particles with $0.578\:\mathrm{cm} \leq r \leq 0.582\:$cm for $\lambda = 10^{-5}\:\Delta x$ and $0.567\:\mathrm{cm} \leq r \leq 0.571\:$cm for $\lambda = 10\:\Delta x$ (grey areas in Fig.\:\ref{temperature_neq}\:(a) and (b)). The distributions of the particle velocities are plotted in Fig.\:\ref{distributions_neq_h} and \ref{distributions_neq_10} for  $\lambda = 10^{-5}\:\Delta x$ and  $\lambda = 10\:\Delta x$, respectively. Before interacting with the shock (i.e. at $t \leq 3.6\:$ms for $\lambda = 10^{-5}\:\Delta x$ and $t \leq 3.2\:$ms for $\lambda = 10\:\Delta x$), they follow MB with $v_{l,\mathrm{rms}} =  v_{t,\mathrm{rms}} = 1 \: \mathrm{cm/s}$, which corresponds to the SIE in zone-1 in the initial conditions. For $\lambda = 10^{-5}\:\Delta x$, the particles interact with the shock at $t \sim 4.0\:$ms. At this time, their velocity distribution has a two-component structure - one component being the shock-heated matter and the other corresponding to the cold matter that has not equilibrated yet. The velocity peak at $v_{l/t} = 0$ is the non-equilibrium feature marked by an arrow in Fig\:\ref{fv_lambda_40}. Only a short time later, at $t=4.6\:$ms, all particle velocities have equilibrated and follow a MB with $v_{l/t,\mathrm{rms}} \sim 30.8\:$cm/s. Due to the larger shock width and earlier onset of particle acceleration, the non-equilibrium effects are more pronounced for $\lambda = 10 \: \Delta x$. At $t = 3.6\:$s, we can see the formation of high-velocity tails in the MB distributions caused by inflowing fast particles. As shown in Fig.\:\ref{distributions_neq_10}, the longitudinal velocity tail contains higher velocities than the transverse one, which explains the larger values of $\xi_l$ in comparison to $\xi_t$ found in Figs.\:\ref{fv_lambda_40} and \ref{fv_lambda_180}. Similar to $\lambda = 10^{-5}\:\Delta x$, the velocity distributions at $t = 4.0\:$ms have a two-component structure but equilibrate to one MB at $t= 4.6\:$ms.\\
\newline
In summary, in our simple implosion simulations we find non-equilibrium features that are also seen in other kinetic shock wave studies. One being the overshoot of the root-mean-square velocity component longitudinal to the shock direction of motion, the other manifesting itself in non-Maxwellian velocity distributions such as a tail of fast particles and a two-component structure due to the mixing of cold and hot matter. A detail in our simulations that will need further attention is the seemingly different equilibration of the longitudinal and transverse velocities for large mean-free-paths after the shock wave passage. It could be related to the expansion of matter behind the shock or be due to the simple treatment of the particle mean-free-paths in our simulation setup. 
\section{3-Zone Implosion Simulations}
\label{implosion_perturbation}
The previous section explored the implosion of a disk with initial homogeneous mass density. While it is a good benchmark setup, this configuration is very different from an ICF capsule. The latter typically contains low-density fusion fuel gas, enclosed by at least one shell of dense matter (e.g. D/T ice, plastic, glass). In this section, we will therefore follow the work of Joggerst et al.\:\cite{Joggerst2014}, who performed hydrodynamic 2D implosion simulations of disks that are divided into three zones with different densities. Similar configurations were previously used by Youngs and Williams \cite{Youngs08} to study turbulent mixing in spherical implosions. In this 3-zone setup the most inner region of the disk (zone-1) has a low mass density and intermediate SIE. It is surrounded by a dense shell (zone-2) with the same pressure but low SIE, while the outer layer (zone-3) has an intermediate mass density and very high SIE. Unlike the 2-zone configuration, this setup allows the development of fluid instabilities in the presence of seeds \cite{Youngs08, Casner12}. The implosion is driven by a time-dependent input of SIE $e_\mathrm{int}(r,t)$ and radial velocity $v_\mathrm{rad} (r,t)$ in a defined boundary region located in zone-3. Joggerst et al.\:\cite{Joggerst2014} apply different hydrodynamic codes and test the formation and evolution of fluid instabilities. One of the codes is \textsc{RAGE} and we will use the published results for comparison. The fluid instabilities originate either from numerical artifacts or from imposed perturbations, which are seeded on the interface between zone-1 and 2. Our simulations are done with the same setup and we want to see whether we find the same or similar behavior as the hydrodynamic code. As we have seen, kinetic simulations are prone to creating fluid instability seeds due to statistical noise (note however that the fluctuation in DSMC can correspond to real thermal fluctuations when each particle represents a single actual molecular particle \cite{Garcia86, Gallis16}) and one possible consequence is that the resulting instabilities will dominate the ones arising from imposed perturbations. 
\subsection{Unperturbed Simulation}
\label{inst_int_eng_in}
\begin{figure}
\begin{center}
\includegraphics[width = 0.45\textwidth]{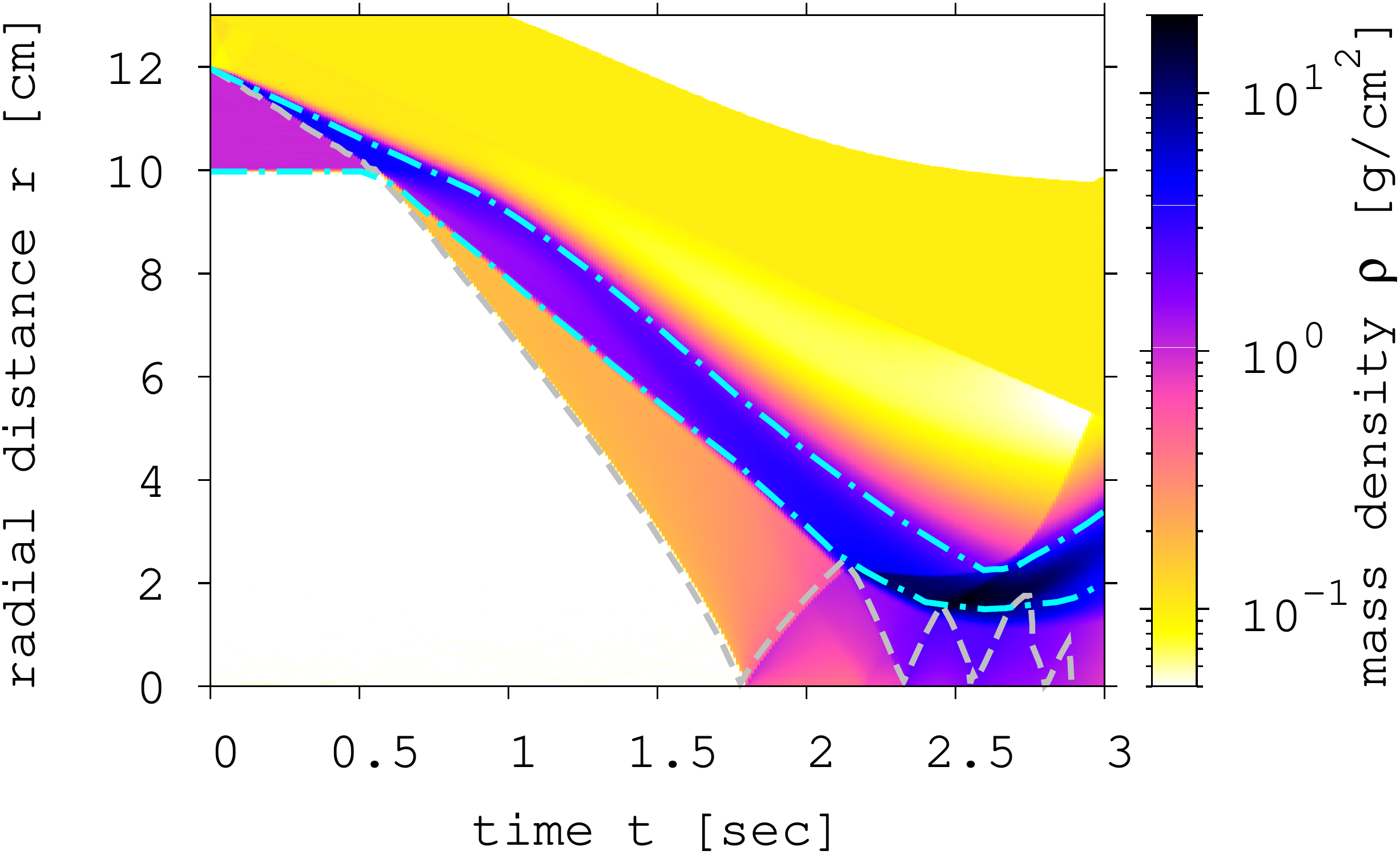}
\caption{Time evolution of the mass density $\rho$ radial profile in the 3-zone implosion. For comparison, we add the inner and outer edge of zone-2 (dashed-dotted lines) and the shock location (dotted line) for a 1D \textsc{RAGE} simulation with the same setup \cite{Joggerst2014}.}
\label{mode0_time}
\end{center}
\end{figure}

The disk has initially a radius of 15\:cm, whereas zone-1 extends from 0\:cm to 10\:cm, zone-2 from 10\:cm to 12\:cm, and zone-3 from 12\:cm to 15\:cm. Particles in each zone are initialized according to the SIE and mass density in table \ref{disk_parameters}. To simulate an ideal gas with $\gamma = 5/3$ as in \cite{Joggerst2014}, we allow particles to have three velocity degrees of freedom. This is different from the 2-zone simulations where particles only have x- and y-velocities, corresponding to the directions in which they propagate. Now, although we still restrict the particle motion to the x-y plane, particles have x-, y-, and z-velocities, which are updated according 3D kinematics in each collision \cite{Sagert14}. Furthermore, instead of initializing particles with the same mass but different number densities we choose a homogeneous particle distribution with zone-dependent masses. This is done for computational reasons. Since the density in zone-2 is twenty times higher than in zone-1, we would have to place twenty times more particles in each computational cell. As a consequence, the search for a collision partner would require a long time. With a homogeneous initial particle distribution, particle masses are assigned according to the zone-dependent mass density $\rho_Z$ (see table \ref{disk_parameters}) while their positions are chosen randomly in a disk with radius 15\:cm. In pure DSMC simulations, one must be cautious when assigning different masses to particles as these enter directly in the determination of interaction probabilities and number of interacting particles \cite{Boyd96}. Large differences in the masses can result in individual particle interactions being non-conservative with regard to energy and momentum \cite{Boyd96}. This problem should not occur in our simulations, since the selection of scattering partners is not directly dependent on masses and the interaction in the center-of-mass frame explicitly conserves energy and momentum.\\
\newline
For the implosion, the time-dependent SIE and radial velocity are imposed in a boundary region defined by $r \geq R_\mathrm{bd}$ with $R_\mathrm{bd} (t) = R_0 (1 - u_\mathrm{bd} t )$, $R_0 = 13\:$cm, and $u_\mathrm{bd} = 0.2\:$s$^{-1}$. The radial velocity is $v_\mathrm{rad} (r, t)= - u_\mathrm{bd} \: {R_0 \: r}/ { R_\mathrm{bd} (t)}$, while the SIE is kept constant at $e_\mathrm{int}(r,t)= 150\:$erg/g for $t \leq 0.5$\:s. For $t > 0.5\:$s, it is decreased linearly with time to $0.15\:$erg/g at $t = 3.0\:$s. Furthermore, the mass density in the boundary region is kept constant at $\rho_\mathrm{bd} = 0.10\:\mathrm{g/cm^2}$. In our simulations, the above conditions are implemented in the following way:  At the beginning of each iteration, we determine the radius $R_\mathrm{bd}$. Particles with $r \geq R_\mathrm{bd}$ are assigned new random positions in the boundary region to achieve a constant density $\rho_\mathrm{bd}$. Each particle is then given a new velocity with thermal components according to $e_\mathrm{int} (r, t)$ and radial components according to $v_\mathrm{rad} (r, t)$. We use $N = 4.0 \times 10^7$ particles in a simulation space with $0 \leq x,y \leq 40\:$cm and the center of the disk at $x = y = 20\:$cm. 
\begin{table}
\centering
\begin{tabular}{| c | c | c | c |}                    
\hline
Zone Z& Radial size $r$ & Density $\rho_\mathrm{Z}$   & SIE $e_\mathrm{int,Z}$ \\
\hline 
\hline 
1      & (0 - 10)\:cm          &  0.05\:g/cm$^2$          & 3.00\:erg/g    \\
2      & (10 - 12)\:cm        & 1.00\:g/cm$^2$           & 0.15\:erg/g    \\
3      & (12 - 15)\:cm        & 0.10\:g/cm$^2$           &150.0\:erg/g     \\
\hline  
\end{tabular}
\caption{Initial conditions of the 3-zone implosion.}
\label{disk_parameters}
\end{table}
The boundary conditions are free everywhere and the simulations are run in the continuum limit with $\lambda = 10^{-5}\:\Delta x$.\\
Fig.\:\ref{mode0_time} shows the evolution of the mass density
for a simulation time of three seconds. The shock wave forms at the outer boundary of zone-2 due to the high pressure exerted by matter in zone-3. Unlike the 2-zone setup, it is now the imposed SIE and radial velocity and not the rocket effect that compress the capsule. Because all particles in the boundary region are directed inward, they cannot leave the disk surface and the entire configuration stays compact. The shock breaks out from zone-2 into zone-1 at $t \sim 0.5\:$s and rebounds at $t\sim 1.8\:$s. As it reaches a distance of $r \sim 2.1\:$cm at $t\sim 2.1\:$s, the shock encounters the converging dense shell. Although the shell decelerates due to the interaction, it continues to move inwards and requires several encounters with the shock to come to a halt. To compare to \textsc{RAGE}, we add the dense shell profiles and shock positions of a 1D \textsc{RAGE} implosion simulation from Joggerst et al.\:\cite{Joggerst2014}. All in all, both calculations are very similar. We notice that the boundaries of zone-2 in the kinetic study are smeared out. This is due to the formation of RTIs which will be discussed in the next section. Furthermore, in the 1D \textsc{RAGE} implosion, the shock is a bit ahead of the kinetic simulation. This difference is either due to the formation of fluid instabilities in the kinetic study, the higher resolution in \textsc{RAGE}, or could originate from small differences in 1D vs. 2D simulations. 
\subsection{Induced Perturbations}
\begin{figure*}
\begin{center}
\includegraphics[width = 0.31\textwidth]{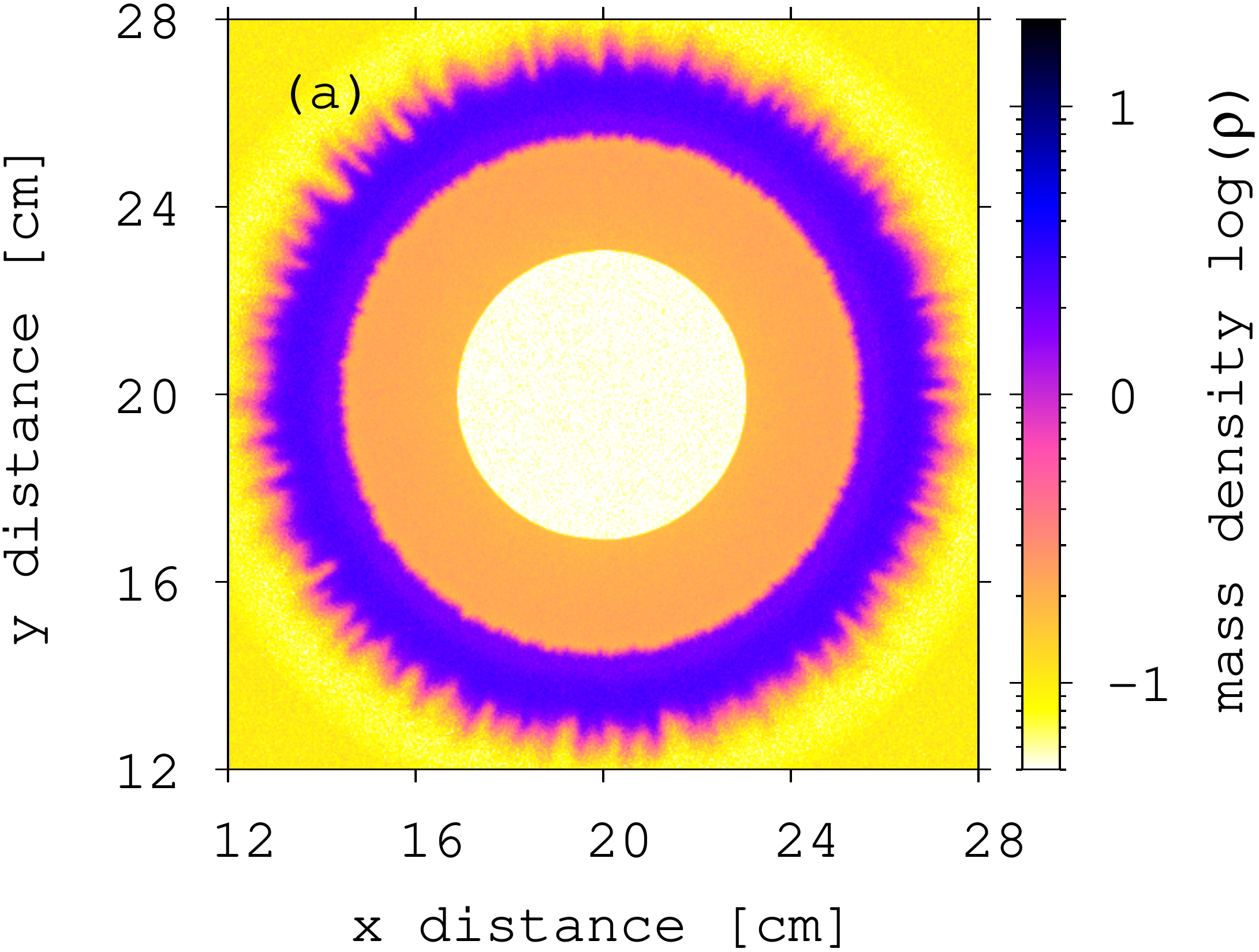}\hfill
\includegraphics[width = 0.31\textwidth]{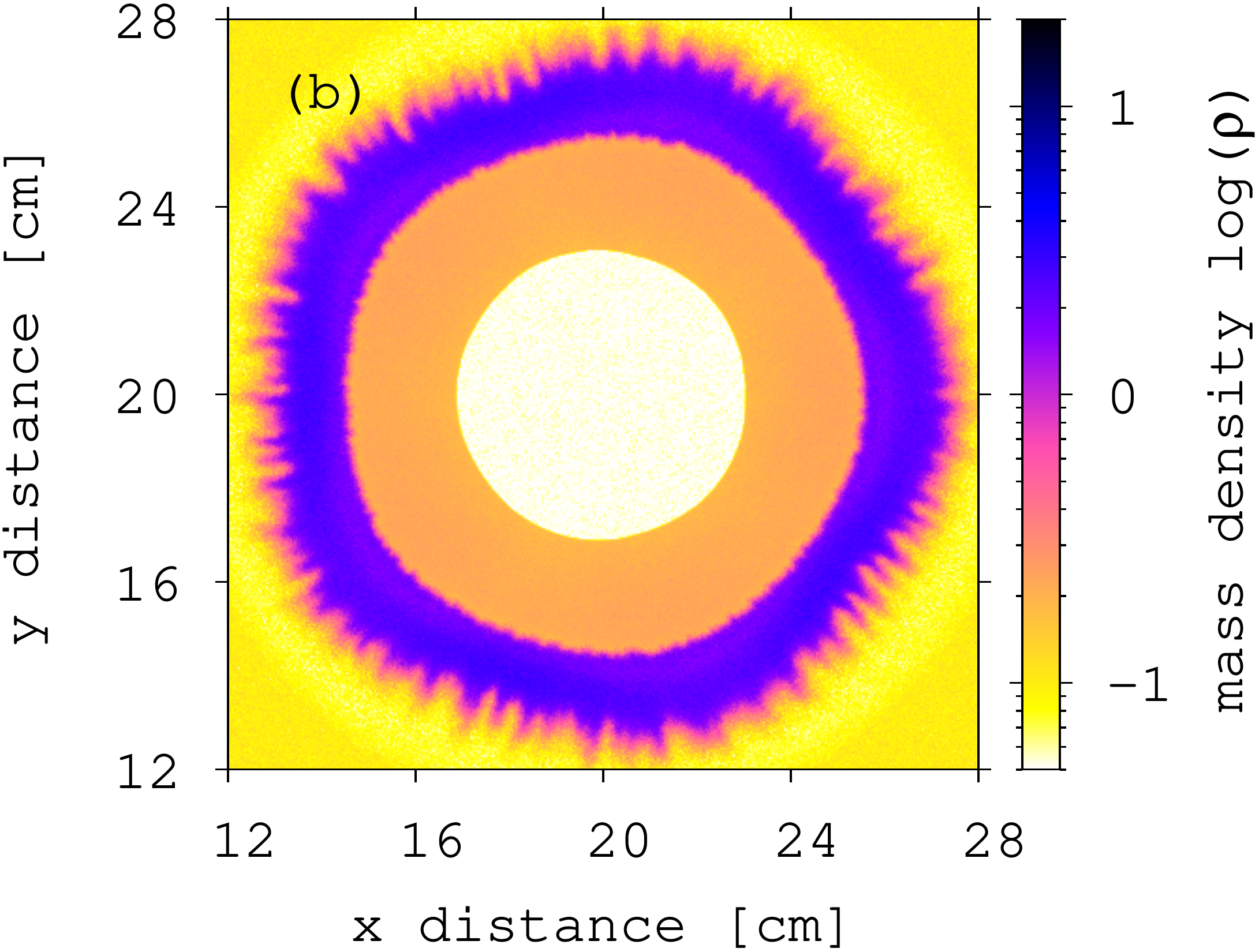}\hfill
\includegraphics[width = 0.31\textwidth]{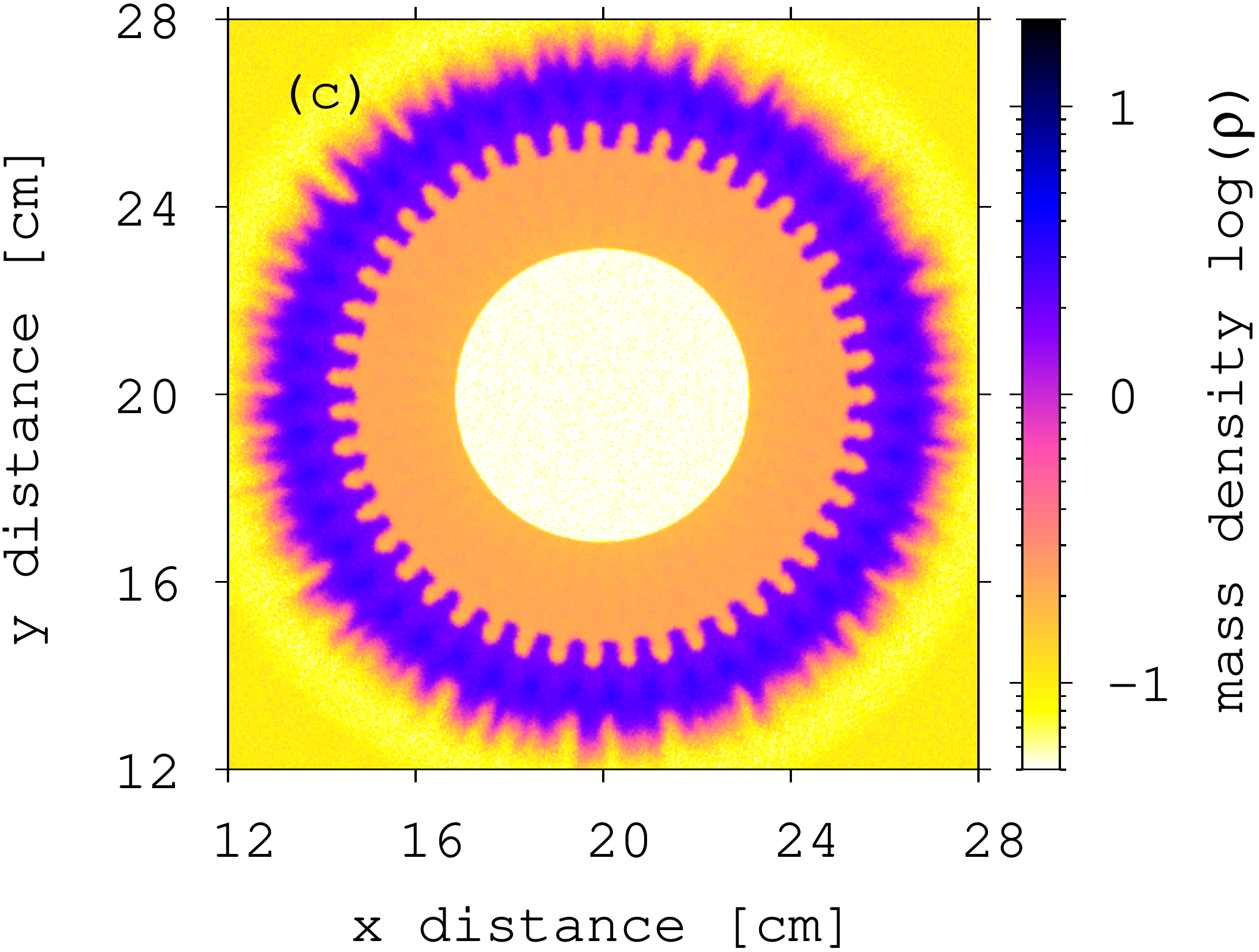}\hfill
\includegraphics[width = 0.315\textwidth]{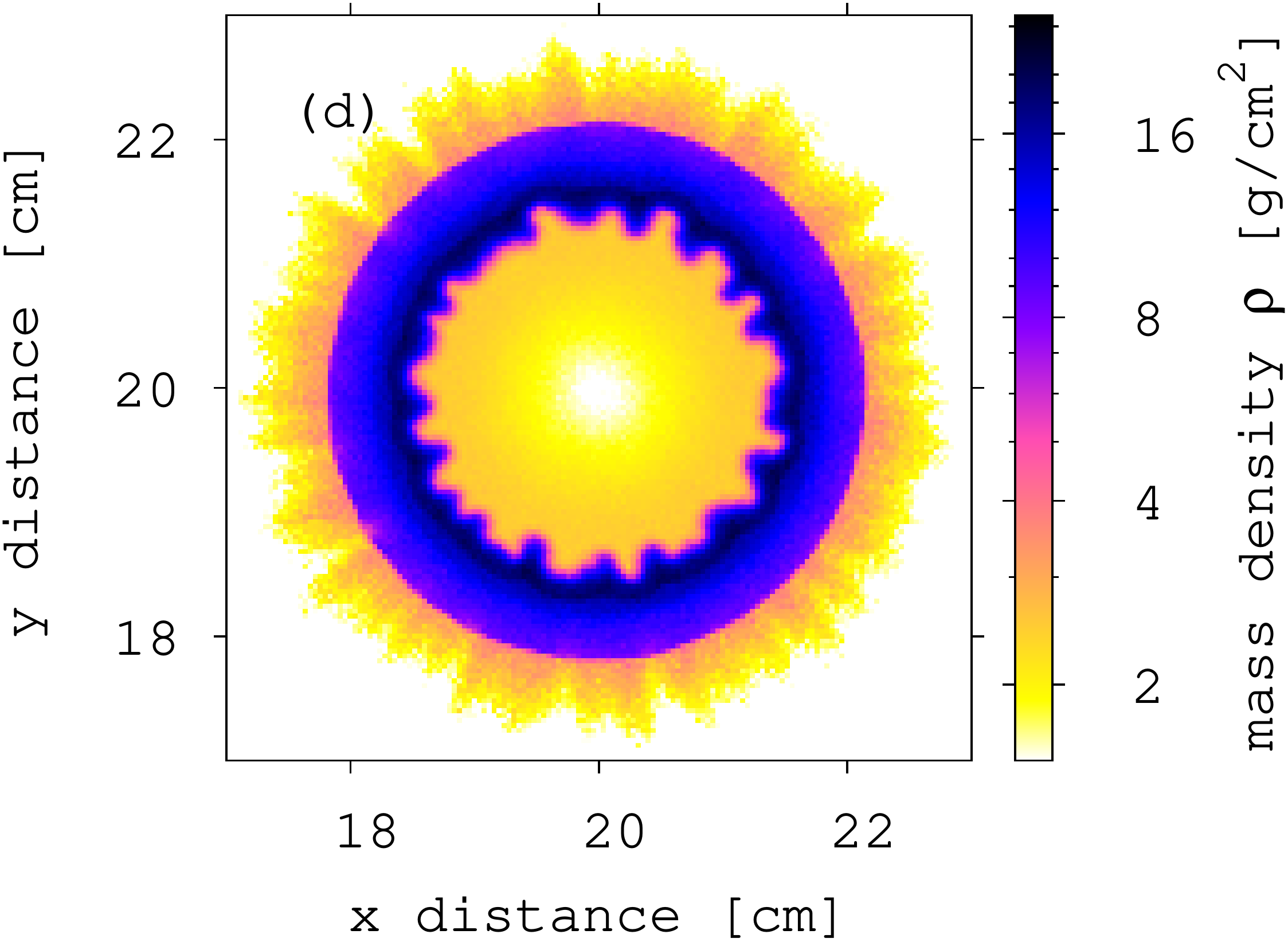}\hfill
\includegraphics[width = 0.315\textwidth]{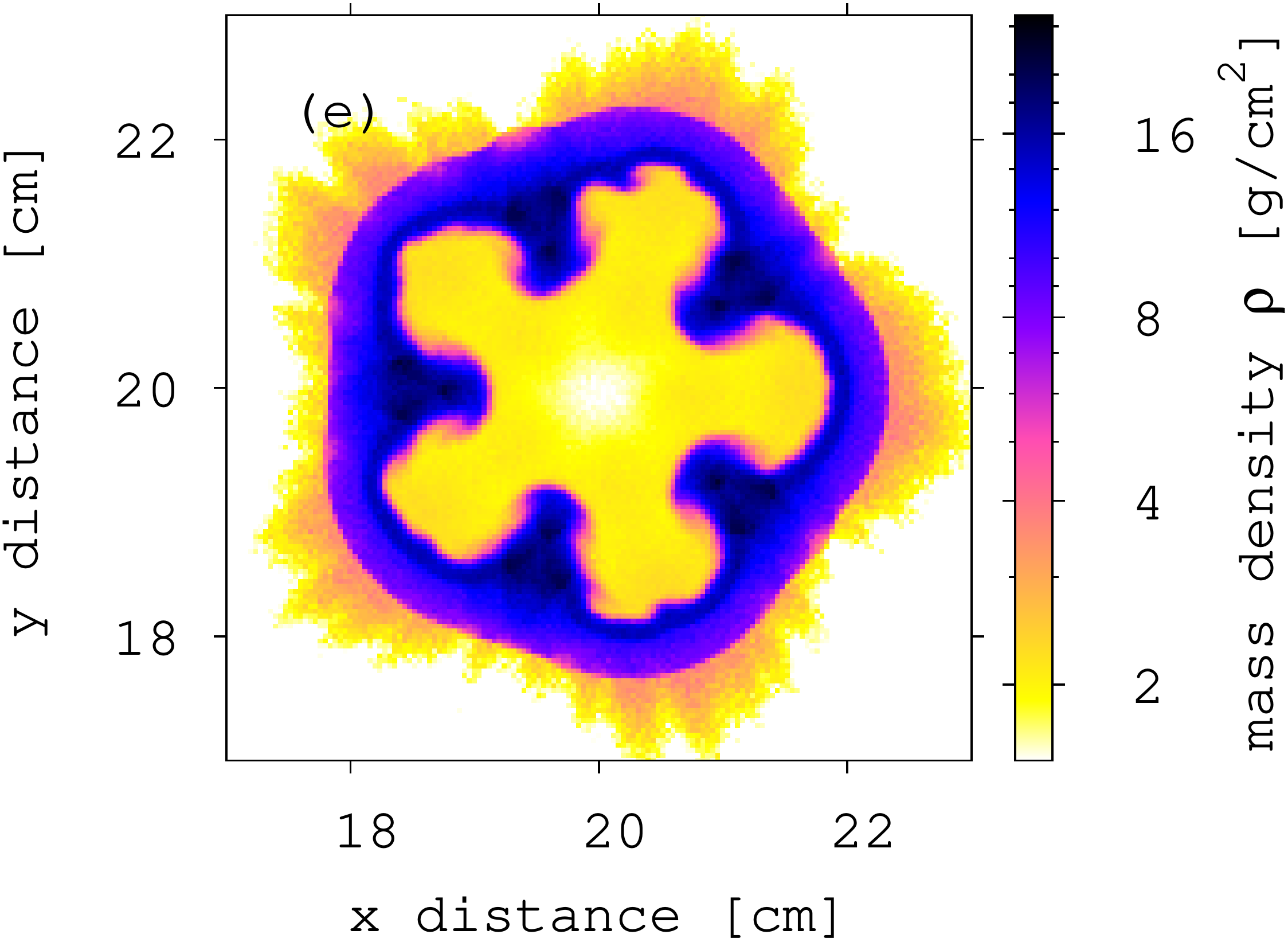}\hfill
\includegraphics[width = 0.315\textwidth]{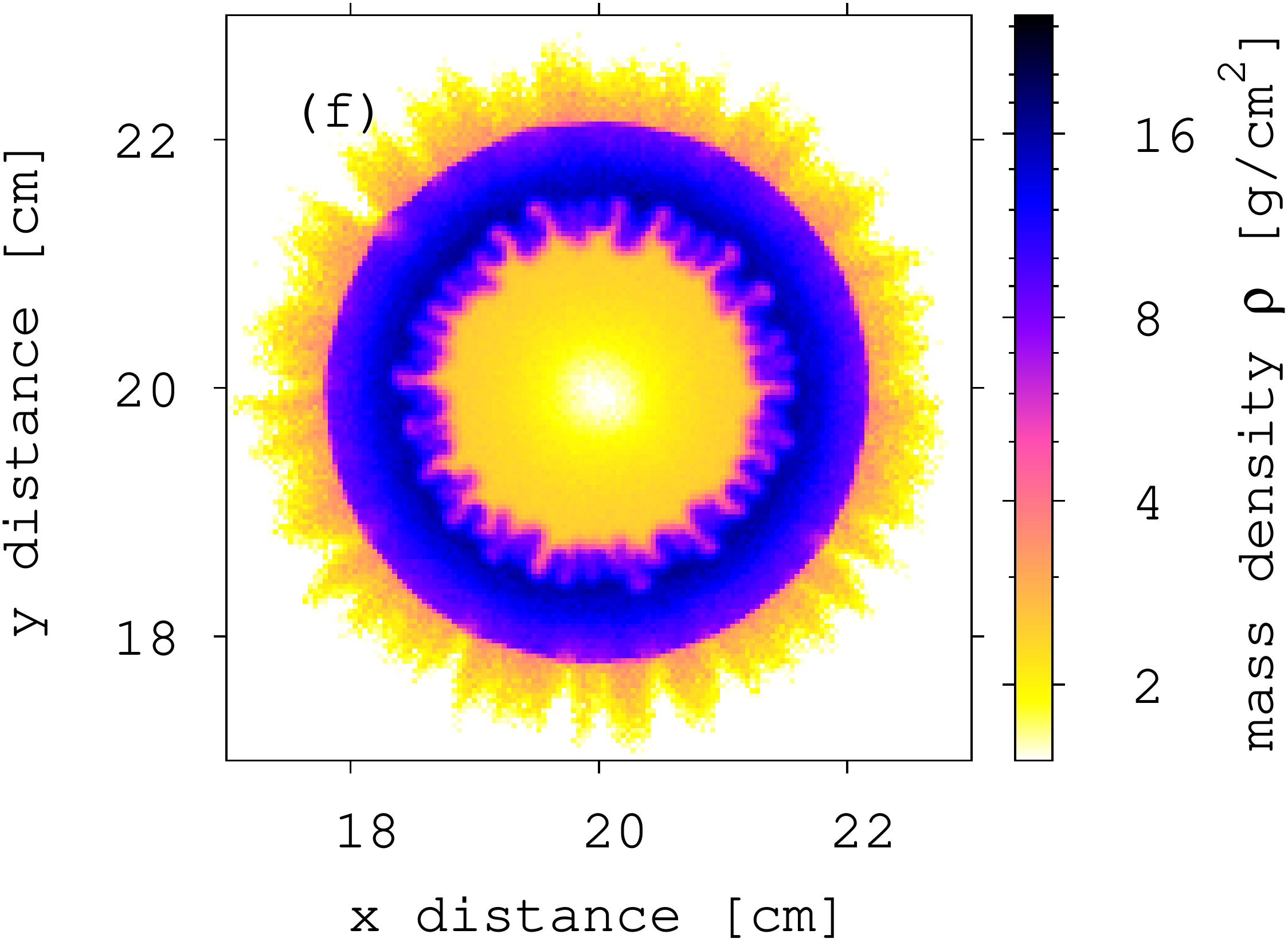}\hfill
\caption{Snapshots of the 2D density distribution for implosions with seeded perturbations with mode 0 (a,\:d), mode 5 (b,\:e) and mode 47 (c, \:f) at $t = 1.5 \: \mathrm{sec}$ (upper row) and $t = 2.5 \: \mathrm{s}$ (lower row) (see text for details). The simulations use $4.0 \times 10^7$ particles and apply the same setup as in Joggerst et al.\:\cite{Joggerst2014}.}
\label{imp_pert_3d}
\end{center}
\end{figure*}

\begin{figure}
\begin{center}
\includegraphics[width = 0.47\textwidth]{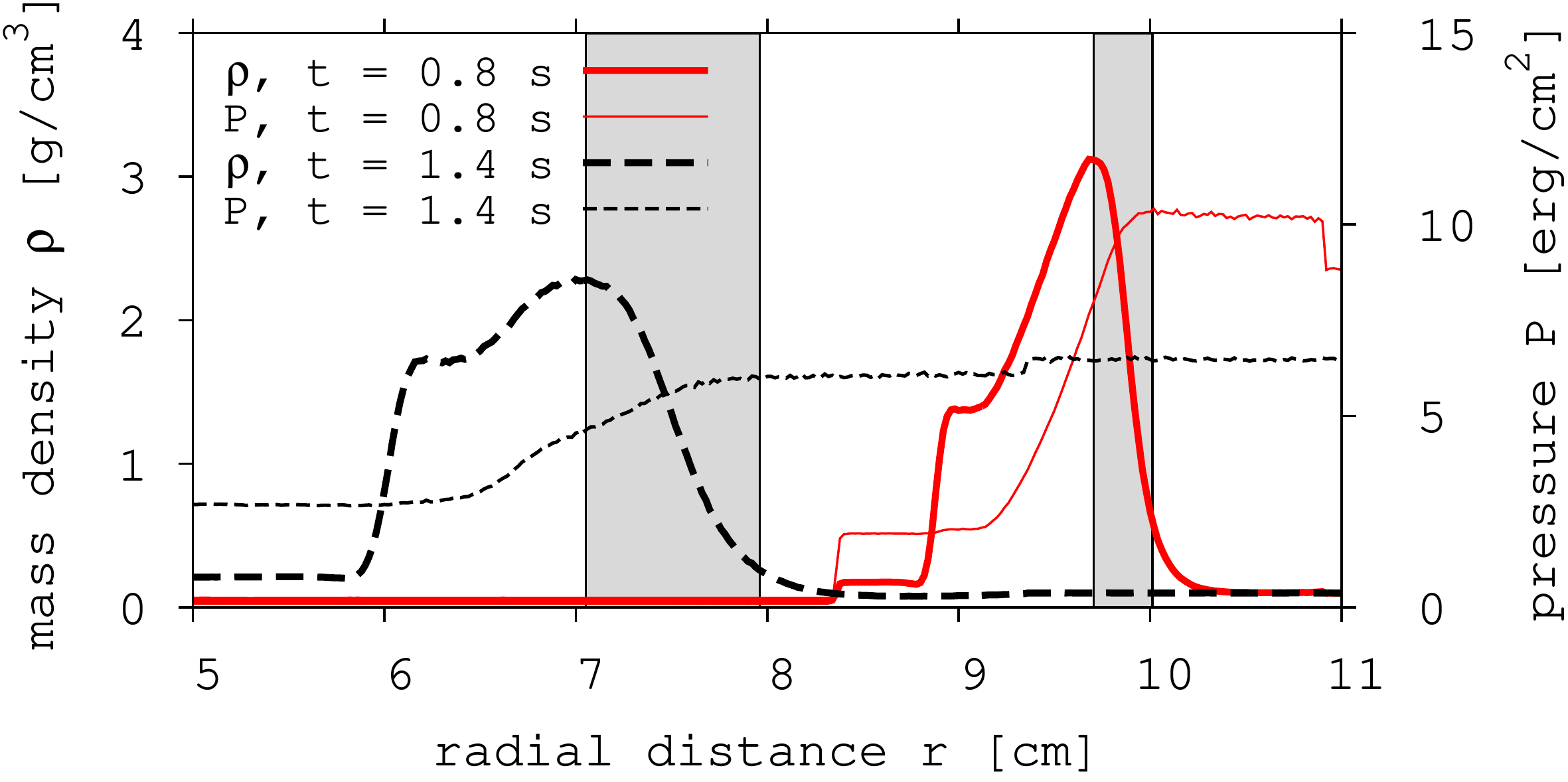}\hfill
\caption{Density $\rho$ (thick lines) and pressure $P$ (thin lines) radial profiles at $t = 0.8\:$s (solid lines) and $t = 1.4\:$s (dashed lines). Gray bands indicate regions with opposite pressure and density gradients that are unstable with respect to RTIs.}
\label{density_pressure_comp3}
\end{center}
\end{figure}

\begin{figure}
\begin{center}
\includegraphics[width = 0.45\textwidth]{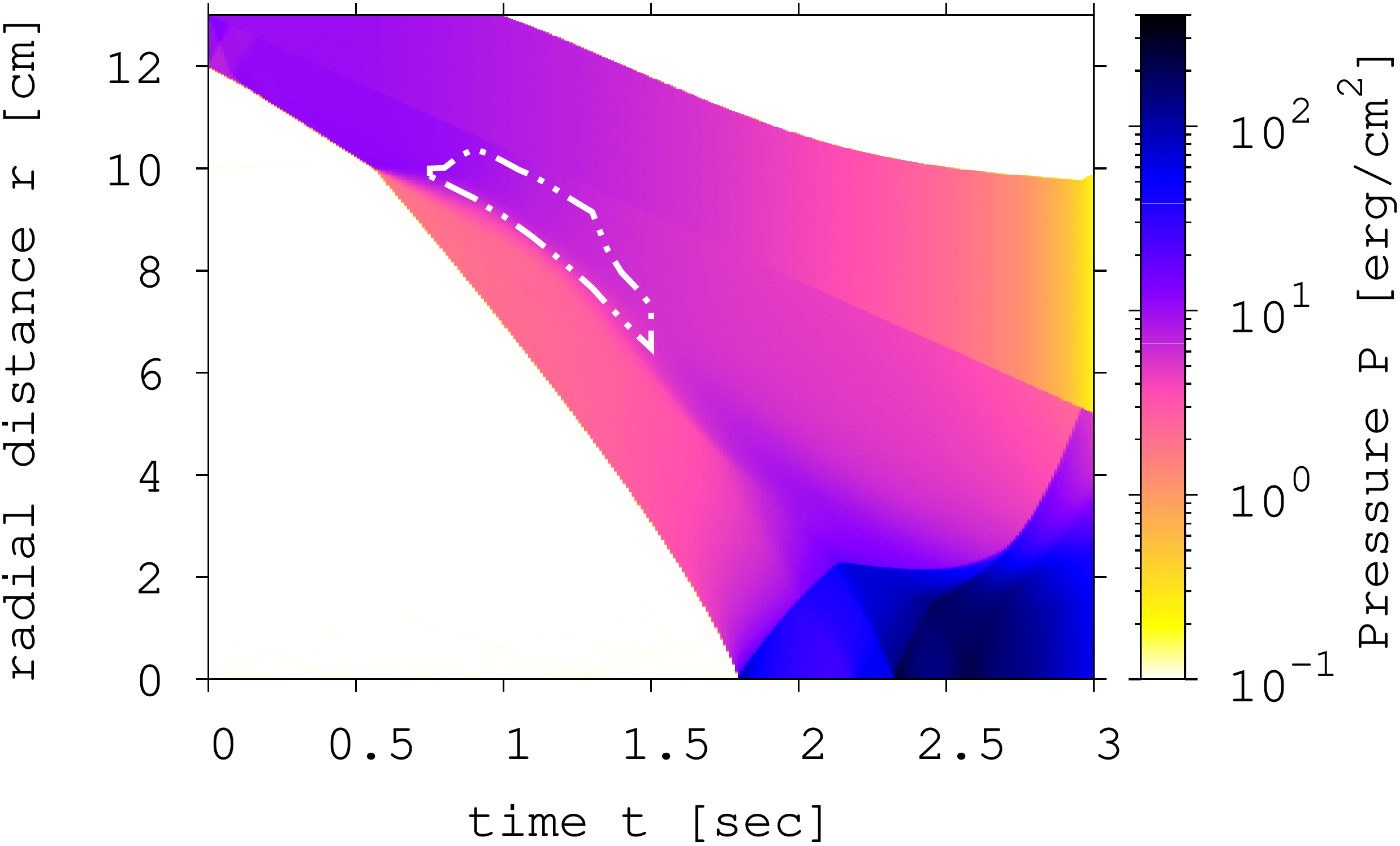}\hfill
\caption{Time evolution of the pressure radial profile in the 3-zone implosion. The dashed-dotted-dotted line marks a region with opposite density and pressure gradients.}
\label{mode0_temp_time}
\end{center}
\end{figure}

To study the evolution of fluid instabilities, we modify the interface of zone-1 and 2 by adding single-mode perturbations with $A_\mu \cos \left( \mu \: \theta \right)$, where $\theta$ is the angle, $A_\mu = 0.125\:$cm the amplitude, and $\mu$ the mode. Chosen values are $\mu = 5$ for long-wavelength perturbations and $\mu = 47$ for short wavelengths. The previously discussed unperturbed simulation corresponds to $\mu = 0$. We find that the general dynamical evolution is similar in all cases, with shock creation, breakout, and rebound at roughly the same times. In Fig.\:\ref{imp_pert_3d}, we plot 2D density distributions for all modes at $t=1.5\:$s and $2.5\:$s. The imposed instabilities are most pronounced for $\mu=47$, which is due to the exponential dependence of the RTI growth rate on the wavenumber \cite{Chandra61, Frieman54}. \\
\newline
We can compare Fig.\:\ref{imp_pert_3d} to Figs.\:4 and 5 in Joggerst et al.\:\cite{Joggerst2014} and find that the general disk configurations are similar. However, in all kinetic simulations, the outer edge of the dense shell has filament-like structures that are not present in \textsc{RAGE}. They result from RTIs that are associated with a rarefaction wave. The wave is created at shock breakout from zone-2 into zone-1. As it propagates outwards it carries perturbations from the inner to the outer edge of zone-2, also known as feedout \cite{Agli10}. The perturbations are seeds for RTIs that grow due to opposite pressure and density gradients. As an example of the latter, we plot the radial profiles of $\rho$ and $P$ for $t = 0.8\:$s and $t = 1.4\:$s in Fig.\:\ref{density_pressure_comp3} and mark the corresponding unstable regions by gray areas. In addition, the entire unstable area is given via a dashed-dotted line in Fig.\:\ref{mode0_temp_time}, which 
shows the pressure as a function of time. The question arises why we don't see the filament-like structures in \textsc{RAGE}.\ Comparing the $t=1.5\:$s disk configurations in Fig.\:\ref{imp_pert_3d} to the \textsc{RAGE} simulation \cite{Joggerst2014}, we find that in the latter zone-2 has very smooth edges. This is most likely due to the higher resolution of \textsc{RAGE}, which is in contrast to our kinetic calculations that are accompanied by statistical noise. As a consequence, although the opposite $\rho$ and $P$ gradients should exist in the \textsc{RAGE} simulations, the higher resolution might prevent RTI seeds from forming.\\
However, different to the outer edge, the inner edge of zone-2 seems to have numerically induced small-scale instabilities in \textsc{RAGE}. They are best visible for $\mu = 0$ at $t = 2.5\:$s \cite{Joggerst2014}. The instabilities are also present in the kinetic study (see e.g. Fig.\:\ref{imp_pert_3d}(d)), but \textsc{RAGE} resolves them to a much smaller wavelength. For the kinetic studies, the instabilities can be linked to RMIs that are created during shock breakout from zone-2 into zone-1 or be a result of so-called feedthrough from the outer edge of zone-2 inwards \cite{Casner12}. They are barely visible at $t=1.5\:$s but, by $t = 2.5\:$s, have been amplified via interactions with the reflected shock. For $\mu=47$, they are least pronounced, being dominated by the induced fluid instabilities.\\
At late times, the short-wavelength fluid instabilities ($\mu=47$) interact with each other. While in \textsc{RAGE}, they form a mixing layer with many small-scale structures, the corresponding region in the kinetic simulation looks quite different. Here, due to the lower resolution of the kinetic code, the structures are much coarser and their amplitude smaller. The large-wavelength instabilities for $\mu=5$, on the other hand, are well resolved by the kinetic code. This is the reason 
\begin{figure}
\begin{center}
\includegraphics[width = 0.47\textwidth]{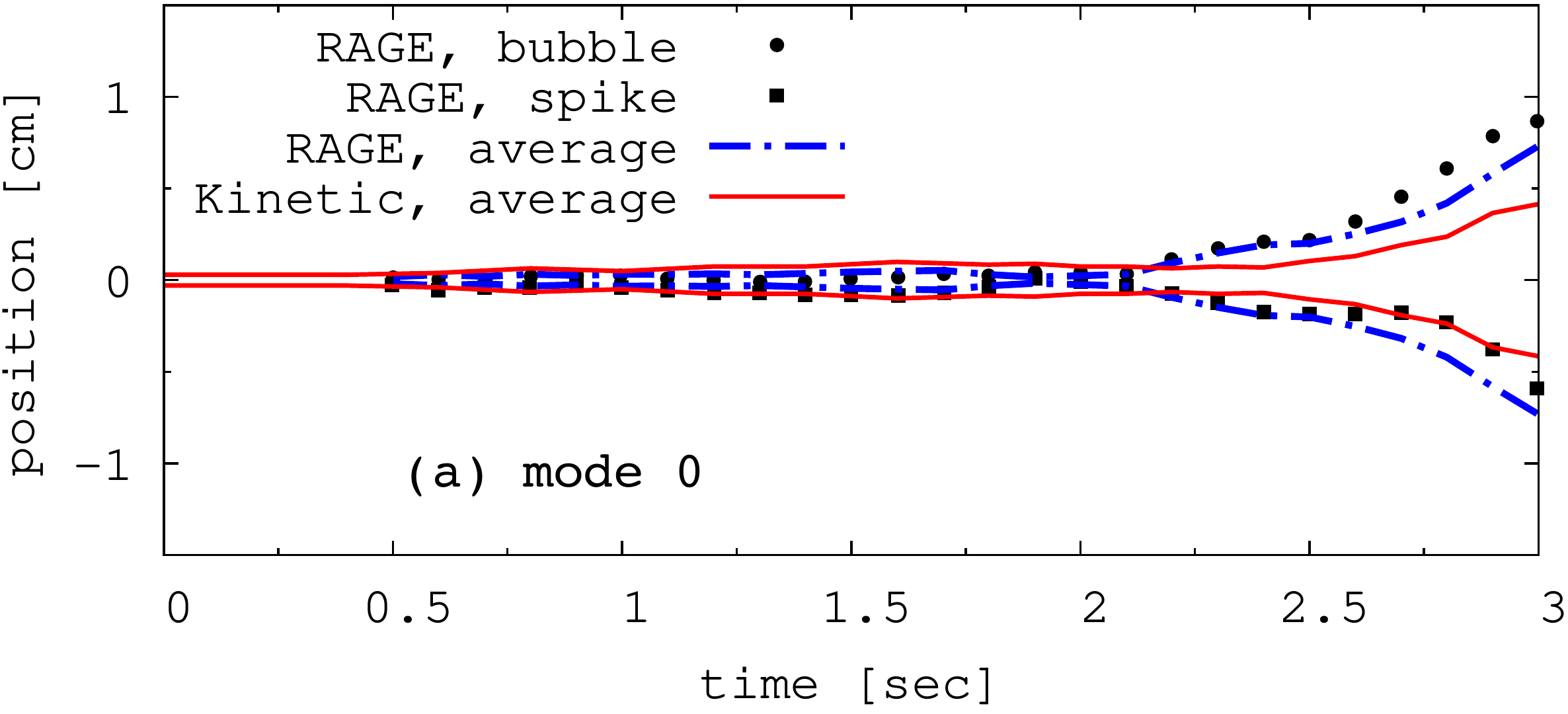}
\includegraphics[width = 0.47\textwidth]{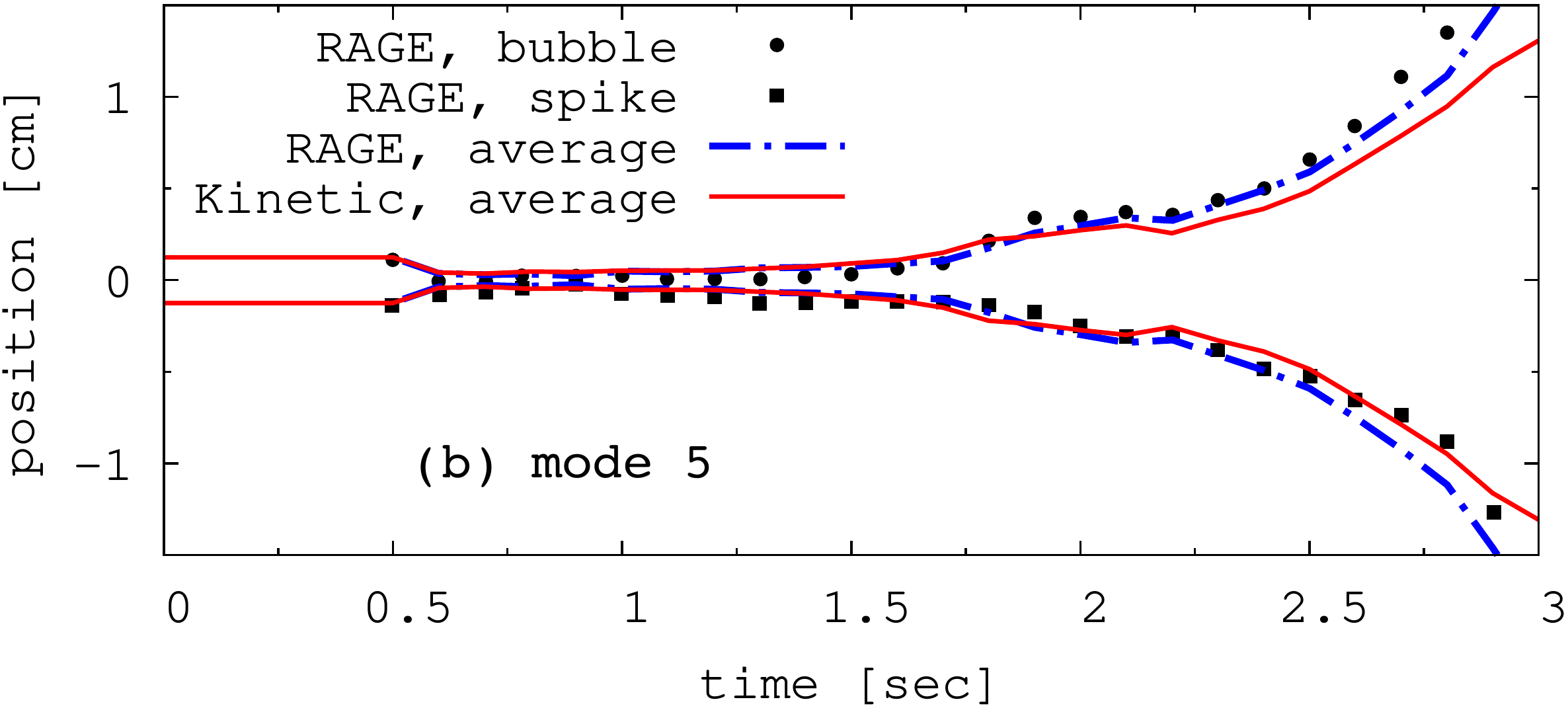}
\includegraphics[width = 0.47\textwidth]{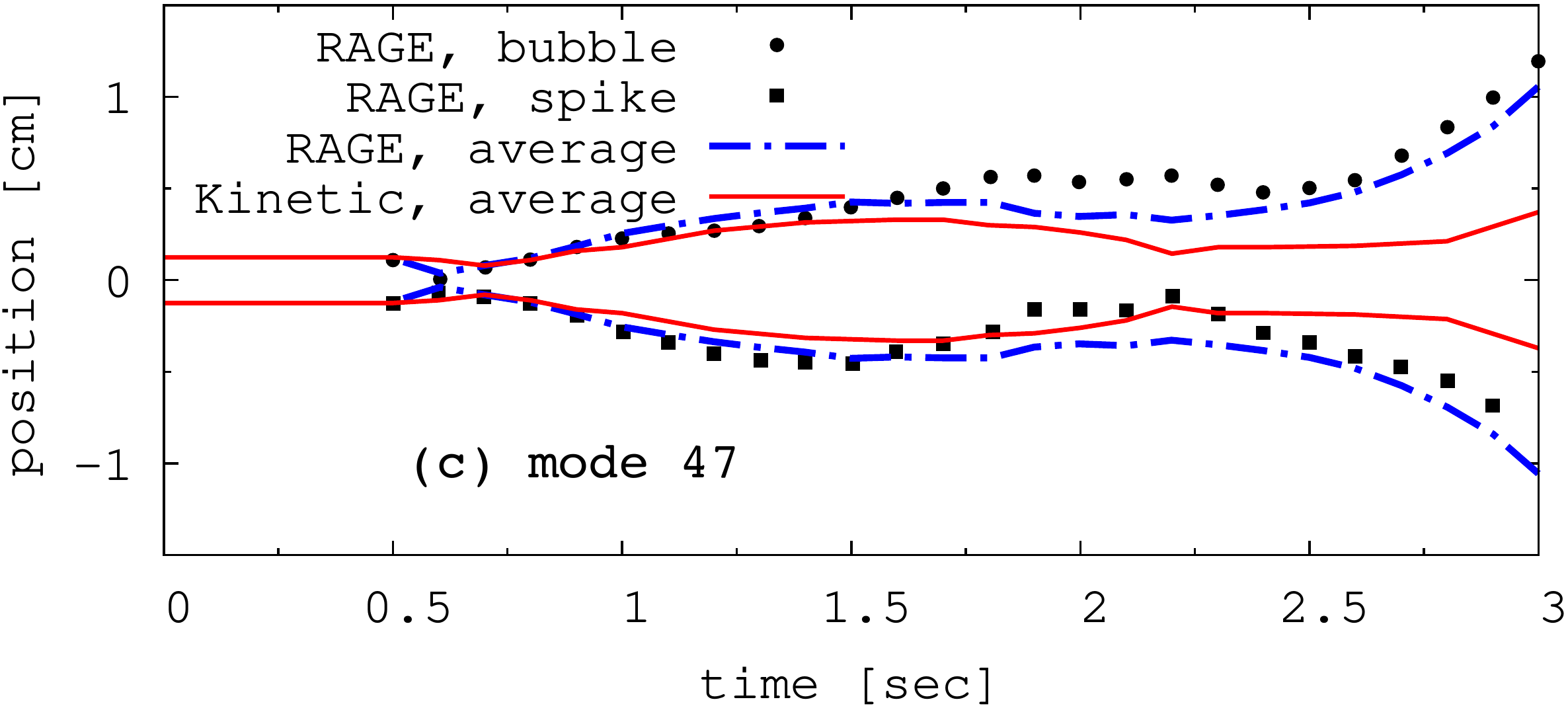}
\caption{Bubble ($>0\:$cm) and spike ($<0\:$cm) heights during the 3-zone implosion. Points show data taken from \cite{Joggerst2014}, where the bubble and spike heights are given with respect to the interface between zone-1 and 2 from a 1D simulation. In the kinetic simulations (solid line), we determine the fluid interface by averaging over the bubble and spike heights. For better comparison with \textsc{RAGE}, we perform a similar averaging for its data points (dashed-dotted line).}
\label{bubble_spike}
\end{center}
\end{figure}
why, for this setup, its agreement with \textsc{RAGE} is best.\\
\newline
To do a more qualitative comparison of the instabilities, we plot the bubble and spike heights in Fig.\:\ref{bubble_spike}. The values for \textsc{RAGE} are taken directly from Joggerst et al. \cite{Joggerst2014}. They are obtained by measuring the bubble and spike positions relative to the interface of zone-1 and 2 in an unperturbed 1D simulation. In the kinetic studies, we determine the location of the interface by averaging between the bubble and spike positions. The distance of the bubbles and spikes from that interface is given by: 
\begin{align}
r_i = \sum\limits_i^B {\left| \vec{r}_{bi} - \vec{r}_c \right|}/{2B} - \sum\limits_j^S {\left| \vec{r}_{sj} - \vec{r}_c \right|}/{2S}, 
\label{bubble}
\end{align}
where $\vec{r}_{bi / si}$ is the position of the $i$th bubble/spike, $\vec{r}_c$ the location of the disk center, $B$ the number of bubbles, and $S$ is the number of spikes. The resulting bubble/spike height is $r_{b/s} = \pm r_i$. Note that with these definitions the bubbles and spikes have the same amplitude while for \textsc{RAGE} they can be different. For a better comparison, we therefore determine average bubble and spike heights in \textsc{RAGE} via $\tilde{r}_i = \tilde{r}_b - 0.5 (\tilde{r}_b + \tilde{r}_s)$, with $r_{b/s} = \pm \tilde{r}_i$, $\tilde{r}_b$ and $\tilde{r}_{s}$ being the measured bubble and spike heights in Joggerst et al. \cite{Joggerst2014}.\\
For $t < 0.5 \:$s, i.e. before shock breakout, the instability heights are similar for \textsc{RAGE} and the kinetic simulations. For $\mu = 0$, they are negligibly small in \textsc{RAGE}, while the particle-caused granularities in the kinetic simulation lead to a finite but small width of the interface between zone-1 and 2. After shock breakout, the instabilities stay small initially. However, as the reflected shock interacts with the converging dense shell ($t \sim 2.1\:$s), they experience a drastic growth. For $\mu = 5$, the behavior is similar, although the bubble and spike growth sets in earlier, at around $t \sim 1.5\:$s. As mentioned before and can be seen in Fig.\:\ref{bubble_spike}, \textsc{RAGE} and the kinetic simulations agree best for $\mu=5$, while the largest differences are found for $\mu = 47$. Both, \textsc{RAGE} and the kinetic simulations see the same trends in instability growth and decrease, however the bubble/spike amplitudes are significantly smaller in the kinetic studies, especially for $t > 2.1\:$s. This is again due to the lower resolution in the kinetic code. While for \textsc{RAGE}, the $\mu=47$ instabilities are amplified by the interaction with the shock, in the kinetic simulations, their structure is much coarser early on and they form a layer that is more compressed than amplified by the reflected shock.\\
\newline
However, in general, the large-scale behavior and time-evolution of the 3-zone kinetic implosion simulation is very similar to \textsc{RAGE}. Differences are the finer details in \textsc{RAGE} and the occurrence of additional fluid instabilities in the kinetic code. Both can be traced back to the higher resolution in \textsc{RAGE} and the presence of statistical noise in the kinetic studies. These differences can be reduced by using a larger number of particles, however, most likely they cannot be completely eliminated.
\section{Summary}
\label{summary}
We present 2D implosion simulations with a kinetic Monte Carlo particle transport code. Its development is motivated by the existence of flows with different Knudsen numbers in a large range of physical phenomena, including inertial confinement fusion capsule implosions. Our code is not an attempt to improve hydrodynamic approaches for matter in the continuum regime. Its target application are systems that contain components at different Knudsen numbers which are usually modeled by different coupled methods. The performed tests in this paper include the implosion of disks with 2-zone and 3-zone configurations. The 2-zone setup is a simple test that we perform with a detailed analysis of the shock propagation and comparison to simulations using the \textsc{RAGE} hydrodynamics code. In the continuum limit, we find very good agreement between the kinetic and the hydrodynamic simulations. The kinetic studies also include simulations with different particle numbers and mean-free-paths to explore the impact of the latter on the implosion dynamics and non-equilibrium phenomena like particle velocity distributions and anisotropies. The 3-zone configuration contains a low-density central zone that is enclosed by a dense shell and an ablator. By imposing single-mode perturbations between the inner region and the dense shell, we induce fluid instabilities and compare their evolution to the corresponding hydrodynamic results. We find that the general dynamical evolution of the implosion agrees well between the \textsc{RAGE} and the kinetic code. Differences originate from the more detailed structures of the fluid instabilities in the hydrodynamic simulations and additional instabilities in the kinetic studies which are seeded by statistical noise. 
\acknowledgments{I.S. acknowledges support through a Director's fellowship from Los Alamos National Laboratory. The authors would also like to thank Mathieu Marciante, Terence J. Tarnowsky, and Joseph M. Smidt for helpful discussions and advice. This research used resources provided by the LANL Institutional Computing Program.} 
\appendix
\section{Kelvin-Helmholtz Instabilities}
\label{kelvinhelmholtz}
\begin{figure*}
\begin{center}
\includegraphics[width = 0.8\textwidth]{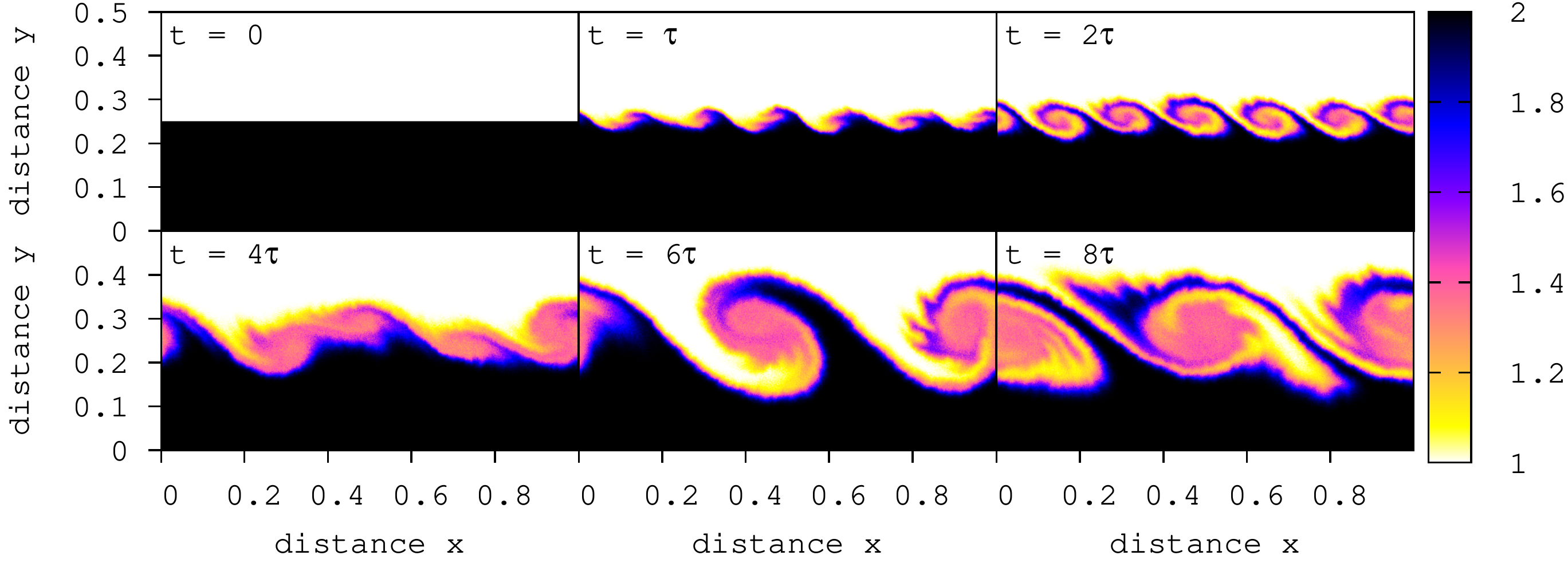}
\caption{Snapshots of the average particle species $s$ during the development of KHIs at different times. The simulation time is given by the characteristic growth time $\tau$. The KHIs are initialized by perturbations of the velocity field given by eq.(\ref{deltavy}).}
\label{kh}
\end{center}
\end{figure*}
To study the development of KHIs with our kinetic code in the continuum limit, we use the same configuration as D.\:J.\:Price \cite{Price08} who applied Smooth Particle Hydrodynamics (SPH) with artificial viscosity and thermal conductivity terms. Our simulation space has the size $0 \leq x \leq 1$ and $0 \leq y \leq 0.5$ and is divided into a lower ( $y < 0.25$) and upper ($y > 0.25$) half. The simulation grid has $4000 \times 2000$ cells with $500 \times 250$ cells for the output. The units in these simulations are given by the dimension of length $\hat{L}$, density $\hat{\rho}$, and pressure $\hat{P}$. As a consequence, the velocity is given in units of $(\hat{P}/ \hat{\rho})^{0.5}$ while time is given in units of $\hat{L} (\hat{\rho}/\hat{P})^{0.5}$ \cite{Sagert14}. The boundary conditions are periodic in x-direction and reflective in the y-direction. The entire space is filled with particles. In the lower half, the mass density is $\rho_2 = 2$ while for the upper half it is $\rho_1 = 1$. The pressure is $P = 2.5$ everywhere. Particles in the lower half have a net x-velocity $v_{x,2} = 0.5$ while particles in the upper half stream in the opposite direction with $v_{x,1} = -0.5$. In addition to their masses, particles have a characteristic species type $s$. We set $s = 1$ for particles in the upper half and $s=2$ for particles in the lower half. To create seeds for single-mode KHIs, we perturb the y-velocity with
\begin{align}
\delta = A  \sin \left( \pm \frac{ 2 \pi (x + 0.5)}{l} \right),   \mathrm{for} \: | y \pm 0.25| < 0.025
\label{deltavy}
\end{align}
where $A = 0.025$ is the amplitude and $l = 1/6$ the wavelength of the single-mode instabilities. Linear theory predicts a characteristic growth time $\tau$ \cite{Price08}:
\begin{align}
\tau = l \: \frac{ \rho_1 + \rho_2}{\sqrt{\rho_1 \rho_2} \: | v_{x,1} - v_{x,2} |} . 
\label{tau}
\end{align}
Since $\tau$ is proportional to $l$, small-scale KHIs will appear first followed by instabilities with larger wavelength. Fig.\:\ref{kh} shows 2D snapshots of the species distribution at different simulation times. The emerging structures can be compared to case 4 and 5 in Fig.\:7 of \cite{Price08}. At $t = \tau$, we see the onset of KHIs with $\lambda = 1/6$, with fully developed vortices at $t = 2 \: \tau$. The time scales and shapes of these instabilities are in good agreement with the SPH results. As expected, the KHIs merge into instabilities with $\lambda = 1/2$ by $t = 6 \: \tau$. D.\:J.\:Price \cite{Price08} shows results for $t = 1\:, 2\:, 4$ and $8 \: \tau$. The last snapshot for case 4 agrees with our results at $t = 8 \: \tau$ while case 5 it is more similar to our KHIs at $t = 6 \: \tau$. The difference between the two SPH cases is the usage of additional thermal conductivity and artificial viscosity terms in case 5 which might impact details in the KHI evolution. However, we find that the general agreement between KHIs in the kinetic simulation and the SPH study is good.
\bibliography{ref}
\end{document}